\documentstyle[preprint,aps,tighten]{revtex}

\begin{document}


\def \INFZ{{\rm Inf}[(\gamma/Z)_{\rm L},\nu]}
\def \SSQW{\sin^2\theta_W}
\def \SSQWSQ{\sin^4\theta_W}
\def \a{{+}}
\def \b{{-}}
\def \AB{$({+}{-})$-maximized}
\def \A00B{$[({+}{0})+({0}{-})]$-maximized}
\def \ZHtrans{$ZH$-transverse}
\def \ZHlong{$ZH$-longitudinal}
\def \BETA{\beta_Z}
\def \GAMMA{\gamma_Z}
\def \GINV{\sqrt{1{-}\BETA^2}}
\def \ts{\thinspace}
\def \longitudinal{longitudinal}
\def \amp{{\cal A}}
\def \Ihat{\widehat{\cal I}}
\def \thetas{\theta^{*}}
\def \T{{\cal S}}
\def \G{{\cal G}}
\def \eebar{e^{+}e^{-}}
\def \WWbar{W^{+}W^{-}}
\def \ginv{\sqrt{1{-}\beta^2}}
\def \GeV{{\rm \enspace GeV}}
\def \beq{\begin{equation}}
\def \eeq{\end{equation}}
\def \beqa{\begin{eqnarray}}
\def \eeqa{\end{eqnarray}}
\def \cxi{\cos\xi}
\def \sxi{\sin\xi}
\def \cth{\cos\theta^{*}}
\def \sth{\sin\theta^{*}}
 
\def	\bra	{\langle}         
\def 	\ket	{\rangle}
\def	\sp	#1#2{\mbox{$\bra #1 #2 \ket$}}    
\def	\cp	#1#2{\mbox{$  [  #1 #2 ]$}}       
\def	\rsp	#1{\mbox{$\,\vert #1 \ket$}}      
\def	\lsp	#1{\mbox{$\bra  #1 \!\vert\,$}}   
\def	\sppr	#1#2{\mbox{$\bra #1 \!\vert #2 \ket$}}  


\draft
\preprint{
  \parbox{2in}{Fermilab--Pub--98/077-T \\
  McGill/98--6 \\
  hep-ph/9803410
}  }

\title{Deconstructing Angular Correlations in\\
$ZH$, $ZZ$, and $WW$ Production at LEP2}
\author{Gregory Mahlon  \cite{GDMemail}}
\address{Department of Physics, McGill University, \\
3600 University St., Montr\'eal, QC  H3A 2T8 \\
Canada }
\author{Stephen Parke \cite{SPemail}}
\address{Department of Theoretical Physics\\
Fermi National Accelerator Laboratory \\
P.O. Box 500, Batavia, IL 60510 \\
USA }
\date{March 20, 1998}
\maketitle
\begin{abstract}
We apply a generalized spin-basis analysis to associated Higgs 
production and gauge boson pair production at LEP.  
This framework allows us to identify a choice of spin axes for 
the processes $e^{+}e^{-}\rightarrow ZH,ZZ$ which leads to 
strikingly different correlations among the decay products, even 
well above threshold.
This spin basis optimizes the difference in the angular
correlations for these two processes.
In contrast, the same distributions display little contrast when 
the helicity basis is used.  We also apply this technique to the 
case of $W$ boson pair production.
\end{abstract}


\section{Introduction}

The search for a light Higgs boson and
the study of gauge boson pair production~\cite{yellow}
are important physics goals of the LEP2 upgrade at CERN.
If kinematically accessible to LEP2, the Higgs boson will 
be primarily produced via the Bjorken 
process $e^{+} e^{-} \rightarrow ZH$\cite{ZHstd,Barger}.
An irreducible physics background to this processes is the
production of $Z$ boson pairs, 
which is most troublesome when $M_H \sim M_Z$.
The clear separation of these two processes is required for a convincing
Higgs boson discovery or lower bound mass limit.
Kinematics as well as characteristic 
angular correlations will be useful in this separation.
The study of $W$ boson pair production  is also interesting
in its own right\cite{Gunion,Hagiwara,anomal}, as it involves
vertices coupling three gauge
bosons (especially $ZW^+W^-$), vertices which to date are  poorly-probed
experimentally.  
These couplings may be probed with varying degrees of sophistication
by testing the predictions for the 
total cross section, production angle distribution, and correlations
among the vector boson decay products.  
Hence for all of these processes, 
$e^{+}e^{-}\rightarrow ZH,ZZ$ and $W^+W^-$,
we are keenly interested in the angular correlations among 
the final state particles. 
To disentangle these correlations it is very useful to 
understand the spin correlations of the heavy particles produced.
These spin correlations as well as the subsequent
correlations of the decay products are the focus of this paper.

Until recently, most spin-related studies were carried out
within the framework provided by the zero momentum frame helicity basis
(see, for example Refs.~\cite{Barger,Gunion,Hagiwara,anomal}).
For particles which are ultra-relativistic, 
such as at the next generation of linear colliders,
this is appropriate.
However, for particles which are only moderately relativistic,
there is no reason to expect that the helicity basis will
produce the best description of the physics involved.
Indeed, for the case of $t\bar{t}$ production at low energy
$p\bar{p}$ and $e^{+}e^{-}$ colliders, it was shown in 
Refs.~\cite{ttbar1,xibasis,ttbar2} that the helicity basis
is far from optimal.  
In this paper we use the generalized
spin basis introduced in Ref.~\cite{xibasis} to describe
$ZH$, $ZZ$, and $WW$ production and decay at LEP2 energies.
In particular, we write the polarized production amplitudes 
in terms of the most general $CP$-conserving choice of
spin axes, and from these expressions determine the choice
of axes which leads to the most distinctive correlations
among the final state particles.  
We argue that it is advantageous
at LEP2 energies to tune the choice 
of spin axis to the experimental issue
being investigated, rather than to simply use the helicity
basis and try to unravel the contributions from what are in
many cases nearly-equally populated spin states.

For all three processes,
$e^{+}e^{-}\rightarrow ZH,ZZ$ and $W^+W^-$,  
much effort in the literature has been devoted to the
analysis of various anomalous trilinear 
couplings\cite{Hagiwara,anomal,industry},
employing, of course, the helicity basis.
Rather than attempt to rework all of those studies in
terms of the generalized spin-basis, we
have chosen to concentrate on the picture within the Standard
Model.  It is clear from our results, however, that such
a study would be worthwhile, as we expect different spin
bases to be optimal for the extraction/limitation of different
anomalous couplings.  Such an analysis would be a major
extension of this work.

The remainder of this paper is organized as follows.  After
a brief discussion of our notation and conventions 
in Sec.~\ref{XIbasis}, we consider the $ZH/ZZ$ system
within the generalized spin basis framework (Sec.~\ref{ZHandZZ}).
For each process, we consider in detail the polarized
production cross sections, and construct spin bases suggested
by the expressions for the amplitudes.  We consider
the decay of polarized $Z$'s, and then link the production
and decay to study the correlations among the decay products.
We are able to construct a basis in which these correlations
differ significantly for the $ZH$ and $ZZ$ cases, whereas
in the helicity basis the correlations are virtually non-existent.
Sec.~\ref{WWsection} contains the corresponding analysis
for the $W^{+} W^{-}$ system.  In this case, matters are
less clear-cut, and we present several suggestions for
bases which may be useful under different circumstances.
Finally, Sec.~\ref{CONC} contains our conclusions.
Explicit expressions for the polarization vectors we
employ for the massive vector bosons in the generalized
spin basis appear in the Appendix.


\section{Notation and Conventions}\label{XIbasis}

To describe the polarized production cross sections  for $ZH$, $ZZ$,
and $WW$
discussed in this paper,
we employ the generic spin basis
introduced by Parke and Shadmi in Ref.~\cite{xibasis},
as illustrated in Fig.~\ref{XIdef}.  
For $ZZ$ and $WW$, this is the most general basis which 
conserves $CP$.
We label the two particles produced in the collision by
$P_1$ and $P_2$.  The zero momentum frame (ZMF) production angle
$\thetas$ is defined as
the angle between the electron and $P_1$ directions.
The spin states for the first particle 
are defined in the its rest frame,
where we decompose the $P_1$ spin along the direction
$\hat{s}_1$, which makes an angle $\xi$ with the $P_2$
momentum in the clockwise direction.
Likewise, the $P_2$ spin states are defined in 
its
rest frame along the direction $\hat{s}_2$, which
makes the {\it same}\ angle $\xi$ with the $P_1$ momentum, also
in the clockwise direction.  
We denote the two transverse polarization states by $\a$ and $\b$
and the \longitudinal\ state by 0.  Throughout this paper 
we use the terms ``transverse''
and ``longitudinal''  to refer to directions relative to the
spin axis, {\it not}\ the direction of motion of the particle.
A generic vector boson spin will be designated by $\lambda$.
If we sum over all of the polarizations of the
vector boson(s), then the dependence on $\xi$ drops out
of the result.

Note that although the Higgs boson is a scalar, we may still
define a spin axis for it as if it were a vector. 
The spin zero character of the Higgs will be reflected in a lack
of any dependence on the choice of this axis.

Within the generic spin basis framework, specific spin bases
are defined by stating the relationship between $\xi$ and $\thetas$
(and any other relevant event parameters).  For example,
the ubiquitous helicity basis is defined by fixing
\beq
\xi \equiv \pi.
\label{HELICITYdef}
\eeq
In this case, the spins are defined along the directions of
motion of the particles as seen in the ZMF.

Another interesting basis is the beamline basis~\cite{ttbar1},
which is defined by
\beq
\sin\xi = { { \ginv\sin\thetas }  \over { 1-\beta\cos\thetas} };
\qquad
\cos\xi = { {\cos\thetas - \beta} \over { 1-\beta\cos\thetas} }.
\label{BMLdef}
\eeq
Here $\beta$ is the ZMF speed of $P_1$.
In this basis, the spin axis for $P_1$ is the electron direction.
Furthermore, if $P_1$ and $P_2$ have identical mass, then the
spin axis for $P_2$ is the positron direction.

Later in this paper, we will encounter additional bases, whose
definitions are inspired by the form of the matrix elements
for the processes under consideration.

Except for the fermion masses, which we set equal to zero,
all input masses and coupling constants used in the computations
presented in this paper are the central values as reported
in the 1996 Review of Particle Properties\cite{PDB96}.
Furthermore, we also neglect the coupling between the electron and 
Higgs.


\section{The Processes {$\protect\lowercase{e}^{+}
\protect\lowercase{e}^{-}
\longrightarrow  ZH,ZZ$} at LEP}\label{ZHandZZ}

If the Higgs is light enough to be observed at LEP2,
then its production will be dominated by the process
$\eebar \rightarrow ZH$, for which the largest background
is $\eebar \rightarrow ZZ$.  This suggests the potential
for difficulties should the Higgs mass lie too close to the
mass of the $Z$.  In particular, at $\sqrt{s}=192\GeV$,
the tree-level cross section for $ZH$
is only 0.5 pb when $M_H=M_Z$, compared to 1.2 pb for $ZZ$.
As noted by Brown~\cite{Brown}, this situation improves 
if $b$-tagging is employed, as the Higgs decays mainly
to $b\bar{b}$ whereas the $Z$ decays to this final state only
15\% of the time.
Rather than rely on $b$-tagging,
Kunszt and Stirling~\cite{almost} have considered the distribution
of the angle that the decay leptons  in
$\eebar \rightarrow ZH/ZZ \rightarrow \ell\bar\ell\ts\ts {\rm jets}$.
make with the direction of the beam in the ZMF.
They find a significant difference in this distribution in
the $ZH$ and $ZZ$ cases.  Finally, Summers~\cite{Summers}
has added the effect of polarizing the beams on the
distribution of Ref.~\cite{almost}.

In this section we examine the $ZH$ and $ZZ$ processes within
the generalized spin basis framework.  We begin in 
Sec.~\ref{polZHprod} by analyzing the
polarized production
amplitudes for the process $\eebar\rightarrow ZH$ in great
detail.  The expressions for these
amplitudes will show us how to construct a basis in
which only transversely-polarized $Z$'s are produced in association
with the Higgs. In Sec.~\ref{polZZprod} we present
polarized production amplitudes for 
$\eebar\rightarrow ZZ$.   After describing the
decay of polarized $Z$'s (Sec.~\ref{polZdec}), we combine
the production and decay information to discuss the
angular correlations among the decay products in the $ZH/ZZ$
system (Sec.~\ref{ZZcorrel}). 
Our analysis shows that the angular variable suggested by Kleiss and
Stirling~\cite{almost}, while close to optimal, may be
improved upon, especially as the $ZH$ pair is produced further
and further above threshold.
In particular, we find 
that in the basis
constructed in Sec.~\ref{polZHprod}, certain decay angular
distributions are very different for $ZH$ versus $ZZ$.
In contrast, these same distributions are nearly featureless
in the helicity basis.  
We conclude this part of the paper in Sect.~\ref{pseudo}
with a brief look at how the production of a pseudoscalar
Higgs would differ from the Standard Model Higgs.

Although we focus on the case where $M_H\sim M_Z$, our results
are more general, since  the existence of our optimized
basis for $\eebar \rightarrow ZH$ depends only on
that process being kinematically allowed.  It is also 
potentially useful at the Tevatron, because the amplitude
for $q\bar{q}^\prime \rightarrow WH$ has the same spin
structure as the amplitude for $\eebar\rightarrow ZH$.


\subsection{Polarized $ZH$ Production}\label{polZHprod}

To describe the process $\eebar\rightarrow ZH$, we
take particle $P_1$ in Fig.~\ref{XIdef} to be the $Z$,
and particle $P_2$ to be the Higgs.
In general, the Higgs and $Z$ have different masses
($M_H$ and $M_Z$ respectively),
leading to the following connection between the center-of-mass
energy $\sqrt{s}$ and speed $\BETA$ of the $Z$ boson:
\beq
\BETA =
{ 
\sqrt{[s-(M_Z+M_H)^2][s-(M_Z-M_H)^2]}
  \over
{s-M_H^2+M_Z^2}
}.
\label{betaZ}
\eeq
Rather than use Eq.~(\ref{betaZ}) to eliminate one of $s$ or $\BETA$
from our expressions, it is more convenient to use both quantities.

If we neglect the electron mass, there is but a single
diagram for $\eebar\rightarrow ZH$.  It leads to the
differential cross
section\footnote{All cross sections given in this paper
are spin-averaged for the incoming particles.}
\beqa
{
{ d\sigma^{\lambda}(\eebar{\rightarrow}ZH) }
\over
{ d(\cos\thetas) }
} = 
{   {G_F^2 M_W^2} \over {\cos^2\theta_W}   } \ts &&
{   {\BETA\GAMMA M_Z^3} \over {32\pi s^{3/2}}   } \ts
\Biggl(
{  {s+M_Z^2-M_H^2} \over {s-M_Z^2}  }
\Biggr)^2
\cr && \times\biggl\{
 \cos^2 2\theta_W \ts
 [\T_L^{\lambda}(\BETA,\thetas,\xi)]^2
+ 4 \sin^4 \theta_W \ts
 [\T_R^{\lambda}(\BETA,\thetas,\xi)]^2
\biggr\}
\label{ZHsigma}
\eeqa
where $G_F$ is the Fermi coupling constant, $M_W$ is the mass
of the $W$ boson, $\theta_W$ is the
Weinberg angle, and $\GAMMA$ is the usual relativistic factor,
$\GAMMA = (1-\BETA^2)^{-1/2}$.  The two terms in the curly
brackets come from the two possible chiralities of the initial
electron line.
All of the spin information
is contained in the spin functions, $\T_{L,R}^\lambda$, 
which are given by
\beq
\T_L^{\pm}(\BETA,\thetas,\xi) = \T_R^{\mp}(\BETA,\thetas,\xi)
=  {1\over{\sqrt{2}}} \Bigl[ \sth\sxi + \GINV(\cth\cxi \pm 1) \Bigr] 
\label{ZHtransS}
\eeq
and
\beq
\T_L^{0}(\BETA,\thetas,\xi) = \T_R^{0}(\BETA,\thetas,\xi) 
= \GINV\cth\sxi - \sth\cxi.
\label{ZHlongS}
\eeq
Summing over the three possible spins of the $Z$ we obtain the
total (unpolarized) differential cross section
\beqa
\sum_{\lambda}
{
{ d\sigma^{\lambda} }
\over
{ d(\cos\thetas) }
} = 
{   {G_F^2 M_W^2} \over {\cos^2\theta_W}   } \ts &&
{   {\BETA\GAMMA M_Z^3} \over {32\pi s^{3/2}}   } \ts
\Biggl(
{  {s+M_Z^2-M_H^2} \over {s-M_Z^2}  }
\Biggr)^2
\cr && \times\biggl\{
1-4\sin^2\theta_W+8\sin^4\theta_W
\biggr\}\biggl\{
2(1-\BETA^2) + \BETA^2\sin^2\thetas
\biggr\}.
\label{ZHsigmaTOTAL}
\eeqa
Eq.~(\ref{ZHsigmaTOTAL}) is independent of the spin
axis angle $\xi$, as it must be.

Since, in general,
at each value of $\thetas$ we are allowed to choose a different
value of $\xi$, it is useful to plot these amplitudes in the 
$\cos\thetas$-$\cos\xi$ plane.  Consequently,
we define the 
quantity
\beq
f^{\lambda}(\BETA,\thetas,\xi) \equiv
{
\displaystyle{ {d\sigma^{\lambda}(\BETA,\thetas,\xi)}
    \over{d(\cos\thetas)} }
\over
{ \displaystyle\sum_{\lambda^\prime }
 \displaystyle{ {d\sigma^{\lambda^\prime}(\BETA,\thetas,\xi)
 }\over{d(\cos\thetas)} } }
},
\label{ZHfraction-def}
\eeq
which is the fraction of the total cross section coming from
the spin state $\lambda$.  
In Fig.~\ref{ZH-PRODcontours}
we have plotted
$f^{\lambda}(\BETA,\thetas,\xi)$ 
for a collider energy $\sqrt{s}=192\GeV$ and Higgs mass $M_H = M_Z$.
These plots illustrate the features of the amplitude discussed below,
and are an indispensable aid when the amplitude becomes complicated,
as in the $ZZ$ and $WW$ cases.

For $\xi = \pi$, we recover the usual helicity basis expressions
from Eqs.~(\ref{ZHsigma})--(\ref{ZHlongS}):
\beqa
\T_L^{\pm}(\BETA,\thetas,\pi) = \T_R^{\mp}(\BETA,\thetas,\pi)
= && {1\over{\sqrt{2}}} \GINV(\pm 1 - \cth) ; \cr
\T_L^{0}(\BETA,\thetas,\pi) = \T_R^{0}(\BETA,\thetas,\pi) 
= && \sth.
\eeqa
At high energy, the transverse amplitudes die off, leaving
only the longitudinal amplitude.

An examination of Eq.~(\ref{ZHtransS}) reveals
that it is not possible to make both of the transverse spin functions
$\T_L^\lambda$ and $\T_R^\lambda$
vanish simultaneously.
Consequently, it is impossible to make either of the transverse
amplitudes of Eq.~(\ref{ZHsigma}) vanish.  
There is, however, a zero in the longitudinal
contribution, Eq.~(\ref{ZHlongS}), which results from choosing
\beq
\sxi = { {\sth}\over\sqrt{1-\BETA^2\cos^2\thetas} }; \qquad
\cxi = { {\GINV\cth}\over\sqrt{1-\BETA^2\cos^2\thetas} },
\label{ZHideal}
\eeq
{\it i.e.} $\tan\xi = \GAMMA\tan\thetas$.
The basis corresponding to this choice, the \ZHtrans\ basis,
will turn out to be very useful in the study of $ZH$ events.
The nonvanishing spin functions in the
\ZHtrans\ basis are
\beqa
\T_L^{\pm}\Bigl(\BETA,\thetas,\tan^{-1}(\GAMMA\tan\thetas)\Bigr)
&=& \T_R^{\mp}\Bigl(\BETA,\thetas,\tan^{-1}(\GAMMA\tan\thetas)\Bigr)\cr
&=& {1\over{\sqrt{2}}}\Biggl[ 
\sqrt{1-\BETA^2\cos^2\thetas} \pm \sqrt{1-\BETA^2}
\Biggr].
\eeqa
These functions are completely flat in $\cos\thetas$ at
threshold, and become proportional to $\sin\thetas$ for
$\beta_Z \rightarrow 1$.
Note that the existence of
this basis does not depend on either the machine energy or
the Higgs mass:  Eqs.~(\ref{betaZ}) and~(\ref{ZHideal}) remain
well-defined so long as $\sqrt{s} \ge M_Z + M_H$.  In particular,
for $\BETA \rightarrow 1$, we have
\beq
\sxi \rightarrow 1; \qquad \cxi \rightarrow 0,
\label{ZHhighE}
\eeq
that is, $\xi \rightarrow \pi/2$.  This is clearly {\it not}\ the
helicity basis.\footnote{The absence of any \longitudinal\ contribution
whatsoever, even at high energy, does not represent a violation
of the vector boson
equivalence theorem, which is really  a statement about the
helicity basis.} 
The \ZHtrans\ basis is also the basis in which the $+$ and $-$
components are each maximized. 

For completeness, we mention the \ZHlong\ basis, defined by
$\tan\xi = - \GAMMA^{-1}\cot\thetas$.  As its name suggests, 
the \ZHlong\ basis maximizes the fraction of \longitudinal\ $Z$'s.
This basis is potentially useful for large values of $\beta_Z$
since the fraction of \longitudinal\ $Z$'s
with increasing energy approaches 
unity faster than in the helicity basis.

In Table~\ref{ZH-Breakdowns}, we list the contributions to the
total amplitude for each of the three possible $Z$ spins as
measured in the helicity, beamline, \ZHtrans, and \ZHlong\
bases.  We employ a center-of-mass energy $\sqrt{s} = 192 \GeV$
and a Higgs mass $M_H = M_Z$, the value which makes separation
of $ZH$ and $ZZ$ events the most difficult.  Note that in the
helicity basis, the three spin states are populated nearly 
equally.  As we shall see below, this feature makes the helicity
basis a very poor choice for studying the $ZH/ZZ$ system. 

Because the Higgs mass is still unknown, 
we might worry that the predictions for the spin compositions
would depend strongly on $M_H$, rendering this discussion 
pointless.
For
any particular machine energy and Higgs mass, Eq.~(\ref{betaZ})
tells us the appropriate value of $\BETA$ to use.  In
Fig.~\ref{ZHbetaplot} we have plotted the fractional contribution
of each of the three spin components as a function of $\BETA$
for the helicity, beamline, \ZHtrans, and \ZHlong\ bases.
We see that in the \ZHtrans\ basis, the spin composition is
remarkably flat as a function of $\BETA$, except near $\BETA = 1$.
In fact, for $M_H \agt 70 \GeV$,
$\BETA \alt  0.5$ when $\sqrt{s} = 192 \GeV$.
So, for the values of $M_H$ and $\sqrt{s}$ accessible at LEP,
the spin composition in the \ZHtrans\ basis is effectively
constant.  Contrast this to the beamline basis, which actually
coincides with the \ZHtrans\ basis at $\BETA=0$ and the helicity
basis for $\BETA\rightarrow 1$!  In this case, the spin composition is not
particularly stable.

A second reason for preferring a basis in which the spin composition
is stable with respect to changes in $\BETA$ 
relates to the effects
of initial state radiation (ISR).
Although we have neglected these effects
in our computations, it is apparent what should happen
qualitatively:
some of the observed events will 
be produced from $\eebar$ pairs with a center-of-mass
energy lower than the machine energy.  Therefore, the smaller
the dependence on $\BETA$, the less sensitive the breakdowns
will be to the effects of ISR.

Finally,
in Fig.~\ref{ZH-CTH} we compare the production angular distributions
of the polarized cross sections in the helicity
and \ZHtrans\ bases.  We see that in the helicity basis, the
identity of the dominant component depends on $\cos\thetas$,
whereas in the \ZHtrans\ basis, the ratio of the two non-vanishing
amplitudes is essentially constant.


\subsection{Polarized $ZZ$ Production}\label{polZZprod}

We now turn to the process $\eebar\rightarrow ZZ$.
With our comparison to $ZH$ production in mind, we 
identify the $Z$ which decays leptonically with $P_1$ and the $Z$ which 
decays to $b\bar{b}$ with $P_2$ in Fig.~\ref{XIdef}.

The only diagrams for $ZZ$ production are $t$-channel electron exchange 
diagrams, one for each ordering of the two $Z$'s.
For an initial fermion pair with left-handed chirality we have
\beqa
{
{ d\sigma_L^{\lambda\bar\lambda}(\eebar{\rightarrow}ZZ) }
\over
{ d(\cos\thetas) }
}&& = 
{ {G_F^2 M_W^4} \over { 32\pi M_Z^2 } } \ts
{ {\cos^4 2\theta_W} \over {16\cos^4 \theta_W} } \ts
\beta\gamma^2 \ts
\Biggl[
{
{\T^{\lambda\bar\lambda}_L(\beta,\thetas,\xi)}
\over
{1{-}2\beta\cos\thetas{+}\beta^2}
} + {
{\T^{\lambda\bar\lambda}_L(-\beta,\thetas,\xi)}
\over
{1{+}2\beta\cos\thetas{+}\beta^2}
}
\Biggr]^2,
\label{prodZZleft}
\eeqa
while the result for right-handed chirality reads
\beqa
{
{ d\sigma_R^{\lambda\bar\lambda}(\eebar{\rightarrow}ZZ) }
\over
{ d(\cos\thetas) }
}&& = 
{ {G_F^2 M_W^4} \over { 32\pi M_Z^2 } } \ts
{ {\sin^8 \theta_W} \over {\cos^4 \theta_W} } \ts
\beta\gamma^2 \ts
\Biggl[
{
{\T^{\lambda\bar\lambda}_R(\beta,\thetas,\xi)}
\over
{1{-}2\beta\cos\thetas{+}\beta^2}
} + {
{\T^{\lambda\bar\lambda}_R(-\beta,\thetas,\xi)}
\over
{1{+}2\beta\cos\thetas{+}\beta^2}
}
\Biggr]^2.
\label{prodZZright}
\eeqa
These expressions contain the common ZMF speed of the
two $Z$'s, which is connected to the center-of-mass energy by
\beq
\beta  = \sqrt{1-4M_Z^2/s}.
\label{plainBETA}
\eeq
We have not combined the pairs of terms appearing in the square
brackets 
in order to make manifest the similarities to the $W$-pair
amplitudes presented in Sec.~\ref{WWsection}.

All of the spin information is contained in the 
spin functions, $\T_{L,R}^{\lambda\bar\lambda}$, which may
be written as
\beq
\T_L^{\a \b}(\beta,\thetas,\xi) =
2\G_1(\beta,\thetas,\xi)\G_2(\beta,\thetas,\xi)
\label{funcAB}
\eeq
\beq
\T_L^{\a \a}(\beta,\thetas,\xi) =
2\G_3(\beta,\thetas,\xi)\G_4(\beta,\thetas,\xi)
\label{funcAA}
\eeq
\beq
\T_L^{\a {0}}(\beta,\thetas,\xi) =
\T_L^{{0} \b}(\beta,\thetas,\xi) =
\sqrt{2}\bigl[\G_1(\beta,\thetas,\xi)\G_4(\beta,\thetas,\xi)
             -\G_2(\beta,\thetas,\xi)\G_3(\beta,\thetas,\xi) \bigr]
\label{funcAOOB}
\eeq
\beq
\T_L^{{0}{0}}(\beta,\thetas,\xi) =
2\beta\sin\thetas(1-2\beta\cth+\beta^2)
-4\G_3(\beta,\thetas,\xi)\G_4(\beta,\thetas,\xi)
\label{funcOO}
\eeq
where
\beqa
\G_1(\beta,\thetas,\xi) &=& 
    (1+\cos\thetas\cos\xi+\ginv\sin\thetas\sin\xi) 
     - \beta(\cos\thetas+\cos\xi)
\cr
\G_2(\beta,\thetas,\xi) &=& 
    (\sin\thetas\cos\xi - \ginv\cos\thetas\sin\xi)
     +\beta(\sin\thetas+\ginv\sin\xi)
\cr
\G_3(\beta,\thetas,\xi) &=& 
     \sin\xi(\beta - \cos\thetas) + \ginv\sin\thetas\cos\xi
\cr
\G_4(\beta,\thetas,\xi) &=& 
    \sin\thetas\sin\xi - \ginv\cos\xi(\beta-\cos\thetas).
\label{Gfuns}
\eeqa
Since the replacement $\xi\rightarrow\xi{+}\pi$
has the effect of interchanging the $\a$ and $\b$ states,
expressions for the remaining spin combinations may be obtained
from the relation
\beq
\T_L^{\lambda\bar\lambda}(\beta,\thetas,\xi)
 = \T_L^{-\lambda,-\bar\lambda}(\beta,\thetas,\xi{+}\pi).
\label{leftOTHER}
\eeq
The total differential cross section is remarkably simple.
Summing over the spins of both $Z$'s and including both
fermion chiralities we obtain
\beqa
\sum_{\lambda,\bar\lambda,C}
{
{ d\sigma_C^{\lambda\bar\lambda} }
\over
{ d(\cos\thetas) }
}=  &&
{ {G_F^2 M_W^4} \over { 16\pi M_Z^2 \cos^4\theta_W } } \ts
(\cos^42\theta_W + 16\sin^8 \theta_W)  \ts
\cr && \times
\beta(1-\beta^2) \ts
{
{  2(1-\beta^2)^3 + \beta^2[8+(1-\beta^2)^2] - 4\beta^4\sin^4\thetas }
\over
{ [(1-\beta^2)^2 + 4\beta^2\sin^2\thetas]^2 }
}.
\label{ZZtotal}
\eeqa

Of the nine spin configurations, there are only six which are
independent, since $CP$ invariance forces 
the equality of the $(++)$ and $(--)$ components, as
well as the equality of the $(\pm 0)$ and $(0 \mp)$ 
components~\cite{Hagiwara}.
Therefore, for the rest of the paper we will refer to $[(++)+(--)]$,
$[(+0)+(0-)]$, and $[(0+)+(-0)]$ as single components.

One of the ways which
the $C$ and $P$ violation present in the weak interactions
manifests itself as the different prefactors in 
Eqs.~(\ref{prodZZleft}) and~(\ref{prodZZright}).
The spin functions, on the other hand, obey these symmetries, and
are connected by
\beq
\T_R^{\lambda\bar\lambda}(\beta,\thetas,\xi)
 = \T_L^{-\lambda,-\bar\lambda}(\beta,\thetas,\xi).
\label{leftright}
\eeq

Because of the somewhat complicated form of the amplitudes,
it is instructive to plot the amplitudes for
each of the six independent spin
configurations in the $\cos\thetas$-$\cos\xi$ plane.
In Fig.~\ref{ZZ-PRODcontours}, we have plotted the 
quantity\footnote{Implicit in this definition is a sum 
(in both numerator and denominator)
over the chirality of the initial electron line.}
\beq
f^{\lambda\bar\lambda}(\beta,\thetas,\xi) \equiv
{
\displaystyle{ {d\sigma^{\lambda\bar\lambda}(\beta,\thetas,\xi)}
    \over{d(\cos\thetas)} }
\over
{ \displaystyle\sum_{\lambda^\prime \bar\lambda^{\prime} }
  \displaystyle{ 
     {d\sigma^{\lambda^\prime\bar\lambda^\prime}(\beta,\thetas,\xi)
 }\over{d(\cos\thetas)} } }
}
\label{fraction-def}
\eeq
({\it i.e.}\ the fraction of the
total amplitude in a given spin configuration)
for a machine center-of-mass energy of 192 GeV.
Visible in the plots 
is the fact that the $(00)$ contribution
is  exactly twice the $[(++)+(--)]$ contribution for all
values of $\thetas$ and $\xi$ [see Eqs.~(\ref{prodZZleft}),
(\ref{funcAA}) and~(\ref{funcOO})].
Also noteworthy are the broad minima in the
$(+-)$, $(-+)$, $(00)$, and $[(++)+(--)]$ components 
in the vicinity of the diagonal.\footnote{None
of these components actually vanish in this region.}
The $[(+0)+(0-)]$
and $[(0+)+(0-)]$ contributions have corresponding
maxima in this region.

In Table~\ref{ZZ-Breakdowns} we give the fraction of the total
amplitude coming from each of the six independent
spin configurations in the four bases tabulated for $ZH$.
In two of the bases (helicity and \ZHlong), the high
degree of symmetry present in Eqs.~(\ref{prodZZleft})
and~(\ref{prodZZright}) resulting from the two identical
final state particles manifests as the equalities  $(+-) = (-+)$
and $[(+0)+(0-)]=[(0+)+(-0)]$.
It is important to note that we {\it define}\ the \ZHtrans\ basis 
in terms of Eqs.~(\ref{betaZ}) and~(\ref{ZHideal}),
even when we are dealing with $ZZ$ production.  Thus, the
spin breakdowns in this basis depend on the Higgs mass
via the value assigned to $\beta_Z$.   For $\beta_Z \alt 0.5$
(the interesting region), the curve traced out by
the \ZHtrans\ basis in the 
$\cos\thetas$-$\cos\xi$ plane lies near the diagonal,
an area in the $\cos\thetas$-$\cos\xi$ plane where
the contributions from the various spins are relatively stable.
Thus, it is not surprising to find that for $\beta_Z \alt 0.5$
and $\sqrt{s} = 192 \GeV$, there are no significant deviations
from the entries for the \ZHtrans\ basis
in Table~\ref{ZZ-Breakdowns}.

In Fig.~\ref{ZZbetaplot} we have have plotted the fraction
of the total amplitude in each of the spin components
using the helicity and \ZHtrans\ bases as a function
of $\beta$.  
In the helicity basis, we find that the $(\a\b)$ and $(\b\a)$
contributions are always equal, as are the $[(\a 0)+(0\b)]$
and $[(0\a)+(\b 0)]$ contributions.  
Consequently, if we look at the population of the spin states for
one of the $Z$'s while summing over the other $Z$, we find
that the $\a$ and $\b$ states are populated equally.
Furthermore, at threshold, this inclusive breakdown is precisely
$1/3$ for each spin, and this changes relatively slowly with 
increasing $\beta$.
The situation is considerably better in the \ZHtrans\ basis,
where no two of the six spin contributions are equal,
and where there is a reasonable difference in the inclusive
contributions.  While we have used $\beta_Z=\beta$ ({\it i.e.}
$M_H=M_Z$) in preparing these plots, we have verified that
for the range accessible to LEP, the dependence on the Higgs
mass is inconsequential.

In Fig.~\ref{ZZ-CTH} we compare the angular distributions
of the polarized production cross sections in the helicity
and \ZHtrans\ bases at $\sqrt{s} = 192 \GeV$.
In the helicity basis there
is a complicated interplay among the six spin components,
whereas in the \ZHtrans\ basis the two dominant components
appear in approximately the same ratio, independent of $\cos\thetas$.


\subsection{Polarized  Decays}\label{polZdec}

The $Zf\bar{f}$ coupling violates both parity and flavor
universality.  Thus, the angular distributions for the decay
of polarized $Z$ bosons are forward-backward asymmetric, and
depend on which fermions appear in the final state.
Neglecting the mass of the final state fermions,\footnote{Inclusion
of the finite mass effects would result in straightforward but
messy modifications to Eqs.~(\protect\ref{ZT}) 
and~(\protect\ref{ZL}).   These effects are greatest for
$b$-quarks, where they are less than a few percent.}
we obtain the following angular distributions in the rest
frame of the decaying $Z$:
\beq
{ {1}\over{\Gamma_f} }
{ {d\Gamma^{\pm}} \over {d(\cos\chi)} }
= {3\over 8}\Bigl[ \alpha_f  (1\mp\cos\chi)^2 
   + (1-\alpha_f)(1\pm\cos\chi)^2 \Bigr]
\label{ZT}
\eeq
for the transverse polarizations and
\beq
{ {1}\over{\Gamma_f} }
{ {d\Gamma^{0}} \over {d(\cos\chi)} }
= {3\over 4} \sin^2\chi
\label{ZL}
\eeq
for the \longitudinal\ polarization.
These distributions have been normalized
to unit area by inclusion of the partial width $\Gamma_f$
for the decay $Z\rightarrow f\bar{f}$.  We take
$\chi$ to be the angle between the direction of motion of the
fermion and the spin axis as seen in the $Z$ rest frame, 
and absorb the dependence on the
couplings between left- and right-handed chirality fermions
to the $Z$ into the factor 
$\alpha_f$.
For convenience, we have collected the values
of $\alpha_f$ in Table~\ref{ZDecayTable}.   
The corresponding
distributions 
are plotted in Fig.~\ref{ZDECAYDIST}.
Unfortunately, the decays with the most distinct distributions,
$Z\rightarrow \nu\bar\nu$ are invisible.
Furthermore, charge and flavor identification for
decays to quarks
(which appear as two jets) is virtually impossible,
except perhaps for $b$ quarks.
For $Z\rightarrow \ell\bar\ell$, we have a fairly
large overlap between the $+$ and $-$ distributions.
All of these features lead us to conclude that
it may not be possible to distinguish between
the $+$ and $-$ polarizations.
However, as we shall see below, the difference
between the $\pm$ and $0$ states is sufficient for separating
$ZH$ and $ZZ$, provided that the appropriate spin basis is used.

Naturally, since the Higgs is a scalar, its decay angle distributions
will be completely flat.


\subsection{Correlations}
\label{ZZcorrel}

In this section we will put everything together to discuss
the correlations among the $ZH/ZZ$ decay products.
For concreteness, we take the  final state to be
$\ell\bar\ell b\bar{b}$,
the $\ell\bar\ell$ coming from a decaying $Z$,
and the $b\bar{b}$ from the Higgs or other $Z$.
The results we present do not change very much if we 
consider a $q\bar{q}b\bar{b}$ final state:
since it is not possible to tell which light jet is the $q$
and which is the $\bar{q}$, all distributions in the $Z$
decay angle $\chi_Z$ defined below would be symmetric about
$\cos\chi_Z=0$.  On the other hand, $\alpha_\ell$ is only slightly
larger than
$1/2$, leading to small asymmetries between positive and
negative values of $\cos\chi_Z$.

In order to display the advantages of using the \ZHtrans\
basis defined by Eq.~(\ref{ZHideal}), we consider a scatter
plot in the decay angles $\chi_Z$ and $\chi_H$, defined as
follows.  On the $Z \rightarrow \ell\bar\ell$ side of the
event, let $\chi_Z$ be the angle between the $\ell$ emission 
direction and the $Z$ spin axis in the $Z$ rest frame.
On the $Z/H \rightarrow b \bar{b}$ side of the event,
let $\chi_H$ be the angle between the $b$ emission 
direction\footnote{If charge identification is not available
for the $b$ jets, then the distributions presented here
should be folded about $\cos\chi_H = 0$.}
and the $Z/H$ spin axis in the $Z/H$ rest frame.
In Fig.~\ref{ZZ-ZHdecayCONTOURS} we present this distribution
for $ZH$ and $ZZ$ production for the helicity and \ZHtrans\ bases
for $\sqrt{s} = 192\GeV$ and $M_H=M_Z$.  The difference
between the two plots is dramatic:  in the helicity basis,
the two distributions are nearly flat, while there is  significant
structure when the \ZHtrans\ basis is used.   
This difference
is underscored by the plots in Fig.~\ref{ZZvsZH1D}, which 
compare the 1-dimensional distributions in $\cos\chi_Z$ 
and $\cos\chi_H$.   Clearly, the angular information contained
in an analysis employing the \ZHtrans\ basis provides a useful
handle in the separation of $ZH$ and $ZZ$ events, even when
$M_H \sim M_Z$.  In contrast, the helicity angles are virtually
useless in this respect!

Kunszt and Stirling~\cite{almost} have noted significant
angular correlations in the ZMF between the direction of the final
state muons and the beam direction.   Effectively this is the
beamline basis.   
At threshold, the beamline basis coincides
with the optimal (\ZHtrans) basis, and the ZMF coincides with
the $Z$ rest frame.
At LEP2, $\beta \sim 1/3$,
{\it i.e.} not that far above threshold.  
Thus, even though they don't use the optimal basis, 
Kunszt and Stirling do see non-trivial angular correlations.
The largest possible correlations, however, are obtained
using the muon direction relative to the \ZHtrans\ basis
spin axis in the $Z$ rest frame.
Furthermore, the gap between their distribution and the 
optimal one widens as we move further above threshold.

An issue which may be raised concerns the effect of
finite detector resolution on the ability of the experimentalists
to accurately measure the decay angles we have proposed,
especially since they have been defined in the rest frames of
the parent particles, not the lab frame.   To get an estimate
of the size of these effects, we introduced Gaussian
energy smearing with a width set by $50\%/\sqrt{E}$ into
the energies of the charged leptons and the $b$ quarks, and then compared
the decay angles reconstructed solely from the smeared momenta
to the actual decay angles.   We take the root mean square
deviation of this difference to be an estimate of the error
introduced by the smearing.   Since the definition of $\xi$
depends both on the speed of the parent boson as well as the
production angle, the detailed values of our error estimate
will vary with the machine energy, choice of basis, and the
process ($ZH$ or $ZZ$) in which the bosons were produced.
Typical values at $\sqrt{s}=192\GeV$ and $M_H=M_Z$
are 
$\delta(\cos\chi)\sim {\cal O}(0.05)$,
for all three bases.
To see the effects of this angular smearing,
an experimentalist would need
to use bins of approximately this size or smaller.
The total number of events observed will probably not be large enough
to warrant such a large number of bins in the $\cos\chi$ plots. 
Although a more detailed study using the full-blown detector
simulation is clearly needed, it would appear that 
it is possible to measure the indicated angles with
sufficient precision.

The amplitudes for the processes 
$\eebar \rightarrow ZH/ZZ \rightarrow \mu^{+}\mu^{-}b \bar{b}$
can be written 
as a sum over the spin states of the intermediate massive particles 
of the production times the decay amplitudes for each spin component.
In the classical limit there is no interference between these 
amplitudes. 
However, in general,
these amplitudes do interfere.
The size of these interference terms can be used 
as a measure of how well the classical result for a particular basis 
describes the full process:  if the interference terms are small, 
then the classical picture captures nearly all of the physics.  
On the other hand,
if the interference terms are large, a
significant amount of the physics is omitted from the classical
picture.

To quantify these considerations, 
at each kinematic point for a given process, we may define the quantity
$\Ihat$, the value of the interference terms for a given spin basis
divided by the total
$2\rightarrow4$ matrix element squared.  
For a process containing
$N$ independent intermediate spin configurations, $\Ihat$
can range from $-\infty$ (total destructive interference)
to $(N-1)/N$ (total constructive interference).
A plot of $d\sigma/d\Ihat$ displays the relative importance
of the various values of $\Ihat$.  Clearly, in the ideal case
where classical intuition captures all of the physics, 
this distribution will be a delta function centered at $\Ihat=0$.
Fig.~\ref{ZHinf} compares the helicity, beamline ant \ZHtrans\
bases 
for $ZH$ production and decay, whereas Fig.~\ref{ZZinf}
does the same for $ZZ$.  
For the signal ($ZH$), we see that the peaking in $\Ihat$
is about an order of magnitude greater in the \ZHtrans\ basis
than in the beamline basis.  The helicity basis shows essentially
no peaking.  For the background ($ZZ$), the difference is not
as dramatic, but it is still clear that the \ZHtrans\ basis
gives the smallest interference terms, and hence the superior
description.


\subsection{Pseudoscalar Higgs}\label{pseudo}

Although the Higgs boson of the Standard Model is unambiguously
$CP$-even,
for the sake of
comparison, it is useful to consider how the spin correlations
differ for a $CP$-odd Higgs $A$ (commonly referred to as
a pseudoscalar),  a feature of two-doublet Higgs
models.  Although a tree-level $ZZA$ coupling is forbidden,
such a coupling can be generated at the one-loop level~\cite{HH}.
Following Ref.~\cite{Barger}, we write the (effective) $ZZA$ vertex
as
\beq
-{ {igM_Z}\over{\cos\theta_W} }
{ {\eta}\over{\Lambda^2} }
k_1^{\mu} k_2^{\nu} \varepsilon_{\mu\nu\alpha\beta},
\eeq
where $k_1$ and $k_2$ are the 4-momenta of the two $Z$'s,
$\eta$ is a dimensionless coupling constant, and $\Lambda$
is the mass scale at which this vertex is generated.
This coupling leads to the differential cross section
\beqa
{
{ d\sigma^{\lambda}(\eebar{\rightarrow}ZA) }
\over
{ d(\cos\thetas) }
} = 
{   {G_F^2 M_W^2} \over {\cos^2\theta_W}   } \ts &&
{ \Biggl( {M_Z\over\Lambda} \Biggr)^4 } \ts
{   {\BETA^3\GAMMA M_Z} \over {32\pi s^{1/2}}   } \ts
\Biggl(
{  {s+M_Z^2-M_H^2} \over {s-M_Z^2}  }
\Biggr)^2
\cr && \times\biggl\{
 \cos^2 2\theta_W \ts
 [\widetilde\T_L^{\lambda}(\thetas,\xi)]^2
+ 4 \sin^4 \theta_W \ts
 [\widetilde\T_R^{\lambda}(\thetas,\xi)]^2
\biggr\}.
\label{ZAsigma}
\eeqa
In this case the spin functions are independent of energy:
\beqa
\widetilde\T_L^{\pm}(\BETA,\thetas,\xi) = 
\widetilde\T_R^{\mp}(\BETA,\thetas,\xi)
= && {1\over\sqrt{2}} \Bigl( \cth\pm \cxi \Bigr) ; \cr
\widetilde\T_L^{0}(\thetas,\xi) = \widetilde\T_R^{0}(\thetas,\xi) 
= && \sxi
\label{ZAspinfun}
\eeqa
Thus, the total (unpolarized) differential cross section is
\beqa
\sum_{\lambda}
{
{ d\sigma^{\lambda} }
\over
{ d(\cos\thetas) }
} = 
{   {G_F^2 M_W^2} \over {\cos^2\theta_W}   } \ts &&
{ \Biggl( {M_Z\over\Lambda} \Biggr)^4 } \ts
{   {\BETA^3\GAMMA M_Z} \over {32\pi s^{1/2}}   } \ts
\Biggl(
{  {s+M_Z^2-M_H^2} \over {s-M_Z^2}  }
\Biggr)^2
\cr && \times\Bigl(
1-4\sin^2\theta_W+8\sin^4\theta_W
\Bigr)\Bigl(
2 - \sin^2\thetas
\Bigr),
\label{ZAsigmaTOTAL}
\eeqa
in agreement with Ref.~\cite{Barger}.

It is clear from
the especially simple form of the spin functions, Eq.~(\ref{ZAspinfun}),
that the optimal basis for $ZA$ is the helicity basis, independent
of the machine energy.  Only the helicity basis has the property
that one of the three amplitudes (the longitudinal one) vanishes.

Because the $\xi$ dependence of the $ZH$ and $ZA$ amplitudes differs
so greatly, a measurement of the $Z$ spin composition of a Higgs
signal for different choices of $\xi$ could potentially differentiate
between the two cases.  In particular, it would be useful to 
measure the fraction of longitudinally polarized $Z$'s in a sample
of $Z$-Higgs candidates in both the helicity and \ZHtrans\ bases.
For the signal events,
a scalar Higgs should show 
no longitudinal $Z$'s in the \ZHtrans\ basis, while for a 
pseudoscalar Higgs this would occur only in the helicity basis.


\section{The Process 
{$\protect\lowercase{e}^{+}\protect\lowercase{e}^{-}
\longrightarrow W^{+}W^{-}$} at LEP}\label{WWsection}

With the crossing of the $W$-pair production threshold at LEP2, 
it is now possible to probe the electroweak sector of the
Standard Model in new ways.
The tree-level
diagrams for $\eebar\rightarrow\WWbar$ contain
triple gauge boson vertices ($WW\gamma$ and $WWZ$),
allowing direct studies of the Yang-Mills
structure of the theory.  
The precise form of these couplings predicted by 
the $SU(2)\times U(1)$ nature of the underlying theory
is reflected in the various angular distributions associated
with the $WW$ events~\cite{yellow,Hagiwara,anomal}.
If we allow the strengths of the existing
$WW\gamma$ and $WWZ$ couplings to vary, 
or if we introduce new ones,  
then not only will the differential distribution in
the production angle $\thetas$ be affected, 
but the shares of the total cross section coming from
the different spin states of the $W$'s will also vary.
This second effect will show up in
the distributions in the angles between the decay products and
suitably defined spin axes.

As mentioned in the introduction, a complete reworking of the
anomalous trilinear couplings analysis using our generalized
spin basis, while worthwhile, would be a major extension of
this work.  Nevertheless, the general means of proceeding should
be clear:  for each anomalous coupling a computation of the
polarized differential cross sections as a function of $\xi$
is required.  By examining the functional form 
of these cross sections
(with the aid of contour plots in the $\cos\thetas$-$\cos\xi$ plane),
a basis may be constructed which highlights the deviations
introduced by the anomalous coupling.  The decay angular
distributions in this basis then encode the consequences of
the new physics being investigated.

In the Standard Model, 
three diagrams contribute to the total 
$\eebar\rightarrow \WWbar$ 
cross section.
We regroup these three diagrams into three contributions as follows.
First, we note that the two diagrams
involving an $s$-channel photon or $Z$ boson
contribute for both possible
chiralities of the initial fermion line.   Thus, the
first two contributions to the total cross section will be
referred to as the
$(\gamma/Z)_{\rm L}$ and $(\gamma/Z)_{\rm R}$
pieces.\footnote{From our point of view, the interference
between the photon and the $Z$ is trivial in the sense that it 
shows up as a ($W$) spin-independent prefactor:  no choice of
spin basis for the $W$'s can make these two contributions easier or 
harder to separate.  Note that the contribution from the photon 
alone may be determined from Eqs.~(\protect\ref{prodWWleft}) 
and~(\protect\ref{prodWWright}) by taking
the $Z$ mass to infinity, and dropping the last term
of Eq.~(\protect\ref{prodWWleft}).}
The third diagram involves
the $t$-channel exchange of a neutrino and contributes only
when the initial fermion line is left-handed. 
It will form the third contribution, which we will refer to
as the neutrino piece.

Before presenting the differential cross sections
for polarized production,
let us consider the relative size of these three contributions
to the total cross section.
In Table~\ref{DIAGcontribs},
we have tabulated the values for each piece at tree-level
for a machine energy of 192~GeV.  We see that the
$(\gamma/Z)_{\rm R}$ contribution is insignificant.
Furthermore, there is a non-trivial
interference term, which we will denote by $\INFZ$,
between the $(\gamma/Z)_{\rm L}$ and
neutrino contributions.\footnote{Recall that different helicity 
amplitudes do not interfere: hence the absence of interference 
terms between the right-handed $(\gamma/Z)$ contribution
and the left-handed pieces.}
In Fig.~\ref{INFZplot}, we have plotted the 
squares of the
three contributions 
as well as the interference term $\INFZ$
as a function of the zero momentum frame
production angle $\thetas$, which is
taken to be the angle between the momentum of the electron
and the $W^{-}$.  Although it is tempting to attribute the
well-known peaking of the cross section in the forward direction
to the $t$-channel pole in the neutrino diagram, Fig.~\ref{INFZplot}
indicates that this is not the case.  The square of the neutrino
diagram is not only well-behaved near $\cos\thetas=1$,
but even {\it decreases}\ slightly in that region.  Instead,
the shape of
the differential cross section is dominated by the interference
between the $(\gamma/Z)_{\rm L}$ and neutrino pieces,
which changes sign from $-$ to $+$ for increasing $\cos\thetas$.
These general features become even more pronounced at higher
center-of-mass energies.


\subsection{Polarized Production}

To describe the polarized production cross sections
for $\eebar \rightarrow \WWbar$,
we choose to identify the $W^{-}$ with particle $P_1$ 
and the $W^{+}$ with particle $P_2$ in Fig.~\ref{XIdef}.

We now present the differential
cross sections for polarized $\WWbar$ production.
When the initial fermion pair has left-handed chirality, we have
\beqa
{
{ d\sigma_L^{\lambda\bar\lambda}(e\bar{e}{\rightarrow}\WWbar) }
\over
{ d(\cos\thetas) }
}&& = 
\cr 
{ {G_F^2 M_W^2}\over{256\pi} }
\beta\gamma^2 &&
\Biggl\{
{ { 2(M_W/M_Z)^2 - (1-\beta^2)\sin^2\theta_W  }
\over
{ 4(M_W/M_Z)^2 - (1-\beta^2) }
}
\Bigl[
\T_L^{\lambda\bar\lambda}(\beta,\thetas,\xi) 
-\T_L^{\lambda\bar\lambda}(-\beta,\thetas,\xi) 
\Bigr]
\cr &&
-{  {2}\over{1-2\beta\cos\thetas+\beta^2}  }
\T_L^{\lambda\bar\lambda}(\beta,\thetas,\xi) 
\Biggr\}^2.
\label{prodWWleft}
\eeqa
Here $\lambda$ ($\bar\lambda$) is the spin of the $W^{-}$ ($W^{+}$).
The first term in the curly brackets is the  $(\gamma/Z)$
contribution, while the second term comes from the neutrino.
For the other initial fermion chirality (right-handed), we
have simply
\beq
{
{ d\sigma_R^{\lambda\bar\lambda}(e\bar{e}{\rightarrow}\WWbar) }
\over
{ d(\cos\thetas) }
}
= { {G_F^2 M_W^2}\over{256\pi} }
{ {\beta}\over{\gamma^2} }
{
{ \sin^4\theta_W }
\over
{ [ 4(M_W/M_Z)^2 - (1-\beta^2)]^2 }
}
\Bigl[ \T_R^{\lambda\bar\lambda}(\beta,\thetas,\xi) 
- \T_R^{\lambda\bar\lambda}(-\beta,\thetas,\xi) \Bigr]^2.
\label{prodWWright}
\eeq
All of the spin information in contained in the $\T's$, which 
are the same spin functions defined in connection with $ZZ$ production,
namely Eqs.~(\ref{funcAB})--(\ref{leftright}).  Naturally, we
should re-interpret $\beta$ as the common ZMF speed
of the two $W$'s.

The presence of the neutrino contribution to Eq.~(\ref{prodWWleft})
makes the expression for the total differential cross section
somewhat messy: 
\beqa
\sum_{\lambda,\bar\lambda,C}
{
{ d\sigma_C^{\lambda\bar\lambda} }
\over
{ d(\cos\thetas) }
}=  && \cr
{
{G_F^2 M_W^2}
\over
{8\pi} 
} &&\ts\ts
{
{ 2(M_W/M_Z)^4 - 2(M_W/M_Z)^2 (1-\beta^2)\sin^2\theta_W
  + (1-\beta^2)^2\sin^4\theta_W }
\over
{ [4(M_W/M_Z)^2 - (1-\beta^2) ]^2 }
}
\cr &&
\qquad\qquad\qquad\qquad\qquad\qquad
\times\beta^3\ts
\biggl\{ 16 + 
        \Bigl[4\beta^2\gamma^2+3(1-\beta^2)\Bigr]\sin^2\thetas \biggr\}
\cr + {
{G_F^2 M_W^2}
\over
{8\pi} 
} &&\ts
\beta\ts\Biggl\{
2 + {1\over2}\beta^2\gamma^2\sin^2\thetas
+ {
{ 2\beta^2(1-\beta^2)\sin^2\thetas }
\over
{(1-2\beta\cos\thetas+\beta^2)^2 }
}
\Biggr\}
\cr - \ts {
{G_F^2 M_W^2}
\over
{8\pi} 
} &&\ts { 
{ 2(M_W/M_Z)^2 - (1-\beta^2)\sin^2\theta_W }
\over
{ 4(M_W/M_Z)^2 - (1-\beta^2) }
} \cr &&
\times \beta \ts\Biggl\{
(1+\beta^2)[4+\beta^2\gamma^2\sin^2\thetas]
+ {
{ 2(1-\beta^2)[\beta^2\sin^2\thetas - 2(1-\beta^2)] }
\over
{ 1-2\beta\cos\thetas+\beta^2 }
} \Biggr\}.
\label{WWtotal}
\eeqa
The successive terms in Eq.~(\ref{WWtotal}) come
from the square of the $(\gamma/Z)$ contributions (summed
over both $\eebar$ chiralities),
the square of the neutrino contribution, and
the interference term $\INFZ$.

As in the $ZZ$ case, the individual spin-dependent
amplitudes are complicated.  Therefore, in 
Fig.~\ref{PRODcontours}, we have plotted the 
fractional contribution to each of the six independent
spin configurations for a center-of-mass energy of 192 GeV.
From these
plots, we see that the $[(\a\a)+(\b\b)]$ and
$(00)$
contributions never dominate the total amplitude
at 192 GeV, while the $[(\a 0)+(0 \b)]$ or $(\a \b)$
contributions can be made large with the proper choice
of $\xi$.


\subsection{Spin Bases}

From the contour plots and the expressions for the spin components 
we have identified four interesting spin bases: helicity, \AB,
beamline, and \A00B, which we have studied in some detail.
In Fig.~\ref{WWsolns}, we have plotted
the connection between $\xi$ and $\thetas$
at 192 GeV for these four bases.
In Tables~\ref{WW-Breakdowns}--\ref{INFZ-Breakdowns}, 
we indicate the fraction of the total
cross section, $(\gamma/Z)$, $\nu$ and $\INFZ$
contributions broken down into the six independent spin
configurations
for an $\eebar$ collider running at 192 GeV.

We begin our survey with the  helicity basis.
This basis has one noteworthy feature:
the $(\a \b)$ and $(\b \a)$ components of both $(\gamma/Z)$
contributions vanish at all $\sqrt{s}$.  
This is not true for the neutrino
diagram; therefore, a measurement of these spin components 
is a direct measurement of the neutrino contribution.
At $\sqrt{s} = 192$ GeV the sum of these two spin components
is 56\% of the total cross section.

Another useful basis is the \AB\ basis, which
is defined by choosing $\xi$ so that the $(\a\b)$
component is as large as possible.  
In principle, it is a straightforward exercise to derive
an analytic expression connecting $\xi$ and $\thetas$ 
by differentiating $d\sigma^{+-}/d(\cos\thetas)$ with
respect to $\xi$ and setting the result equal to zero.
In practice, however, such an expression would be too complicated
to be illuminating.
Furthermore, it turns out that
we must allow negative values of
$\sin\xi$ in order
to get the largest possible contribution. 
This is equivalent
examining both the $(\a \b)$ and $(\b\a)$ amplitudes, and
taking the larger of the two at each point.   Selecting
$\sin\xi<0$ whenever the $(\b\a)$ amplitude is chosen converts
it to a $(\a\b)$ amplitude, according to Eq.~(\ref{leftOTHER}).
Thus, it is expedient to use a numerical procedure to 
determine $\xi$ in this basis:  the results are plotted
in Fig.~\ref{WWsolns} for $\sqrt{s} = 192\GeV$.
The discontinuity seen in the $\cos\xi$ versus $\cos\thetas$
curve for this basis near $\cos\thetas = -0.8$ is caused by
the presence of two competing maxima.
At $\sqrt{s}= 192\GeV$ the $(\a\b)$ component is 64\% of the total
in this basis.

In the beamline basis, defined by Eq.~(\ref{BMLdef}),
the direction
of the spin axis for the $W^{-}$ ($W^{+}$) viewed
in its rest frame coincides with the 
direction of the electron (positron) in that frame. 
With the choice~(\ref{BMLdef}) for $\xi$, the functions
$\G_1(\beta,\thetas,\xi+\pi)$ and $\G_3(\beta,\thetas,\xi)$ vanish.
This means that the neutrino does not contribute to the
$[(\a \a)+(\b \b)]$, $[(0\a)+(\b 0)]$, and $(\b \a)$
spin configurations~\cite{Pundurs}.  
Unfortunately, the fraction of the total in these spin 
configurations is small, less than 5\% at $\sqrt{s} = 192$ GeV. 
However, the $[(\a0)+(0\b)]$ component is 
80\% of the total at this $\sqrt{s}$.

Finally, we come to 
the \A00B\ basis, which is defined by choosing
the value of $\xi$ which maximizes the fraction of the
amplitude coming from the $[(\a 0)+(0\b)]$ component.
Once again the maximization condition leads to a complicated
expression (it is a quartic equation for $\tan\xi$).
Therefore, we present the numerically derived solution
in Fig.~\ref{WWsolns}.
In this basis, the $[(\a0)+(0\b)]$ component is more than 92\% 
of the total at $\sqrt{s}= 192$ GeV.

For these last three bases, \AB, beamline, and \A00B,
there are large negative interference terms between the 
$\nu$ and $(\gamma/Z)_{\rm L}$ contributions in 
all of the dominant spin components.

Fig.~\ref{WWcomposCTH} shows the angular dependence of the 
polarized production cross sections.  It is clear from these
plots that in the helicity basis, there is a complicated
structure, with four different amplitudes being the largest,
depending on the value of $\cos\thetas$.  On the other hand,
in the \AB\ basis, the $(+-)$ component dominates at all angles.
Likewise, in the beamline and \A00B\ bases, the $[(+0)+(0-)]$
component dominates at all angles.

Finally, because the relative contributions of the polarized
cross sections depend upon the center-of-mass energy, we
display this dependence in the range from threshold to
210 GeV in Fig.~\ref{WWcomposROOTS}.  
As mentioned near the end of Sec.~\ref{polZHprod},
the effect of including initial state radiation 
in the results is to produce events at lower $\sqrt{s}$.
From Fig.~\ref{WWcomposROOTS}, we see that the addition of
such events would slightly increase the fraction of the
dominant spin component in the beamline and \A00B\ bases
and slightly decrease the dominant component in the helicity
and \AB\ bases.   The \A00B\ basis is slightly less sensitive
to small shifts in $\sqrt{s}$ than the beamline basis, while
the \AB\ basis is slightly less sensitive to small shifts in 
$\sqrt{s}$ than the helicity basis.


\subsection{Polarized Decay}

The presence of maximal parity violation in the coupling between
the $W$ boson and fermions results in distinctly different
angular distributions for the $W$ decay products for all
three $W$ polarizations.  
In fact, the decay distributions in the rest frame of the
decaying $W^{-}$ (or $W^{+}$) are the same as for $Z$ decay
[{\it i.e.}\ Eqs.~(\ref{ZT}) and~(\ref{ZL})] with  $\alpha_f = 1$
for all possible final states, leptonic and hadronic.
Thus, they follow the $\nu\bar\nu$ curves presented in 
Fig.~\ref{ZDECAYDIST}.
Note, however, in the case of $W^{+} \rightarrow \bar\ell \nu$,
it is more convenient to use the distribution in the {\it antilepton}\
angle $\bar\chi$:  
this distribution may be generated by replacing $\cos\chi$
with $-\cos\bar\chi$ in Eqs.~(\ref{ZT}) and~(\ref{ZL}).

Because the distributions for transversely-polarized $W$'s
are narrower than the one for \longitudinal ly-polarized $W$'s,
it is easier to identify the $+$ and $-$ polarization states
than the $0$ state.  To demonstrate this, 
suppose we ``tag'' the spin of the parent $W^{-}$ based on
the value of $\cos\chi$ as follows:
the spin is taken to be $\a$ if $\cos\chi < -y$,
$\b$ if $\cos\chi > y$, and $0$ if $\vert\cos\chi\vert \le y$.
The probabilities that these assignments are correct may
be computed from Eqs.~(\ref{ZT}) and~(\ref{ZL}), yielding
\beqa
\hbox{$1\over{12}$}(7+4y+y^2),\qquad && \hbox{$\pm$ states} \cr
\hbox{$1\over{6}$}(3-y^2),\qquad && \hbox{$0$ state}.
\label{Wtagging}
\eeqa
No matter what value is chosen for $y$, the probability of
correct identification in the central region is always less
than 50\%.  On the other hand, the correct identification rate for
the $+$ and $-$ states in their respective regions can be of
order 75\%.


\subsection{Correlations}

We now consider correlations between the decay products
of the $\WWbar$ pair.
For the sake of concreteness, let us consider the case
where the $W^{-}$ decays to $\mu^{-}\bar\nu_\mu$
and the $W^{+}$ decays to $\mu^{+}\nu_\mu$.
Let $\chi$ be the angle between the $\mu^{-}$ emission
direction and the $W^{-}$ spin axis in the $W^{-}$ rest frame, and
$\bar\chi$ be the angle between the $\mu^{+}$ emission
direction and the $W^{+}$ spin axis in the
$W^{+}$ rest frame.
For simplicity we are assuming one can 
determine the direction of the two neutrinos. 
If one of the $W$-bosons decays into two jets, then the 
correlations of the down-quark jet are the same as that of the charged
lepton.
Since determining which jet is the down-type jet is problematic,
the distributions given here will have to be folded such
that they are symmetric when $\cos \chi \leftrightarrow -\cos\chi$.

The first set of correlations we wish to discuss may
be conveniently displayed as
a scatter plot in $\cos\chi$
and $\cos\bar\chi$.  In Fig.~\ref{WWdecayCONTOURS} we have
plotted the predictions for this distribution using the
helicity, \AB, beamline, and \A00B\ bases.
On one hand, the \A00B\ plot is very nearly the pure
distribution for the decay of the $[(+0)+(0-)]$ spin state.
However, the broadness of the decay distributions for the 
longitudinal $W$'s cause the events to be rather spread out
in this plot.  On the other hand, there is a nice contrast
between the maximum and minimum values in the \AB\ plot.
The functional dependence is less simple, however, as fully
three of the six spin configurations contribute at or above
the $10\%$ level.

Although a precision measurement of the complete 2-dimensional
distribution requires a large number of events, it
is still possible to perform some interesting tests with
fewer events.  For example,
if $CP$ is conserved, then the scatter plot will be 
symmetric about the line $\cos\chi = \cos\bar\chi$.
So a measure of the asymmetry in the number of events on
either side of this line provides a simple test of $CP$ symmetry.

In Fig.~\ref{WWinf}, we have plotted the distribution of the
interference terms $d\sigma/d\Ihat$ 
(as described in Sec.~\ref{ZZcorrel})
for $\WWbar$ production and decay using the helicity,
\AB, beamline, and \A00B\ bases.  None of the bases under
consideration (nor any basis which we have found) performs
particularly well in this respect.  The \AB\ basis contains
the smallest interference terms on average, but there is still
a non-negligible component away from $\Ihat=0$.
This bump/peak at positive $\Ihat$ for all spin bases
is caused by the strong azimuthal correlations associated with
the pure $V{-}A$ coupling of the $W$-boson to its decay products.
A similar structure appears in the $\Ihat$ plots for 
$\eebar\rightarrow ZZ$ in the situation where both $Z$'s decay
to $\nu\bar\nu$.

Although there is no obvious spin basis which is ideally suited to the
study of $W$ pair production at LEP, the \AB\ basis is clearly 
better than the helicity basis in many ways. 
The same can be said of the \A00B\ basis compared to the beamline basis.
A detailed study of the correlations in more than one basis is
required to disentangle the spin correlations in this process.



\section{Conclusions}\label{CONC}

If the Higgs mass turns out to be in the vicinity of the $Z$ mass,
then the \ZHtrans\ basis, in which the $Z$ produced in association
with he Higgs is purely in the $\pm$ polarization states, provides
a useful handle with which to distinguish between $ZH$ and $ZZ$
events, based upon the angular distributions of the decay products.
In this situation, the helicity basis supplies no useful information.
Even if the Higgs mass is significantly different from the $Z$
mass, the \ZHtrans\
basis still exists and leads to distinctive angular correlations
among the decay products.  
A comparison between the \ZHtrans\ and helicity descriptions
of the data is useful for distinguishing a scalar from
pseudoscalar Higgs.

For the study of the $WW$ events there is no ideal basis.
However, the
\AB\ basis is better than helicity,
and the \A00B\ basis is better than the beamline basis
at describing the correlations.
Not only is there a greater fraction of the total cross section 
concentrated in the dominant components in these 
preferred bases, but the interference
distributions are somewhat narrower.
The lack of an ideal basis implies that study of the $WW$ system
is best carried out using more than one basis, depending upon
which quantity is being tested.  Furthermore, simply checking
to see that the correlations change in the correct manner as
the basis is varied tests the Standard Model in ways which
cannot be accomplished with the helicity basis alone.


\acknowledgements

The Fermi National Accelerator 
Laboratory (FNAL) is operated by Universities Research Association,
Inc., under contract DE-AC02-76CHO3000 with the U.S. Department
of Energy.
High energy physics research at McGill University is
supported in part by the Natural Science and
Engineering Research Council of Canada. 
We would like to thank Peter Fisher and Philip Bambade 
for illuminating discussions.
GM would like to thank the FNAL theory group for their kind
hospitality during visits to initiate and wrap-up this work.
SP would like to thank the Aspen Center for Physics where a
portion
of this work was completed.


\newpage
\appendix

\section*{Polarization Vectors for Massive Vector Bosons}

Following Kuijf\cite{Kuijf}, we
consider a massive vector boson of momentum $V$ and mass $M$
such that $V^2 = M^2$. Let $S$ be the spin vector associated with
this vector boson such that $S^2 = -1$ and $V \cdot S = 0$.
We need the three polarization vectors $\epsilon_{\lambda}^{\mu}$
such that in the vector boson rest frame the boson has spin
projection $\lambda  =(+,0,-)$ with rest to the spatial part of
the spin vector $S$.

These three polarization vectors are conveniently written in terms of
the two vectors
\beq
  V_1 = \hbox{${1\over 2}$}(V+MS) 
  \quad{\rm and}\quad  
  V_2 = \hbox{${1\over 2}$}(V-MS). 
\eeq 
Note $V_1+V_2=V$, $V_1^2 = V_2^2 = 0$ and $2V_1 \cdot V_2 = M^2$.
The decay products of a transverse vector boson are directly
correlated with $V_1$ and $V_2$, not $V$ and $S$.
Using the spinor notation of Mangano and Parke\cite{mp}, where
\beq
\lsp{V_1\pm} ~\equiv~ \bar{u}(V_1)\hbox{${1\over2}$} (1\mp\gamma_5) 
\quad{\rm and}\quad
\rsp{V_1\pm} ~\equiv~ \hbox{${1\over2}$} (1\pm\gamma_5) v(V_1),
\eeq
we have
\beqa
\epsilon_{\pm}^{\mu} & ~=~ &
{\lsp{V_1\pm} \gamma^{\mu} \rsp{V_2\pm}}
\over
{\sqrt{2}M} \\
\epsilon_{0}^{\mu} & ~=~ &
{
{\lsp{V_1+} \gamma^{\mu} \rsp{V_1+} ~-~ 
\lsp{V_2+} \gamma^{\mu} \rsp{V_2+} } 
\over
{2M}
} ~ \Biggl(  ~=~ 
{  {V_1^{\mu}-V_2^{\mu}} \over {M} } ~ \Biggr) .
\eeqa
The phase factors have been chosen such that 
$\epsilon^{\mu}_0$
is real,
$\epsilon^{\mu}_- = (\epsilon^{\mu}_+)^*$, and 
$\epsilon^{\mu}_- \leftrightarrow \epsilon^{\mu}_+$ if
we interchange $V_1$ and $V_2$.
Our convention is that these polarization vectors are for 
outgoing vector bosons whereas 
for incoming vector bosons we use $(\epsilon_{\lambda}^{\mu})^*$.

These polarization vectors satisfy transversality, orthogonality,
and completeness relations:
\beqa
V \cdot \epsilon_{\lambda} &  ~=~ & 0, \\
\epsilon_{\lambda} \cdot \epsilon^*_{\lambda^{\prime}} &  ~=~ & 
- ~\delta_{\lambda\lambda^{\prime}}, \\
\sum_{\lambda} \epsilon_{\lambda}^{\mu} 
~\epsilon_{\lambda}^{\nu *} &  ~=~ &  -~g^{\mu\nu} 
~+~ { V^{\mu} V^{\nu} \over M^2 }.
\eeqa

The helicity basis is obtained by choosing the spatial part
of the spin vector $S$ to be in the same direction as the spatial
part of the momentum vector $V$ of the vector boson. For example,
if 
\beq 
V = \gamma M (1, ~ \beta \hat{n})
\eeq
using an obvious notation then choose 
\beq
V_1 = {1\over2}\gamma M (1+\beta) ( 1, \hat{n})
\quad{\rm and}\quad
V_2 = {1\over2}\gamma M (1-\beta) (1, -\hat{n}).
\eeq



\begin{figure}[h]

\vspace*{15cm}
\includegraphics{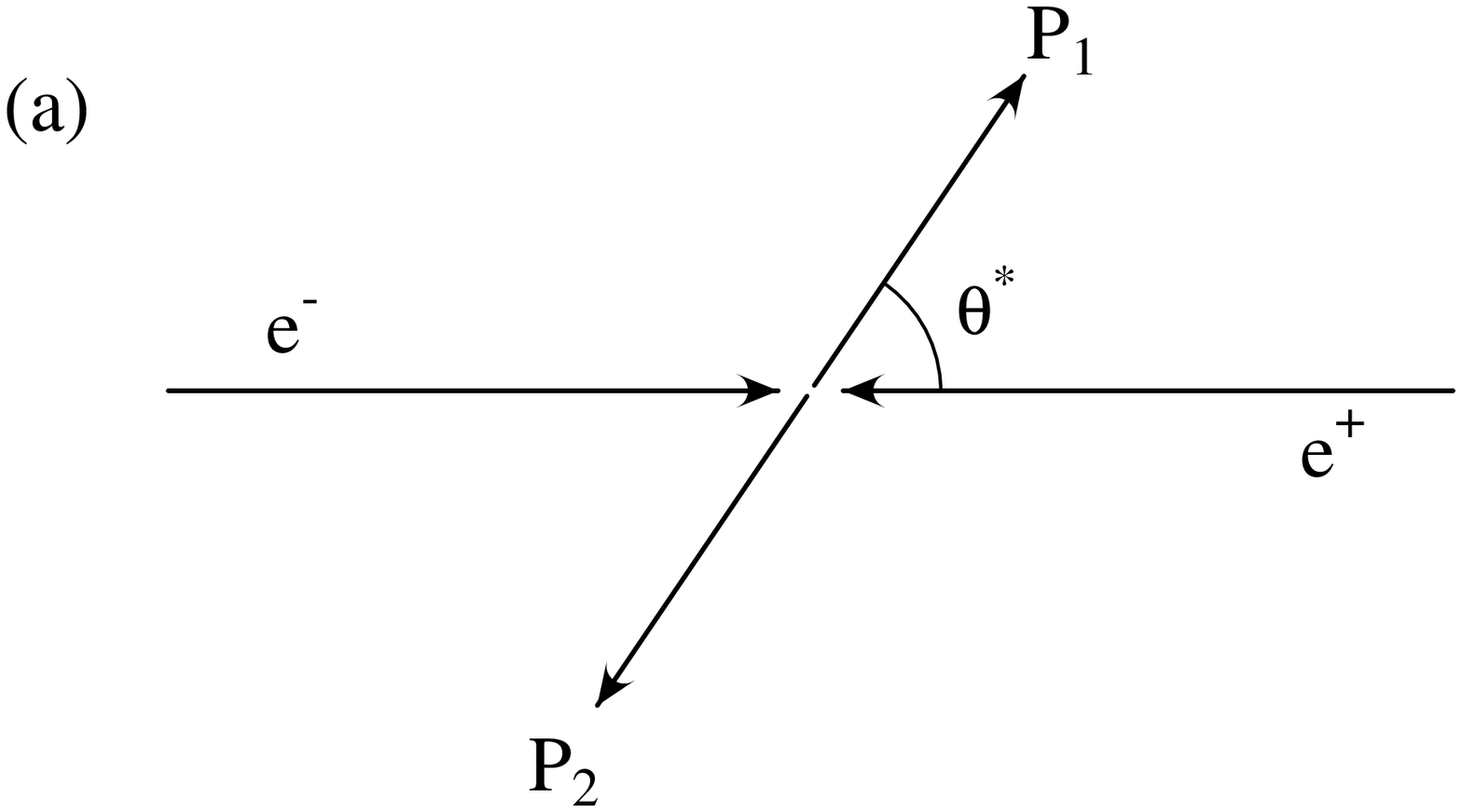}
\includegraphics{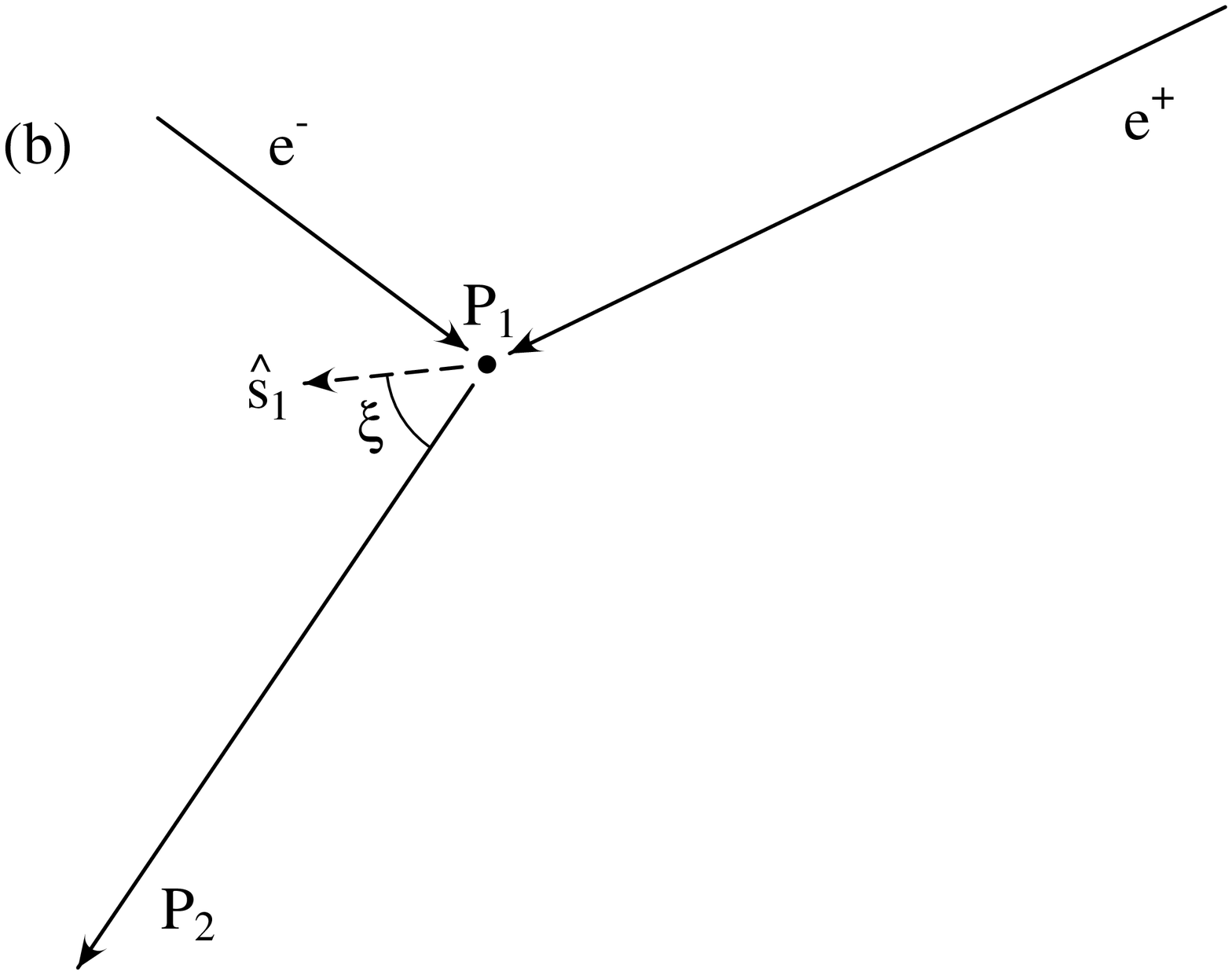}
\includegraphics{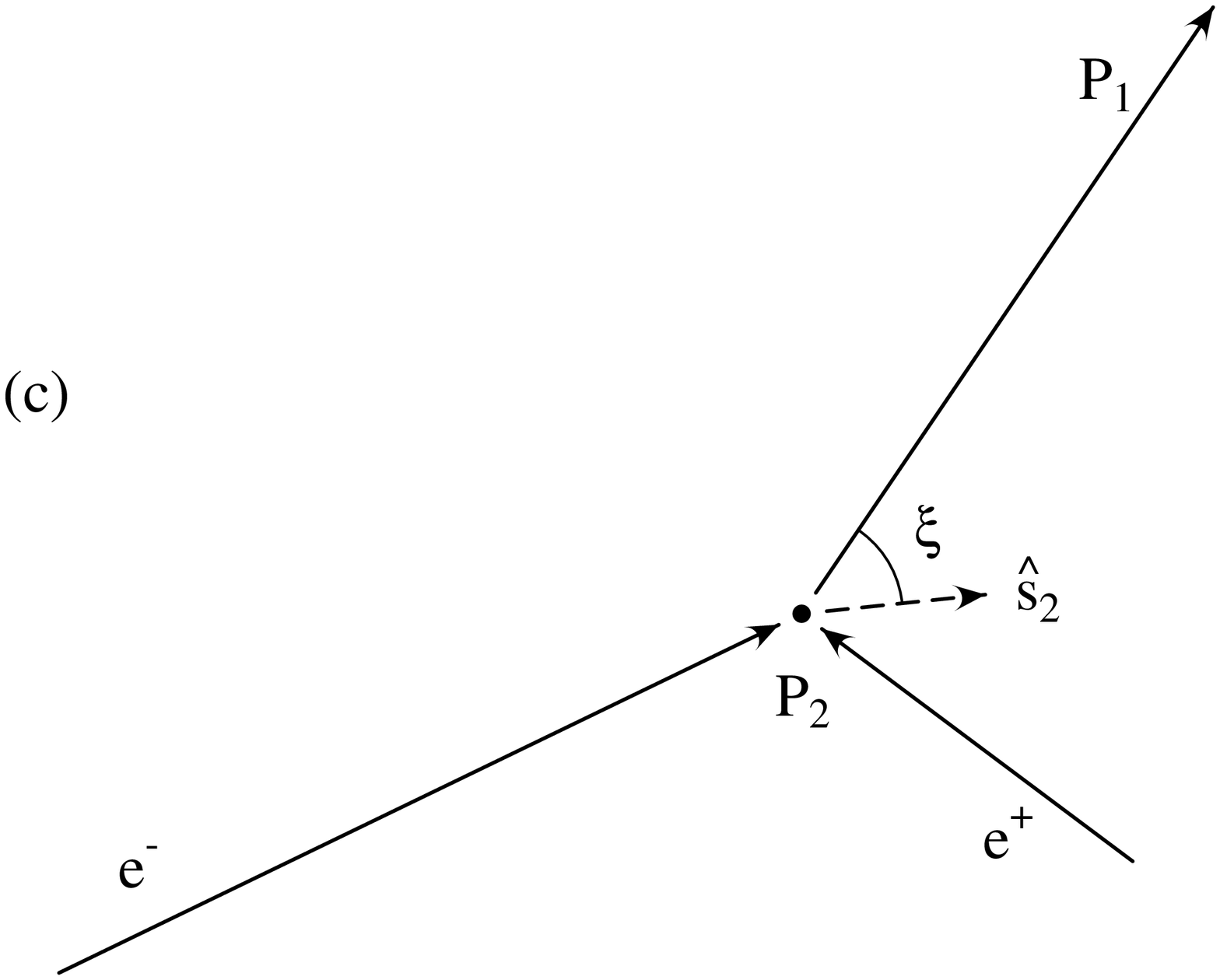}
\vspace{6.0cm}

\caption[]{The scattering process in (a) the zero momentum frame,
(b) the rest-frame of $P_1$ and (c) the  rest-frame of $P_2$.
$\hat{s}_j$ is the spin axis for $P_j$.
}
\label{XIdef}
\end{figure}


\vspace*{1cm}

\begin{figure}[h]

\vspace*{15cm}
\includegraphics{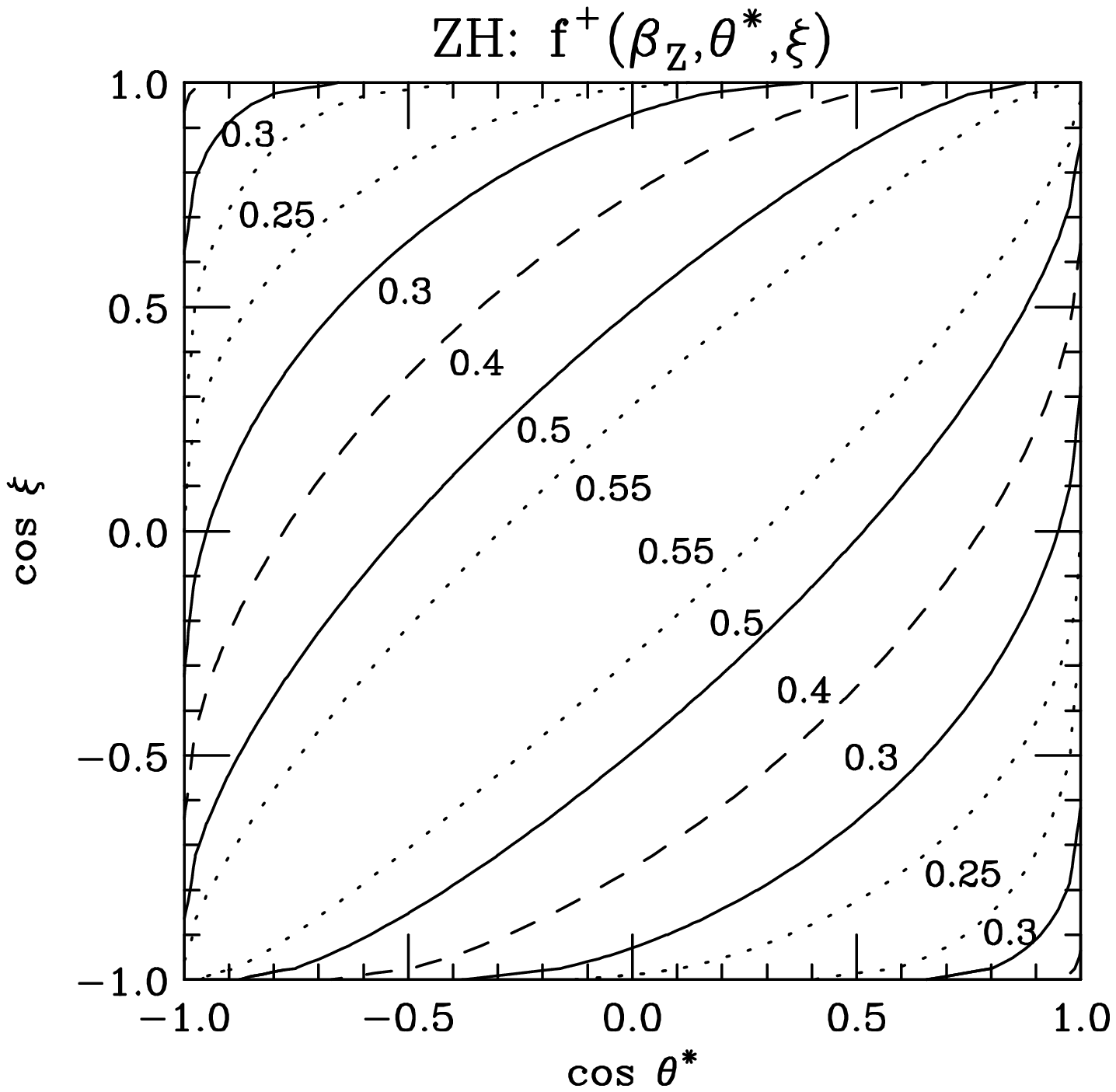}
\includegraphics{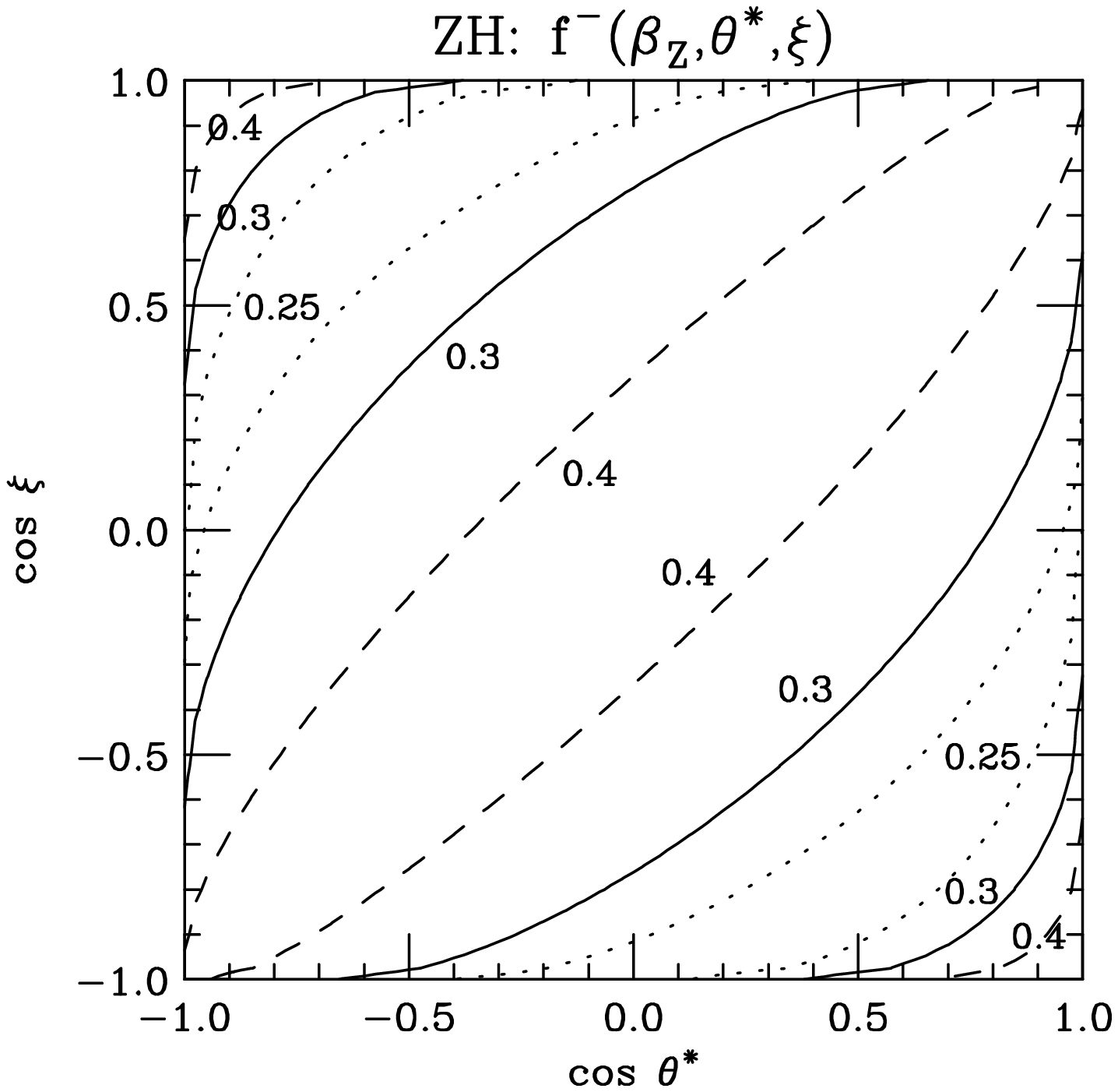}
\includegraphics{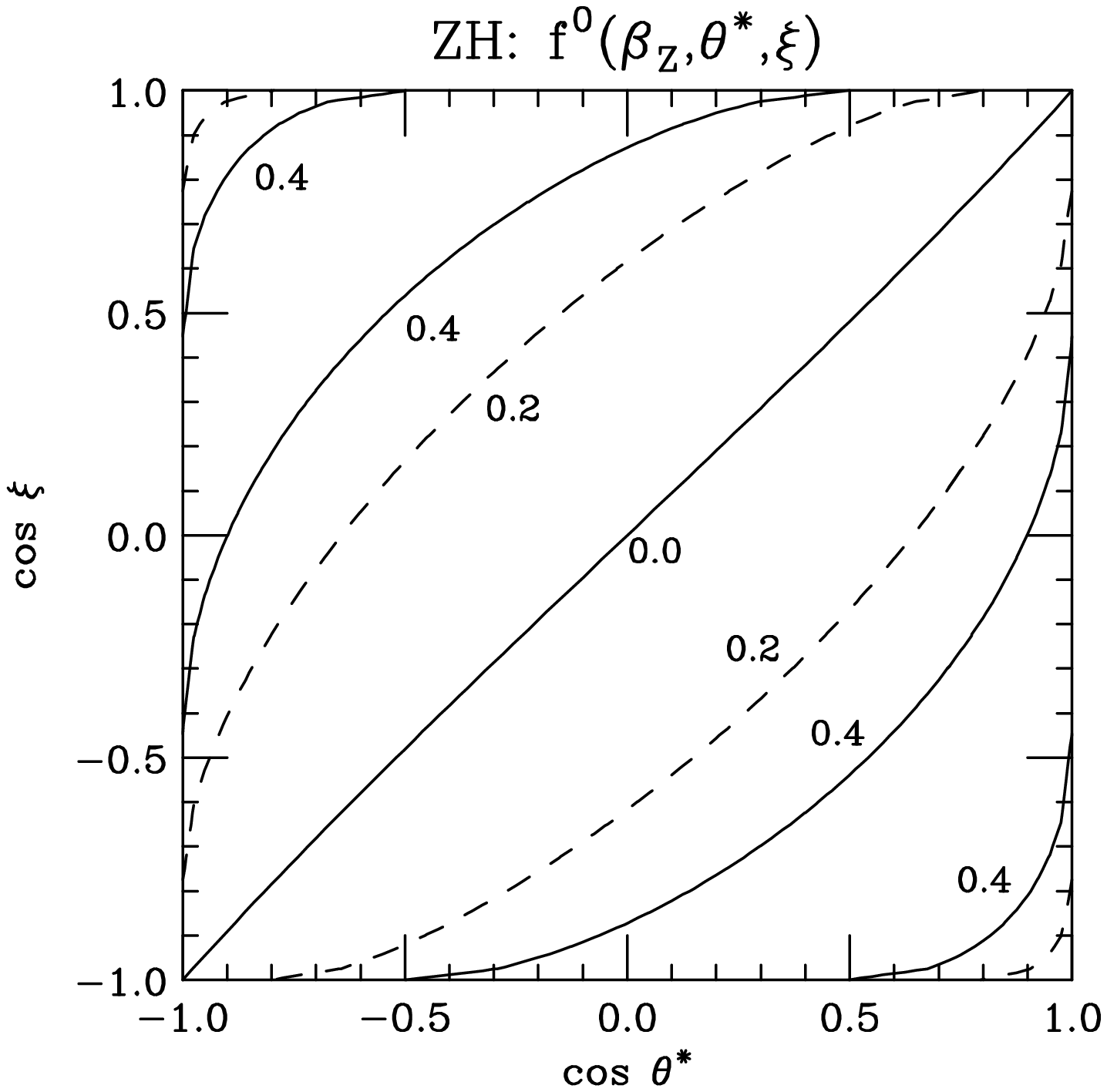}
\vspace{4.0cm}

\caption[]{Structure of the $\eebar\rightarrow ZH$
polarized production amplitudes in the $\cos\thetas$-$\cos\xi$
plane for $\sqrt{s}=192\GeV$ and $M_H = M_Z$ ($\beta_Z=0.313$).  
Plotted is 
$f^{\lambda}(\beta_Z,\thetas,\xi)$,
the fraction of the total amplitude squared in each spin 
configuration [see Eq.~(\protect\ref{ZHfraction-def})].  
In all of the plots, $\sin\xi \ge 0$.
}
\label{ZH-PRODcontours}
\end{figure}


\vspace*{1cm}

\begin{figure}[h]

\vspace*{15cm}
\includegraphics{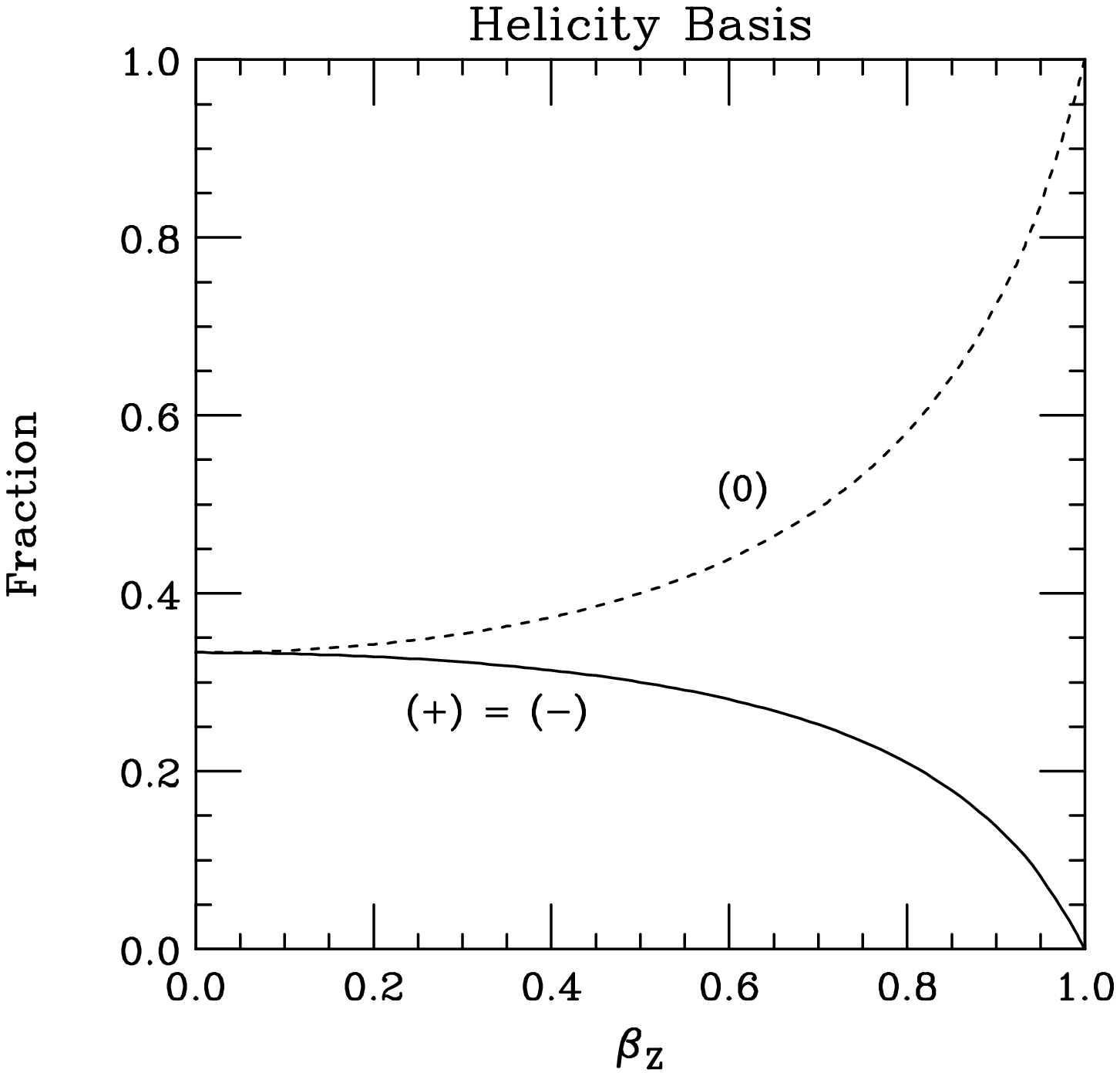}
\includegraphics{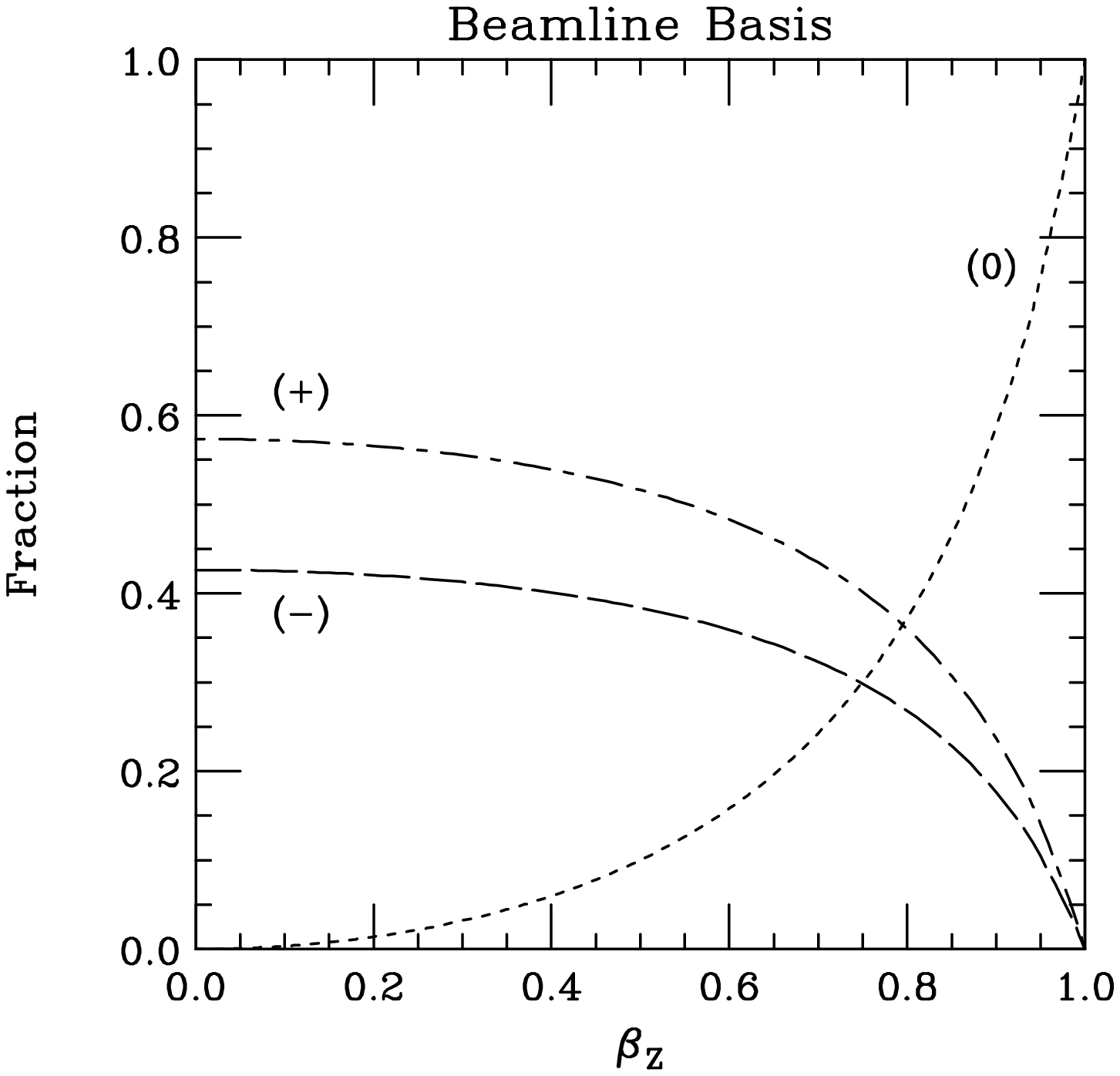}
\includegraphics{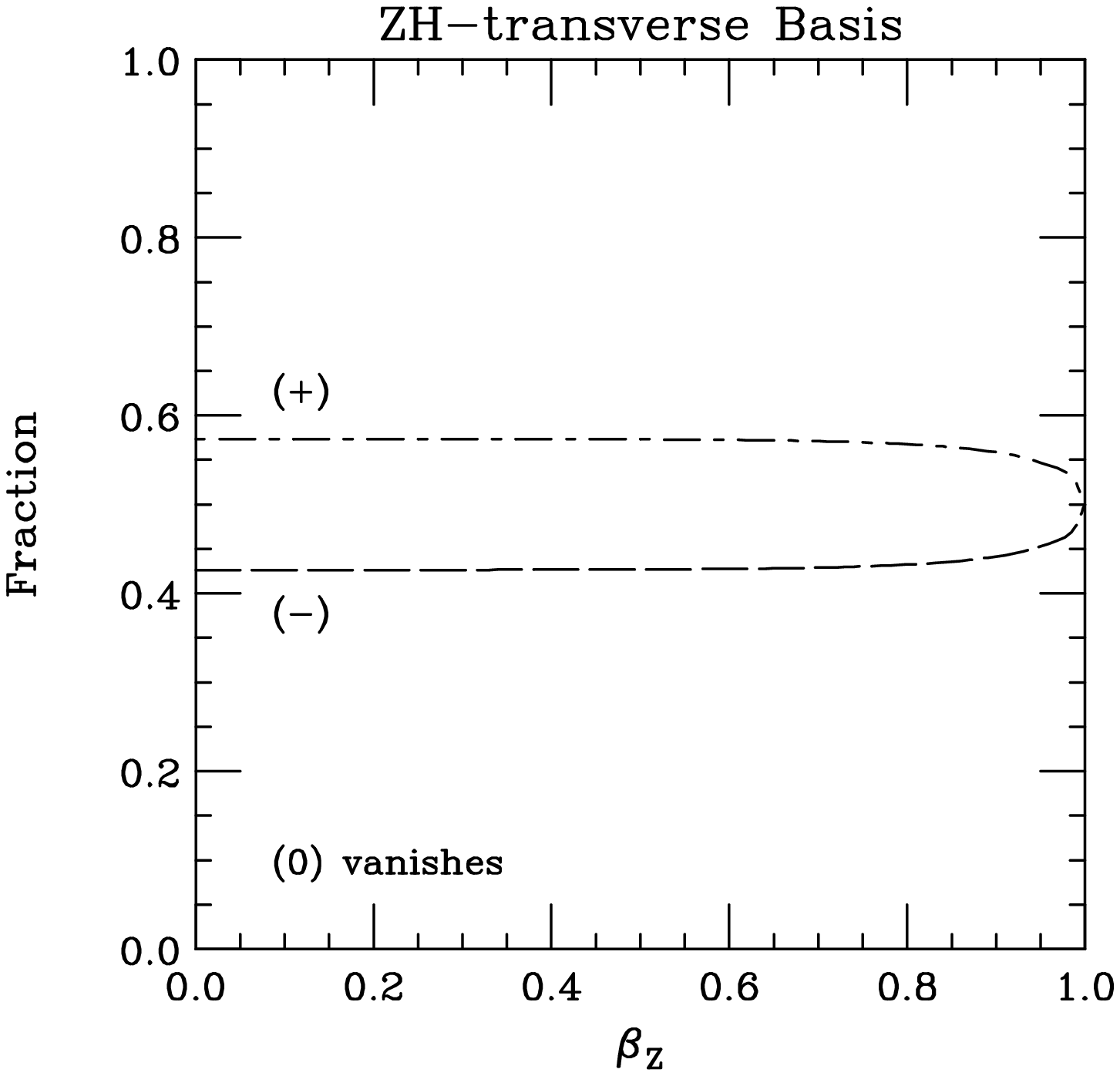}
\includegraphics{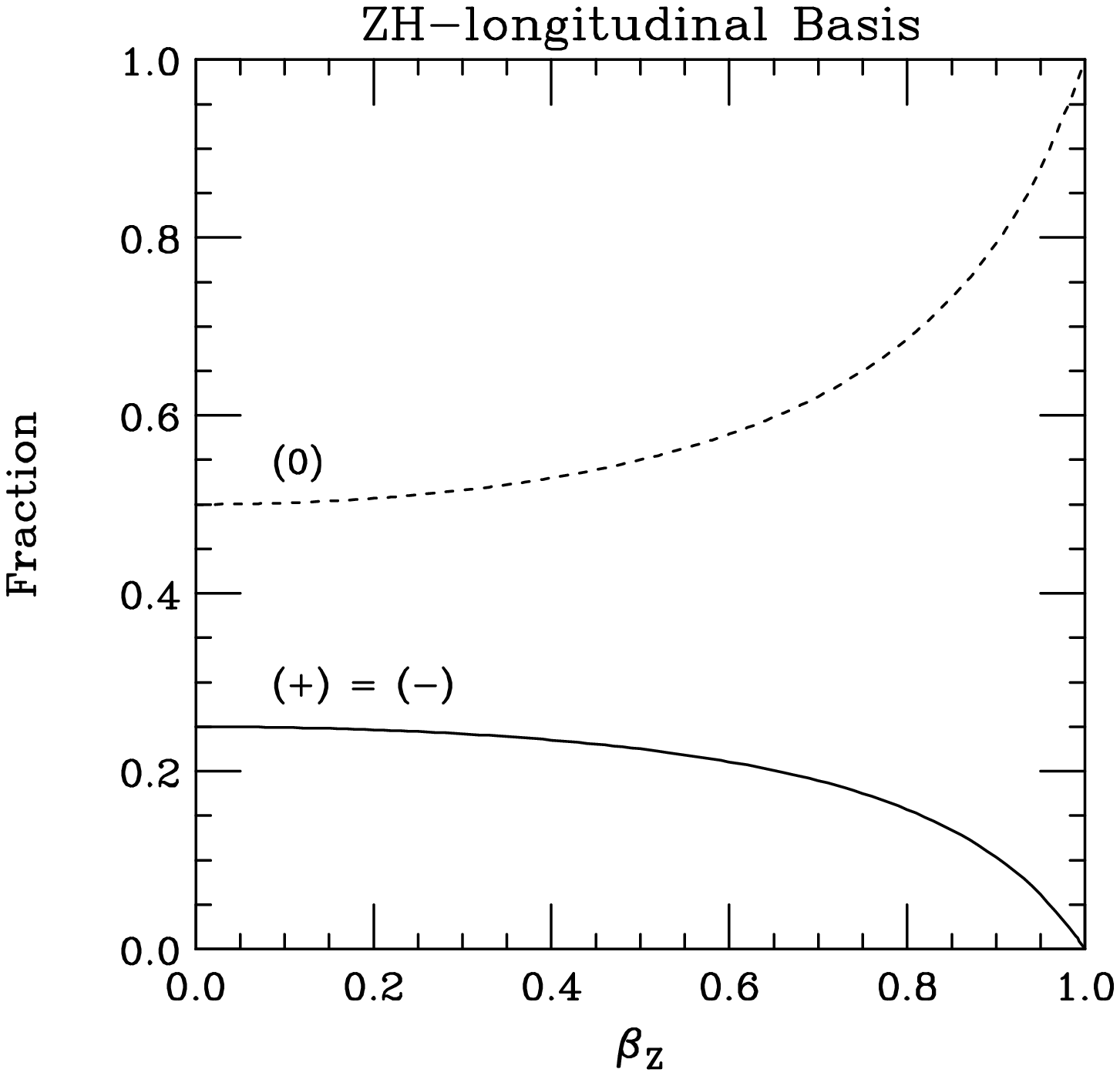}
\vspace{4.0cm}

\caption[]{Spin decomposition of the $e^{+}e^{-}\rightarrow ZH$
cross section as a function of the ZMF speed $\beta_Z$ of the $Z$
boson, assuming that $M_H = M_Z$.  Shown are the fractions of
the total cross section in the $(+)$, $(-)$, and $(0)$ spin states
for the helicity, beamline, \ZHtrans, and \ZHlong\ bases.
}
\label{ZHbetaplot}
\end{figure}


\vspace*{1cm}

\begin{figure}[h]

\vspace*{15cm}
\includegraphics{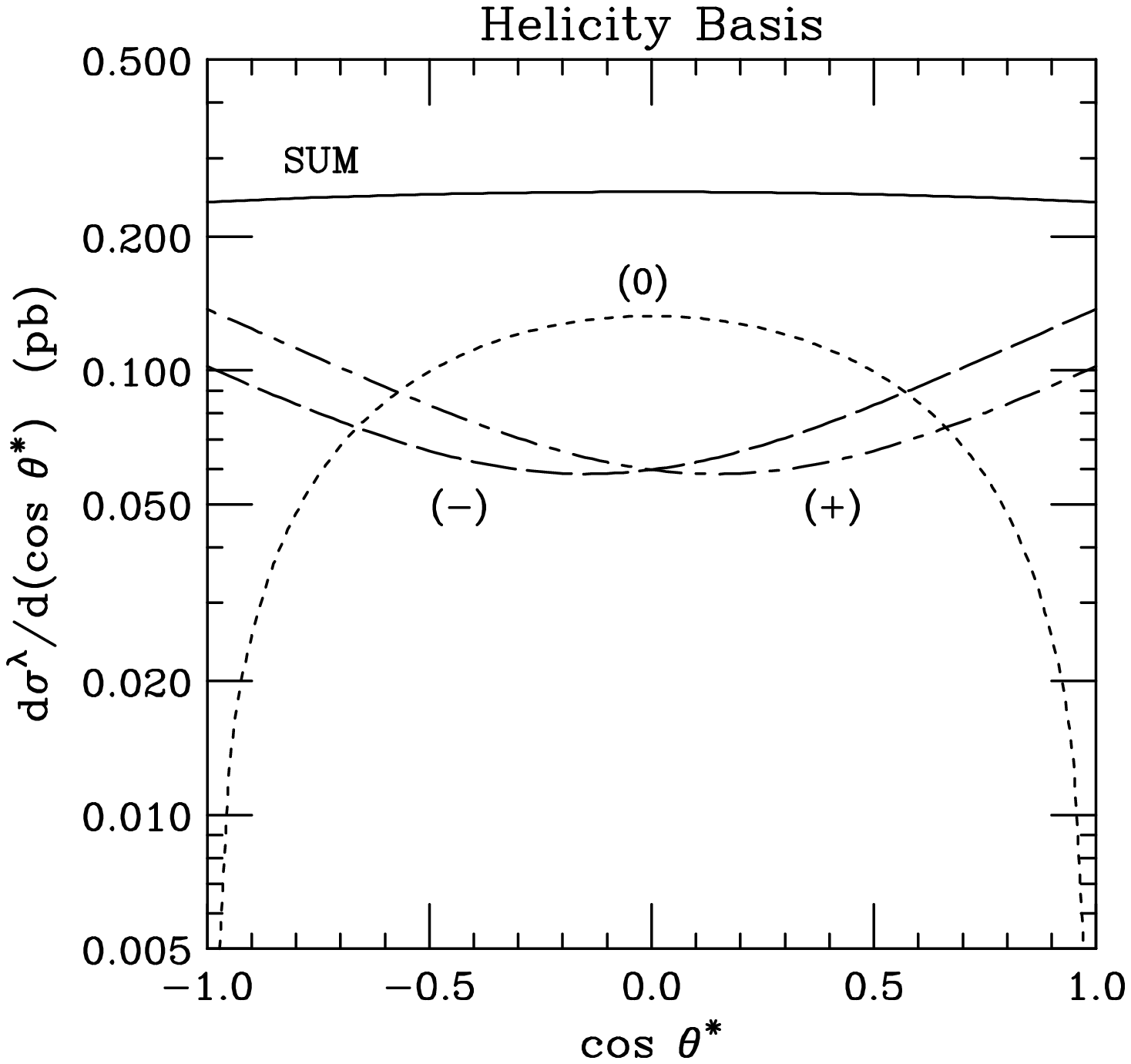}
\includegraphics{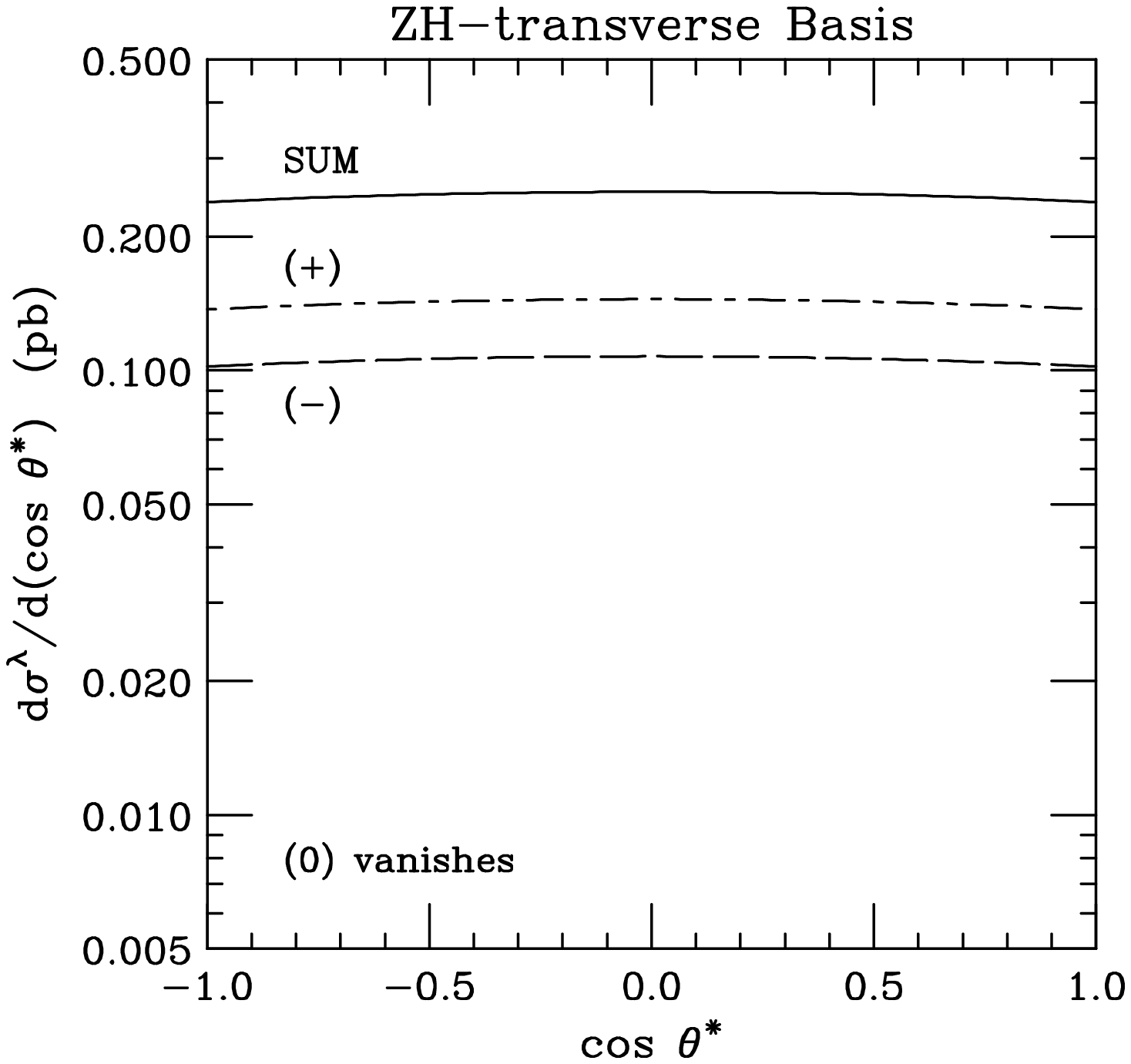}
\vspace{3.0cm}

\caption[]{Distribution in production angle of the 
$\eebar\rightarrow ZH$ cross section 
for $M_H = M_Z$ at $\sqrt{s}=192\GeV$, broken down
into the contributions from the three possible $Z$ spins, for the
helicity and \ZHtrans\ bases.
}
\label{ZH-CTH}
\end{figure}
 

\vspace*{1cm}

\begin{figure}[h]

\vspace*{15cm}
\includegraphics{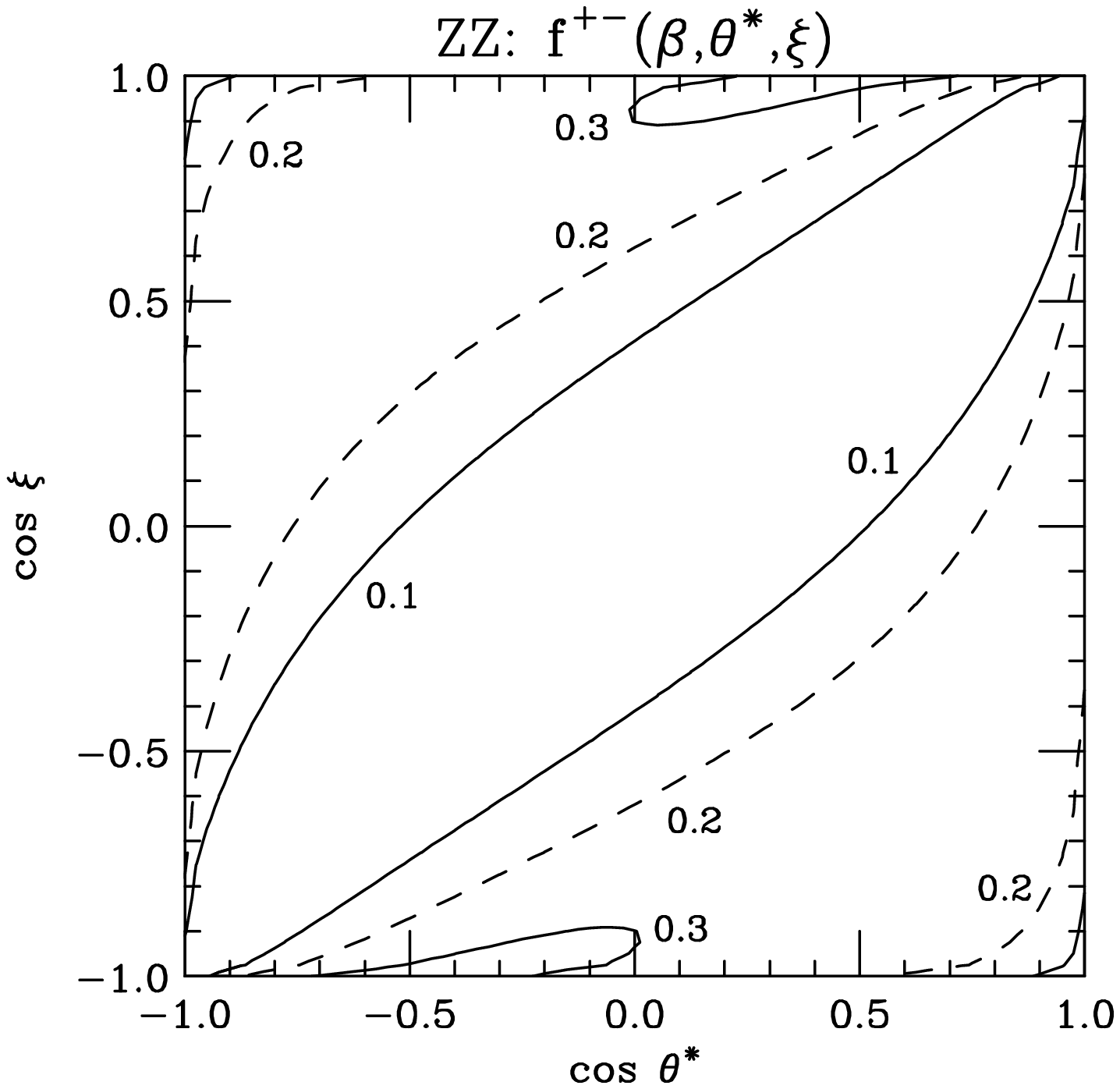}
\includegraphics{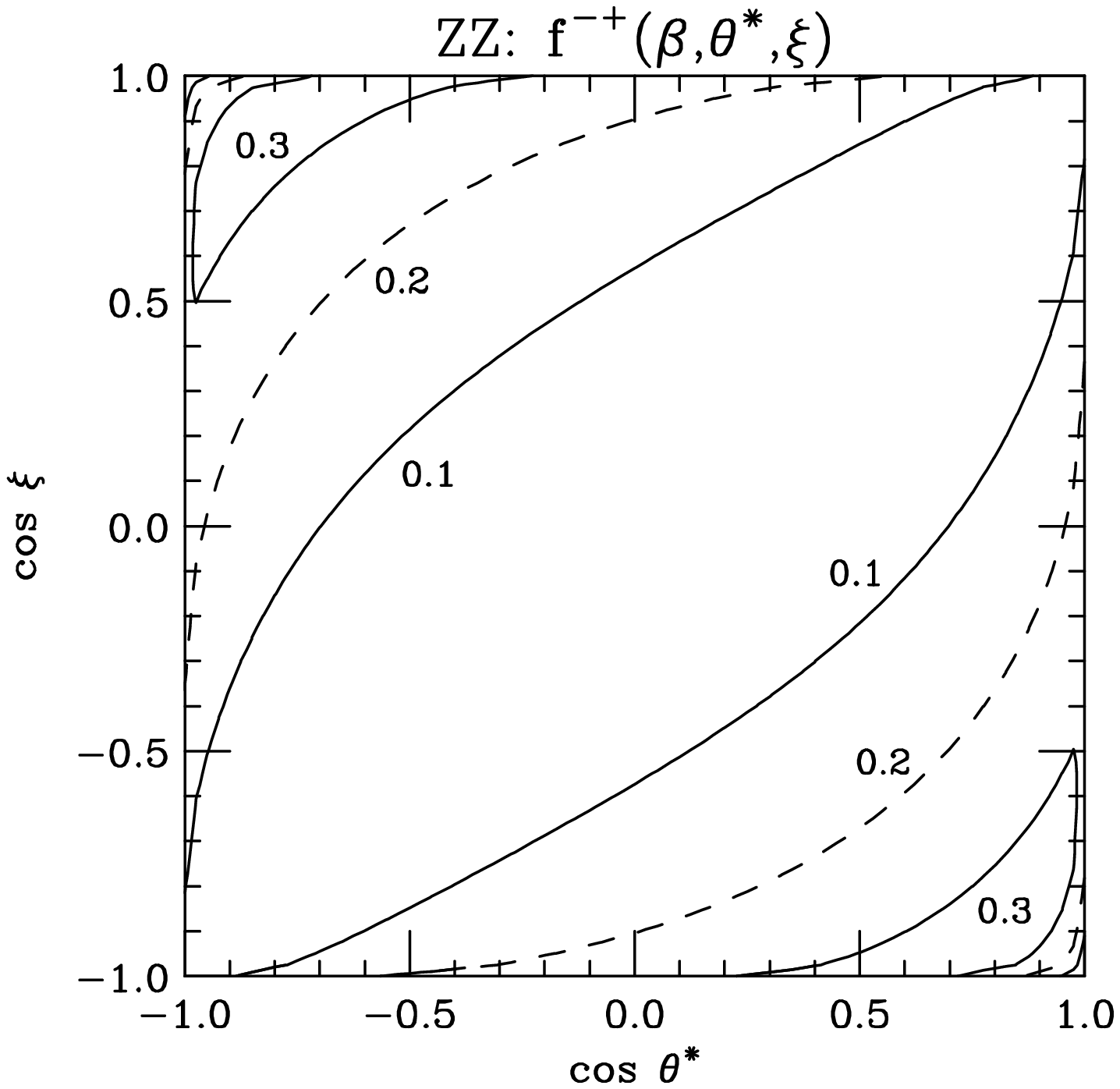}
\vspace{3.0cm}

\caption[]{Structure of the $\eebar\rightarrow ZZ$
polarized production amplitudes in the $\cos\thetas$-$\cos\xi$
plane for $\sqrt{s}=192\GeV$ ($\beta=0.313$).  Plotted is 
$f^{\lambda\bar\lambda}(\beta,\thetas,\xi)$,
the fraction of the total amplitude squared in each spin 
configuration [see Eq.~(\protect\ref{fraction-def})].
In all of the plots, $\sin\xi \ge 0$.
}
\label{ZZ-PRODcontours}
\end{figure}
\eject

\vspace*{1cm}

\begin{figure}[h]

\vspace*{15cm}
\includegraphics{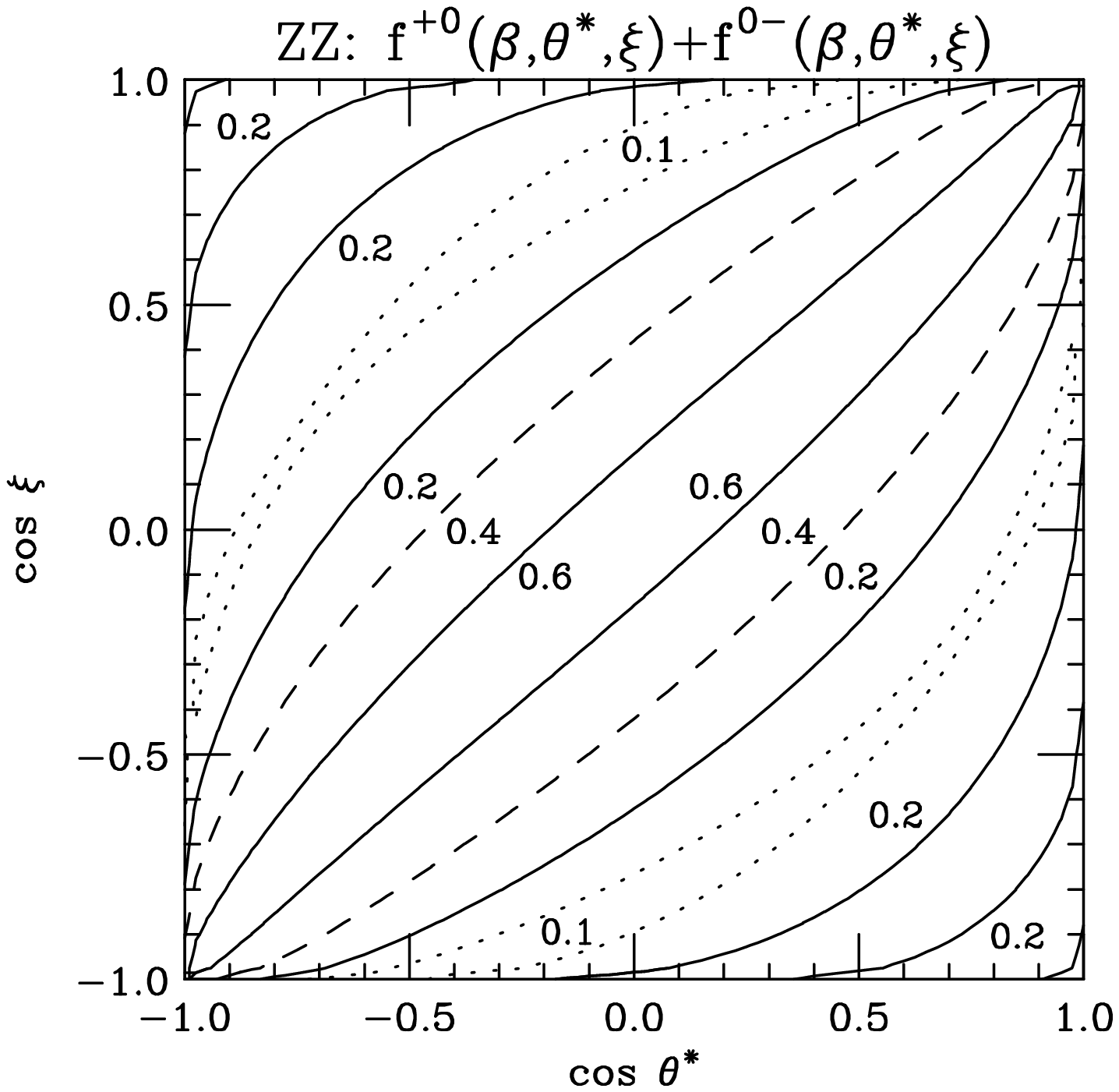}
\includegraphics{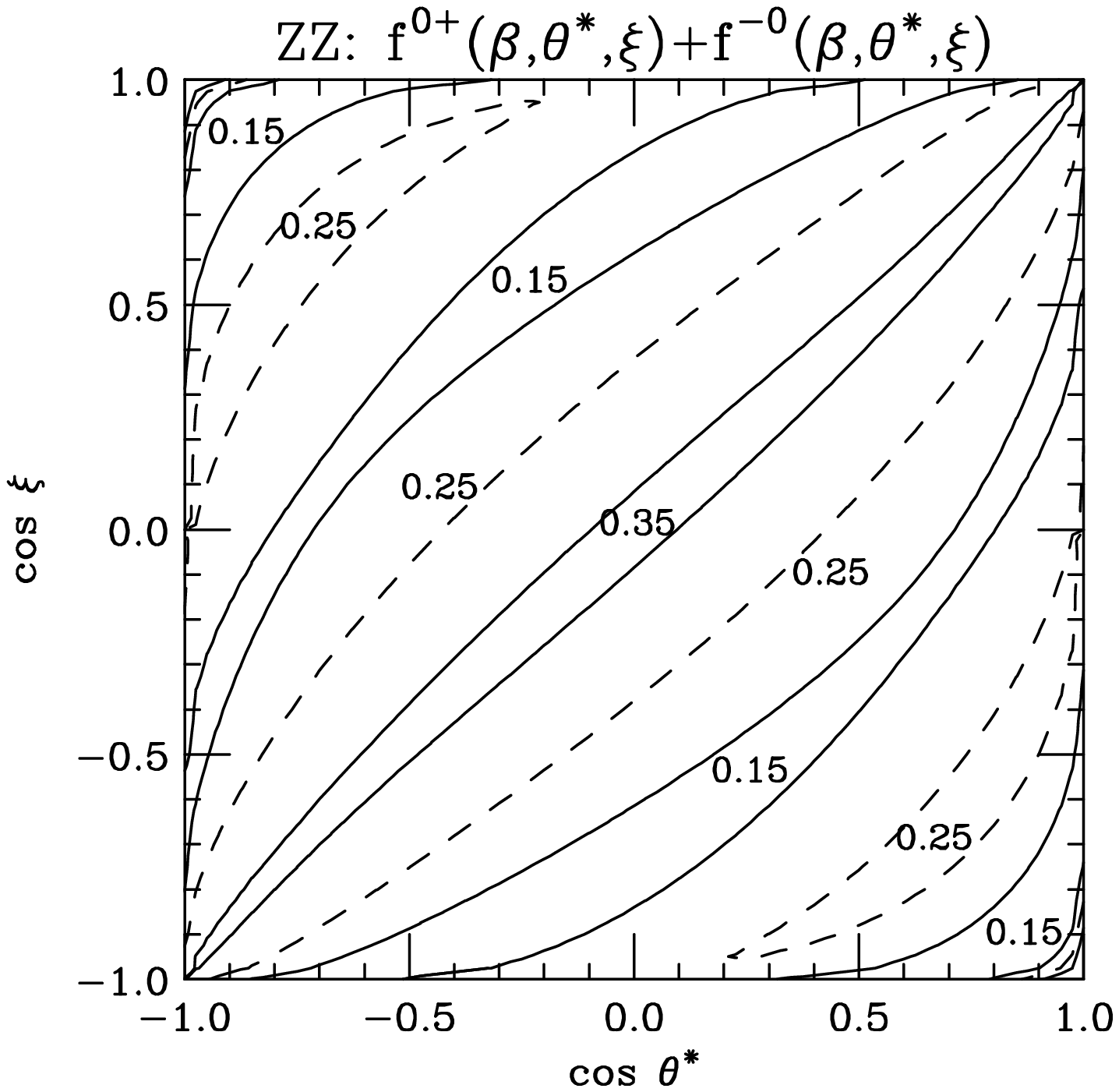}
\vspace{3.0cm}

\end{figure}
\vfill{\hfill FIG. \ref{ZZ-PRODcontours}. (continued)\hfill}
\eject

\vspace*{1cm}

\begin{figure}[h]

\vspace*{15cm}
\includegraphics{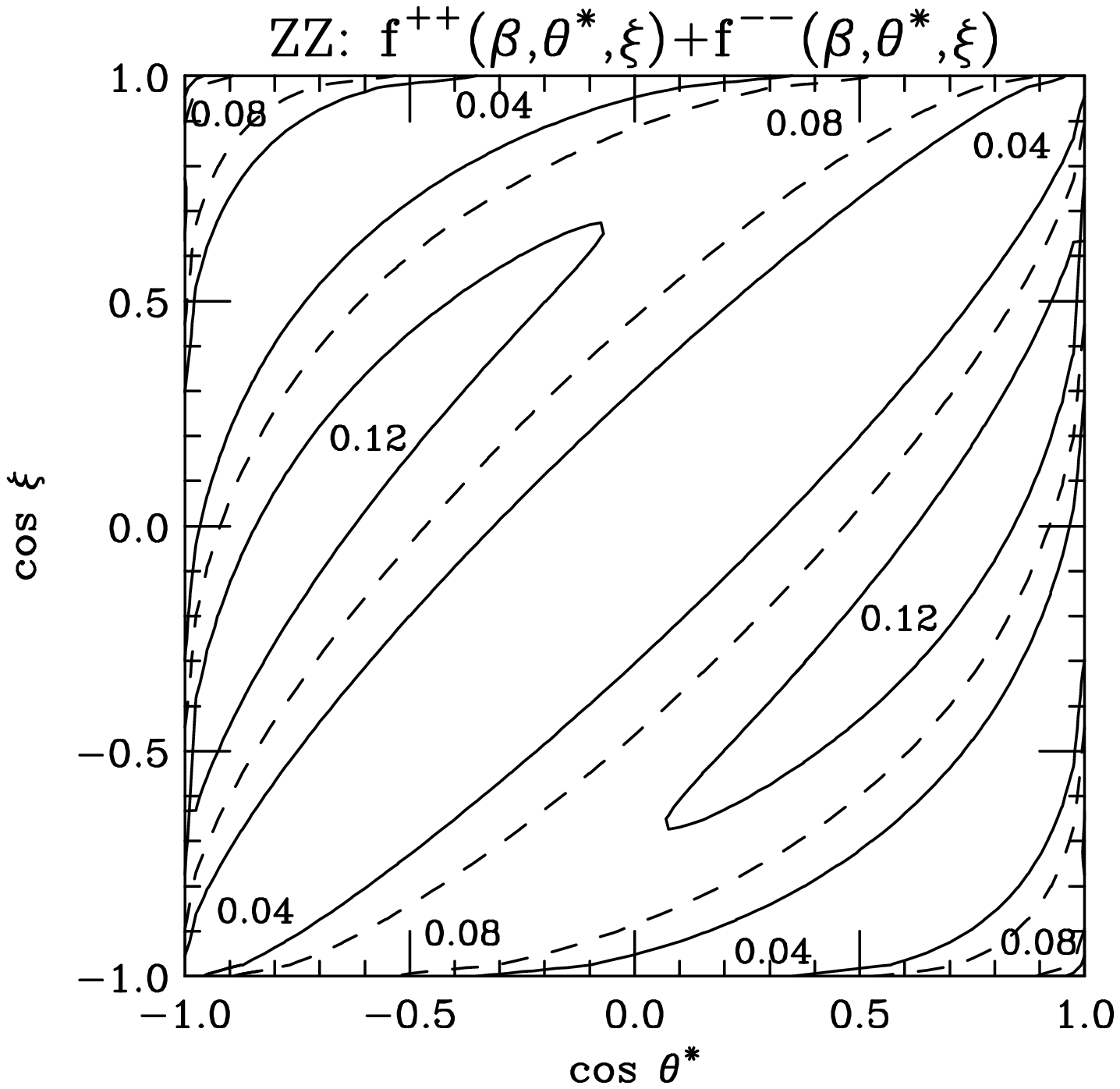}
\includegraphics{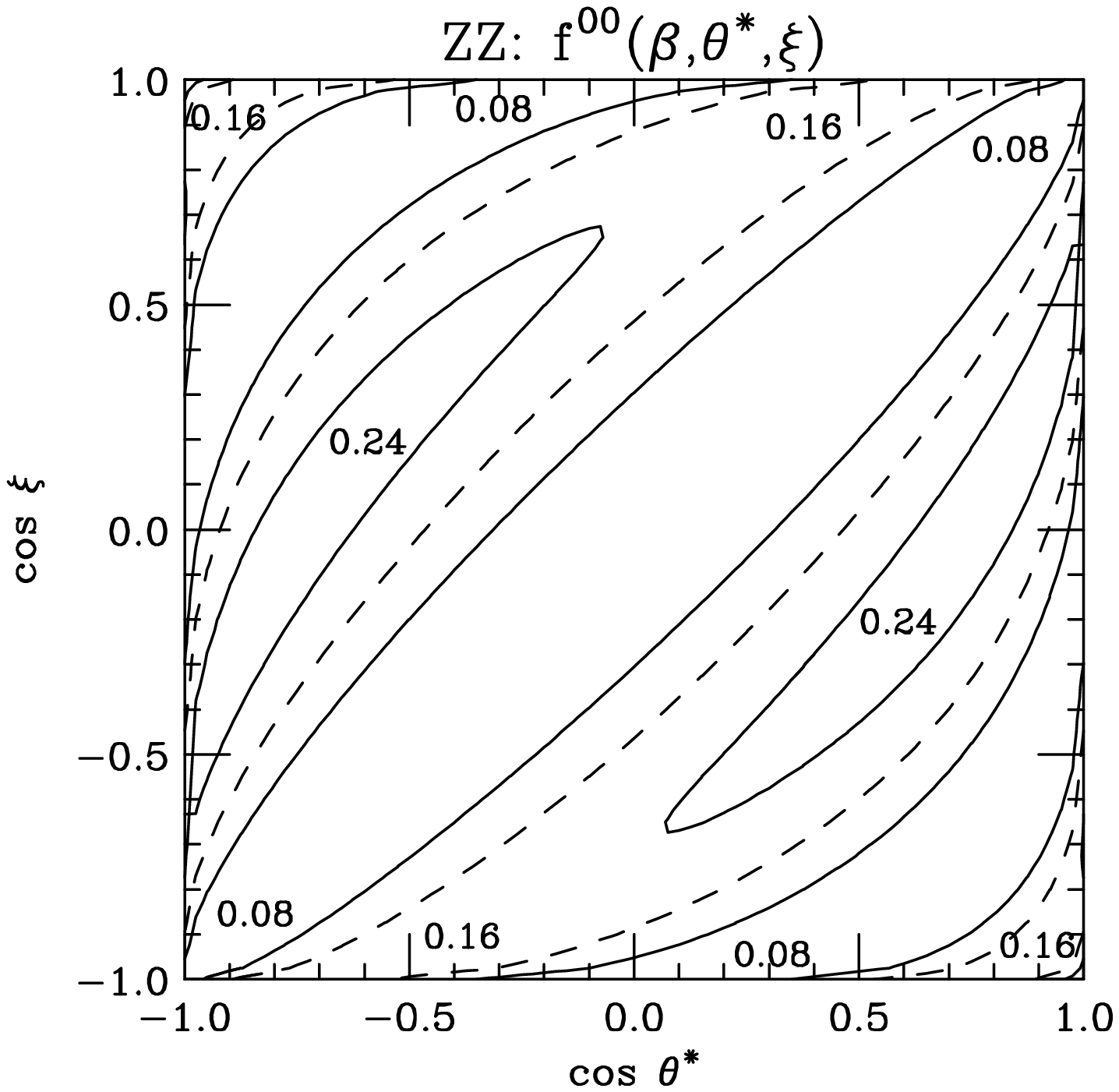}
\vspace{3.0cm}

\end{figure}
\vfill{\hfill FIG. \ref{ZZ-PRODcontours}. (continued)\hfill}
\eject


\vspace*{1cm}

\begin{figure}[h]

\vspace*{15cm}
\includegraphics{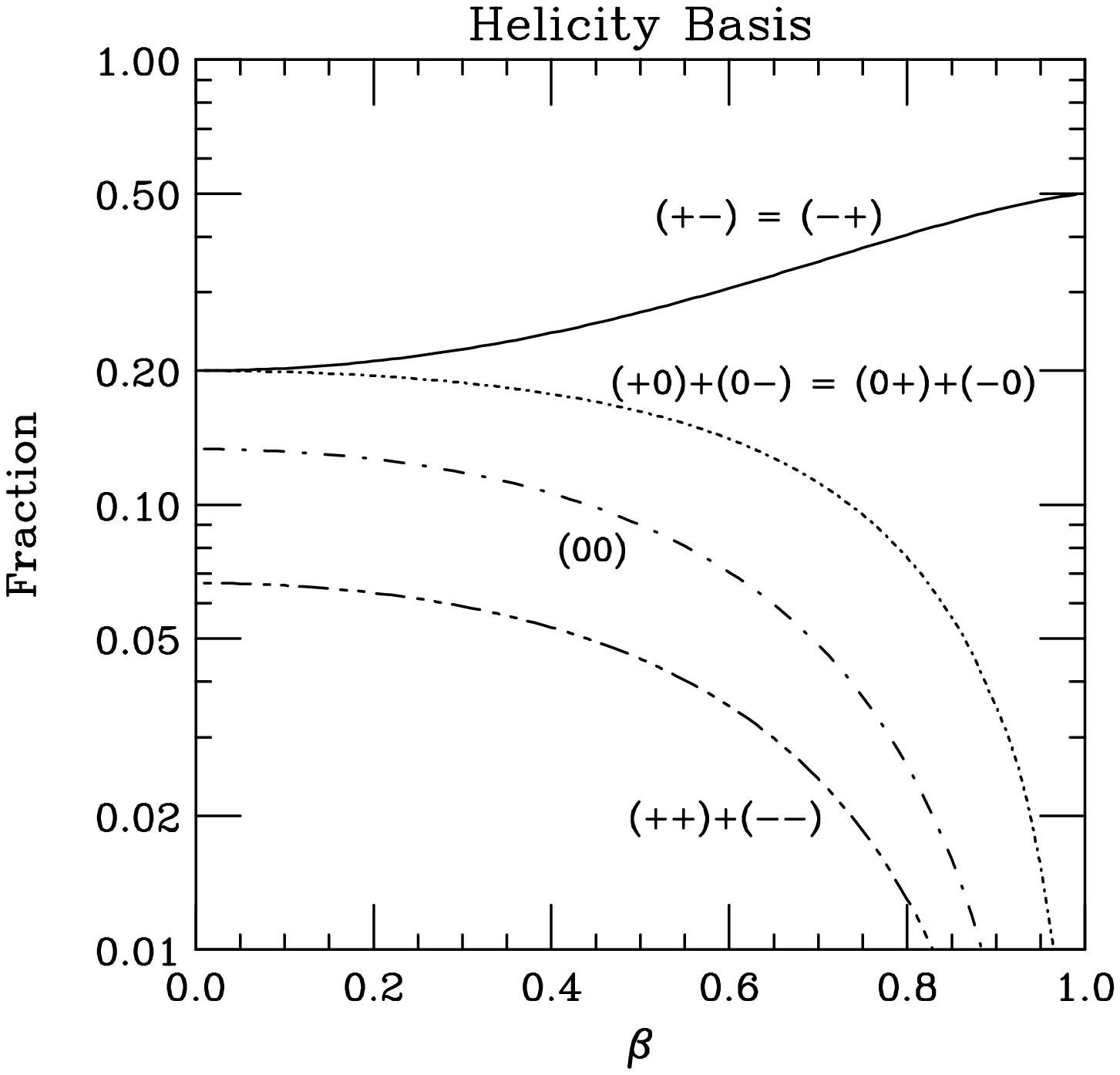}
\includegraphics{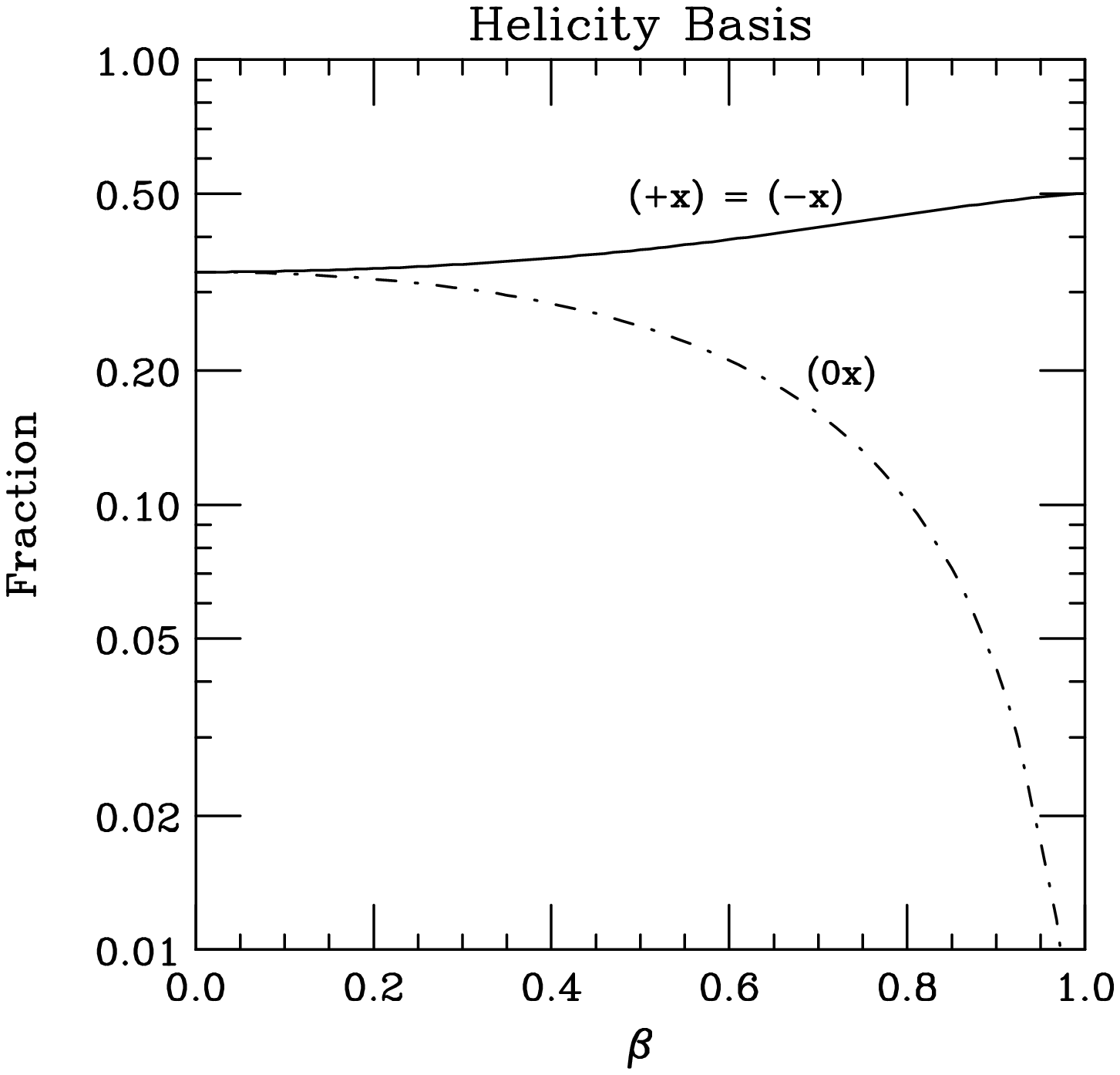}
\includegraphics{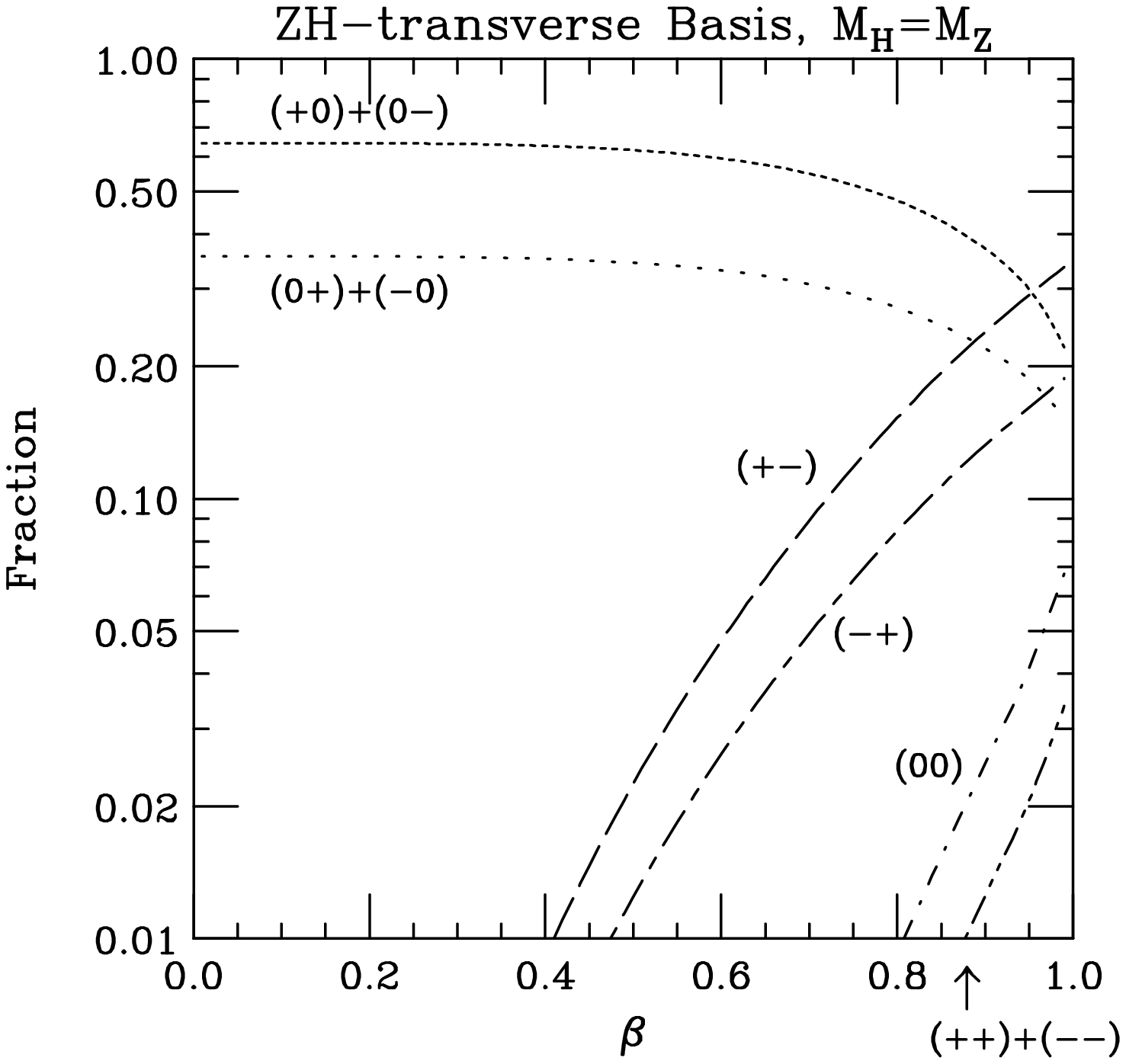}
\includegraphics{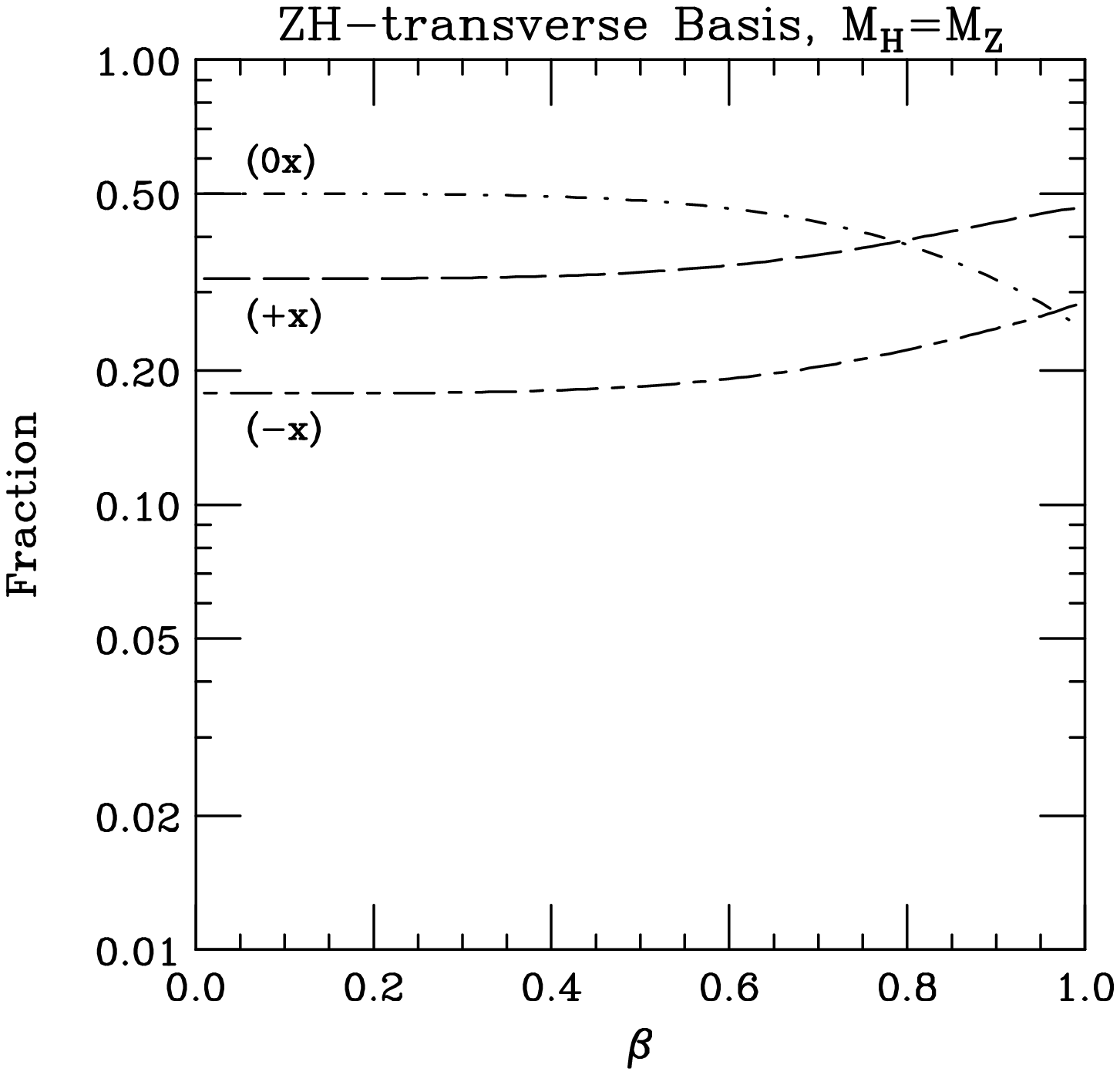}
\vspace{3.0cm}

\caption[]{Spin decomposition of the $e^{+}e^{-}\rightarrow ZZ$
cross section as a function of the ZMF speed $\beta$ of the $Z$
boson.  The plots on the left show the fractions of
the total cross section in the $(00)$, $(+-)$, $(-+)$, 
$[(+0)+(0-)]$, $[(0+)+(-0)]$, and $[(++)+(--)]$ spin states
for the helicity and \ZHtrans\ bases.  The plots on 
the right show inclusive fractions where we have summed over
all possible spins of the other $Z$, {\it e.g.}
$(0x) \equiv (0+)+(00)+(0-)$.   
The curves for the \ZHtrans\
basis are drawn for the case $M_H = M_Z$.
}
\label{ZZbetaplot}
\end{figure}


\vspace*{1cm}

\begin{figure}[h]

\vspace*{15cm}
\includegraphics{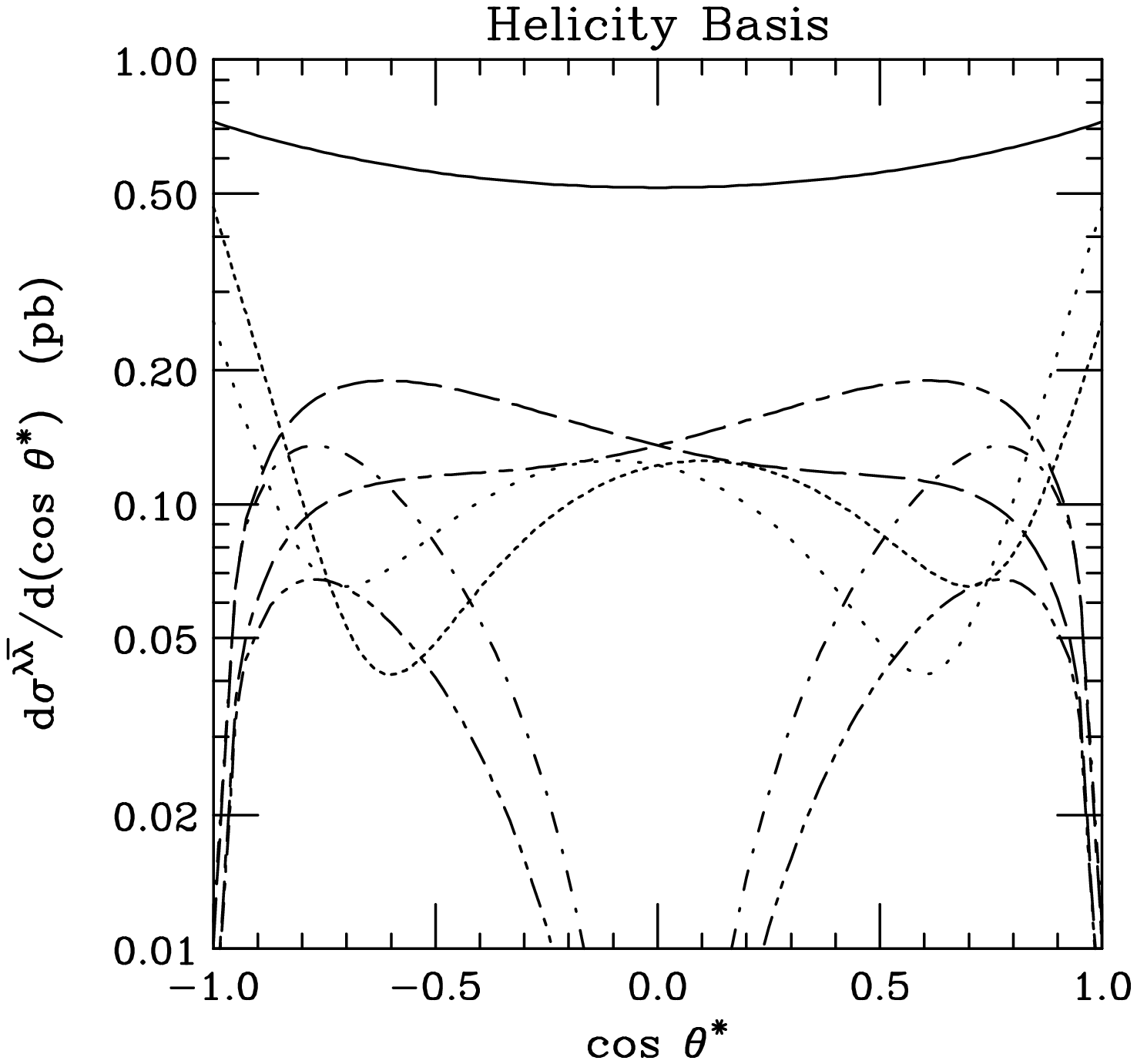}
\includegraphics{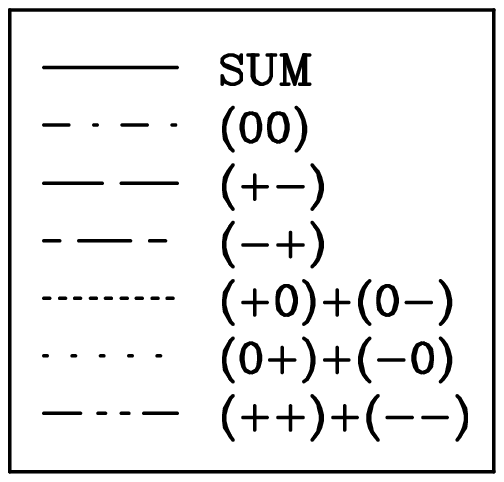}
\includegraphics{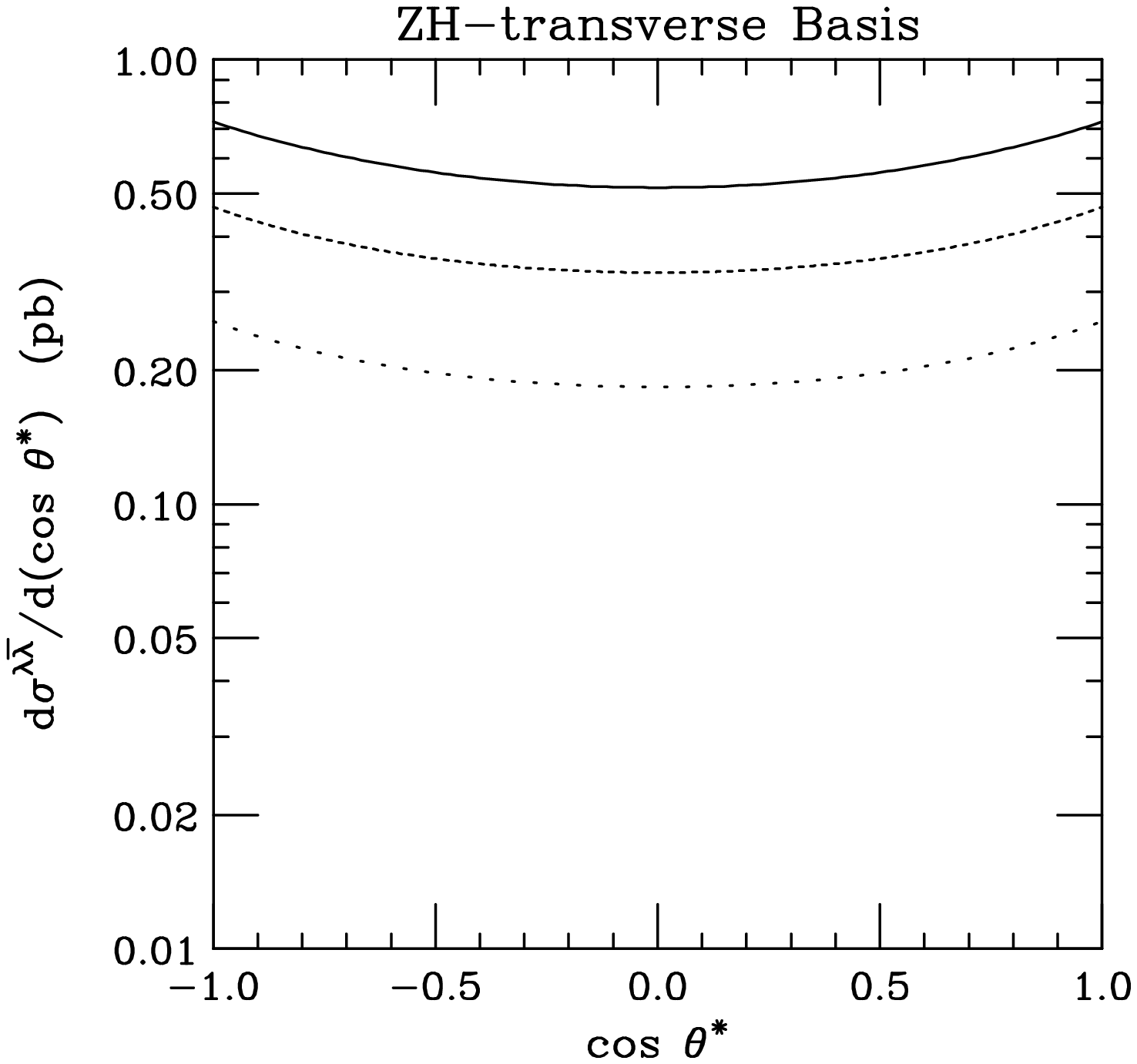}
\vspace{3.0cm}

\caption[]{Distribution in production angle of the 
$\eebar\rightarrow ZZ$
cross section at $\sqrt{s}=192\GeV$, broken down
into the six independent spin combinations for the helicity
and \ZHtrans\ (with $M_H=M_Z$) spin bases.  Only two
spin components contribute
above the 1\% level in the \ZHtrans\ basis.
}
\label{ZZ-CTH}
\end{figure}


\vspace*{1cm}

\begin{figure}[h]

\vspace*{15cm}
\includegraphics{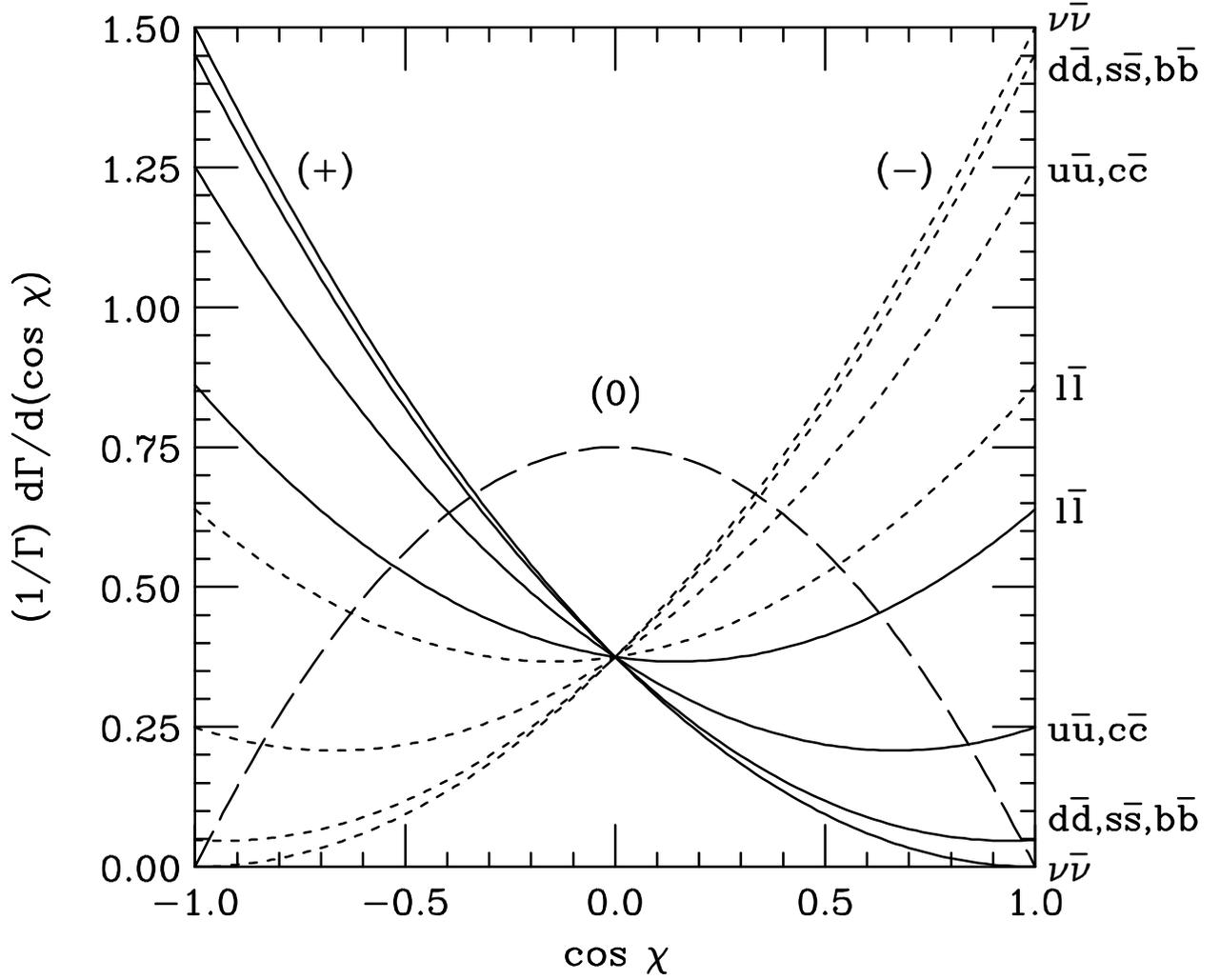}
\vspace{1.0cm}

\caption[]{Angular distributions for the decay of a polarized
$Z$.  $\chi$ is the angle between the direction of the 
fermion ($\nu, e, \mu,\tau,u,d,s,c,$ or $b$) and the chosen spin axis, 
as viewed in the $Z$ rest frame.
}
\label{ZDECAYDIST}
\end{figure}


\vspace*{1cm}

\begin{figure}[h]

\vspace*{15cm}
\includegraphics{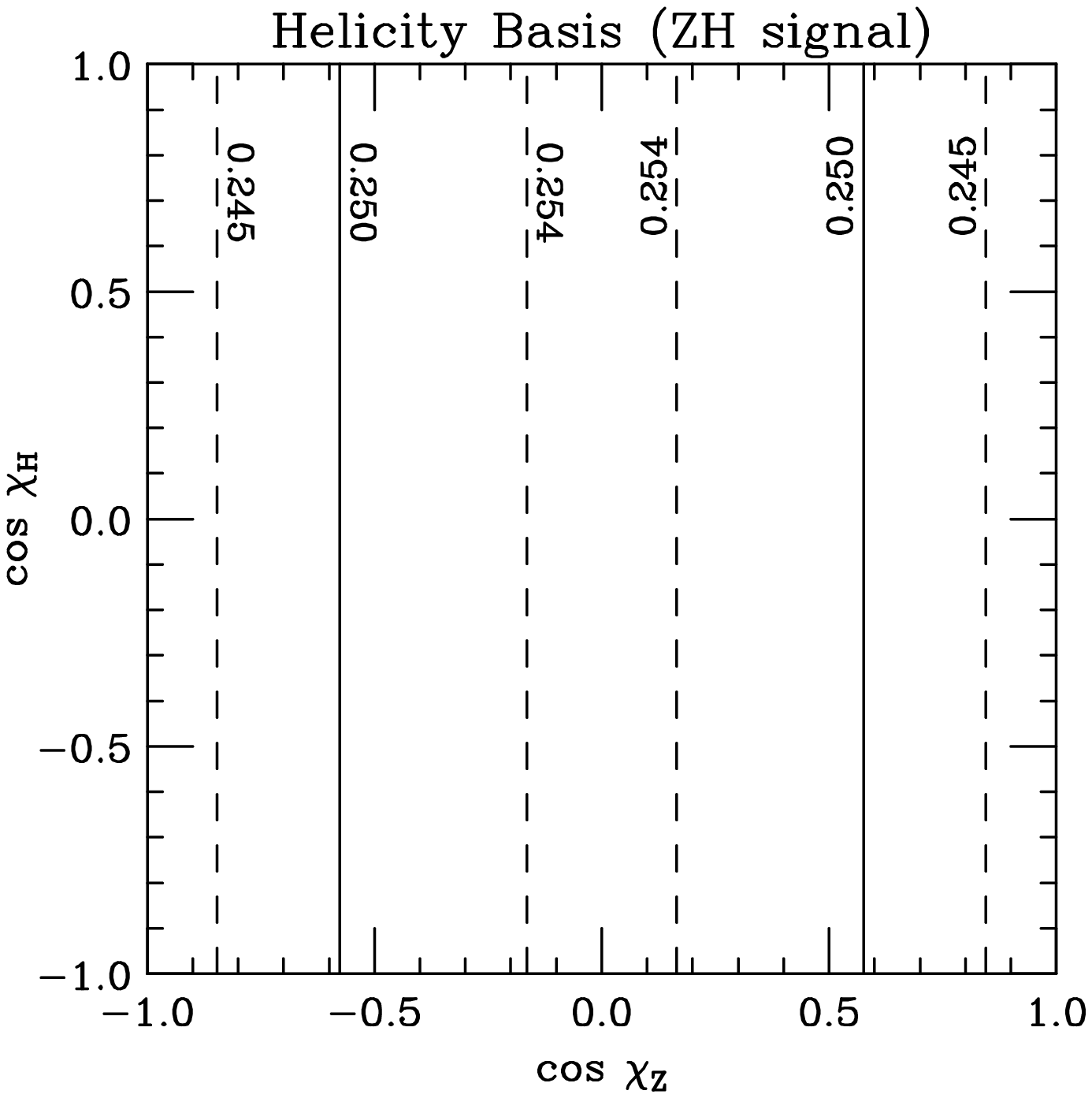}
\includegraphics{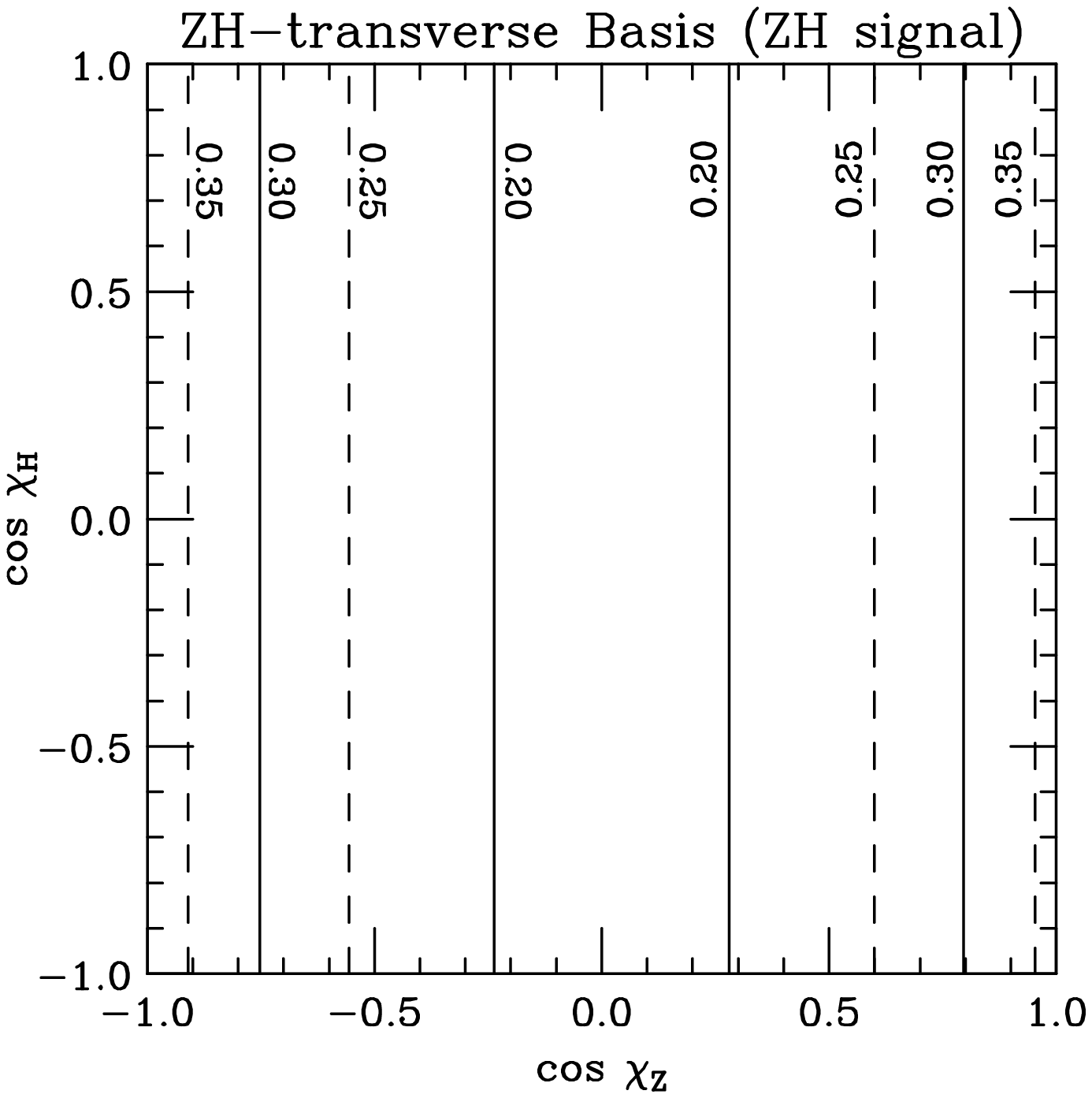}
\includegraphics{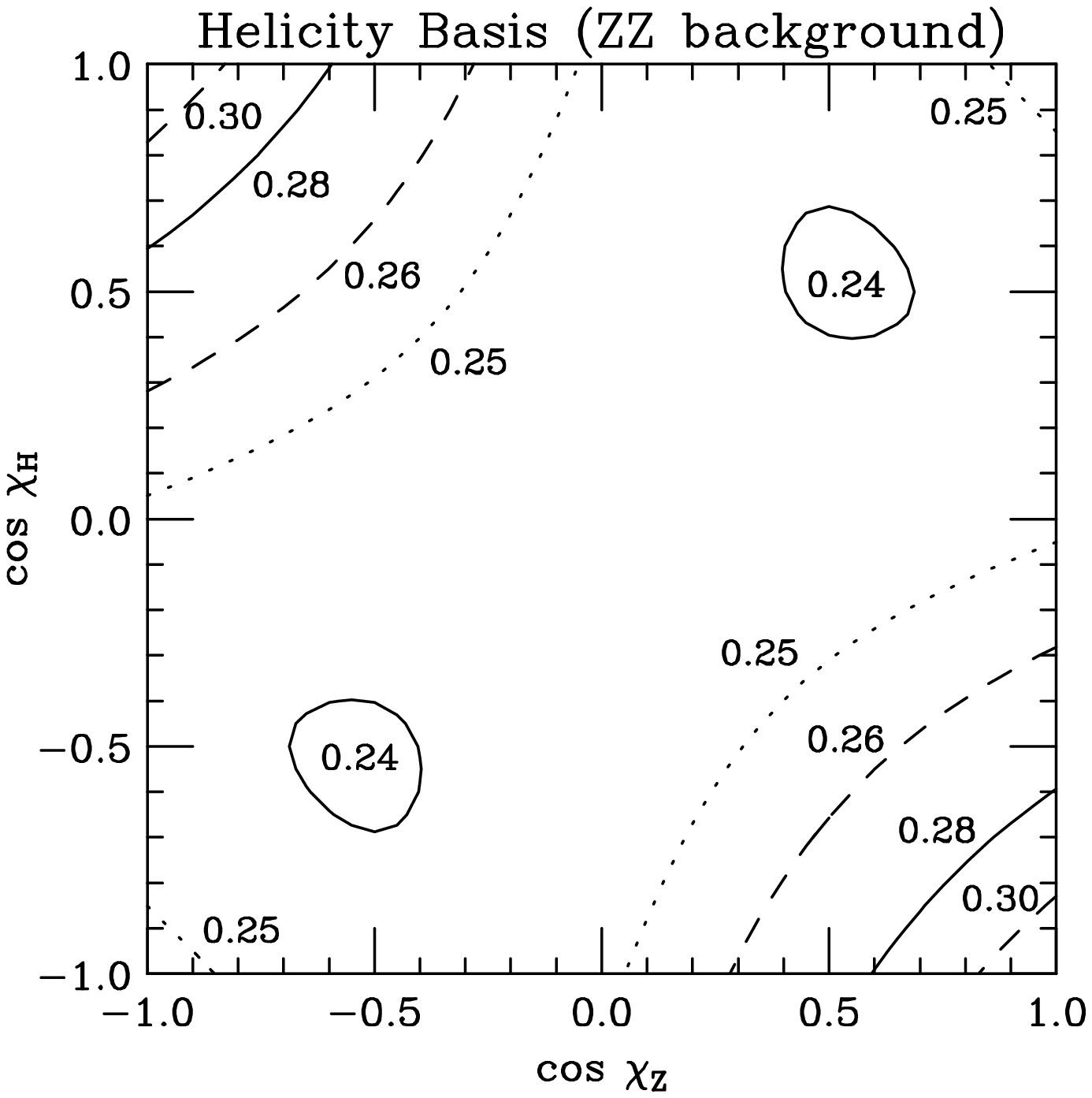}
\includegraphics{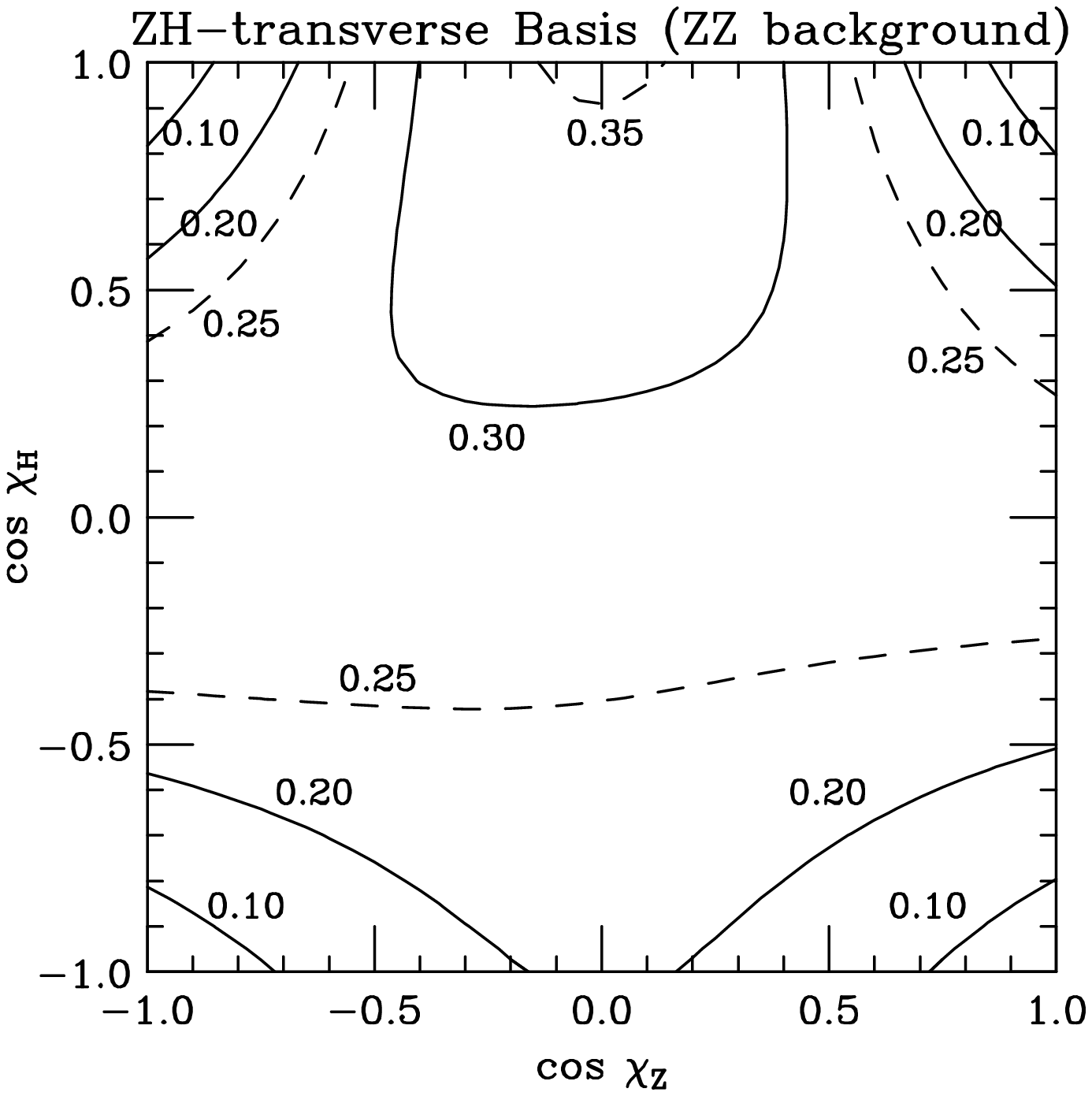}
\vspace{3.0cm}

\caption[]{Double differential decay distributions 
$(1/\sigma)\ts d^2\sigma/[d(\cos\chi_Z)d(\cos\chi_H)]$
for the processes
$e^{+}e^{-} \rightarrow ZH/ZZ \rightarrow \ell\bar\ell b \bar{b}$
in the helicity and \ZHtrans\ bases, using $M_H=M_Z$.  
$\chi_Z$ ($\chi_H$) is the angle between the lepton 
($b$ jet) and the spin axis,
as viewed in the $\ell\bar\ell$ ($b\bar{b}$) rest frame.
In the absence of charge identification for the $b$'s,
these plots should be folded about $\cos\chi_H = 0$.
For completely uncorrelated decays,
this distribution would have a uniform value of 1/4.
}
\label{ZZ-ZHdecayCONTOURS}
\end{figure}


\vspace*{1cm}

\begin{figure}[h]

\vspace*{15cm}
\includegraphics{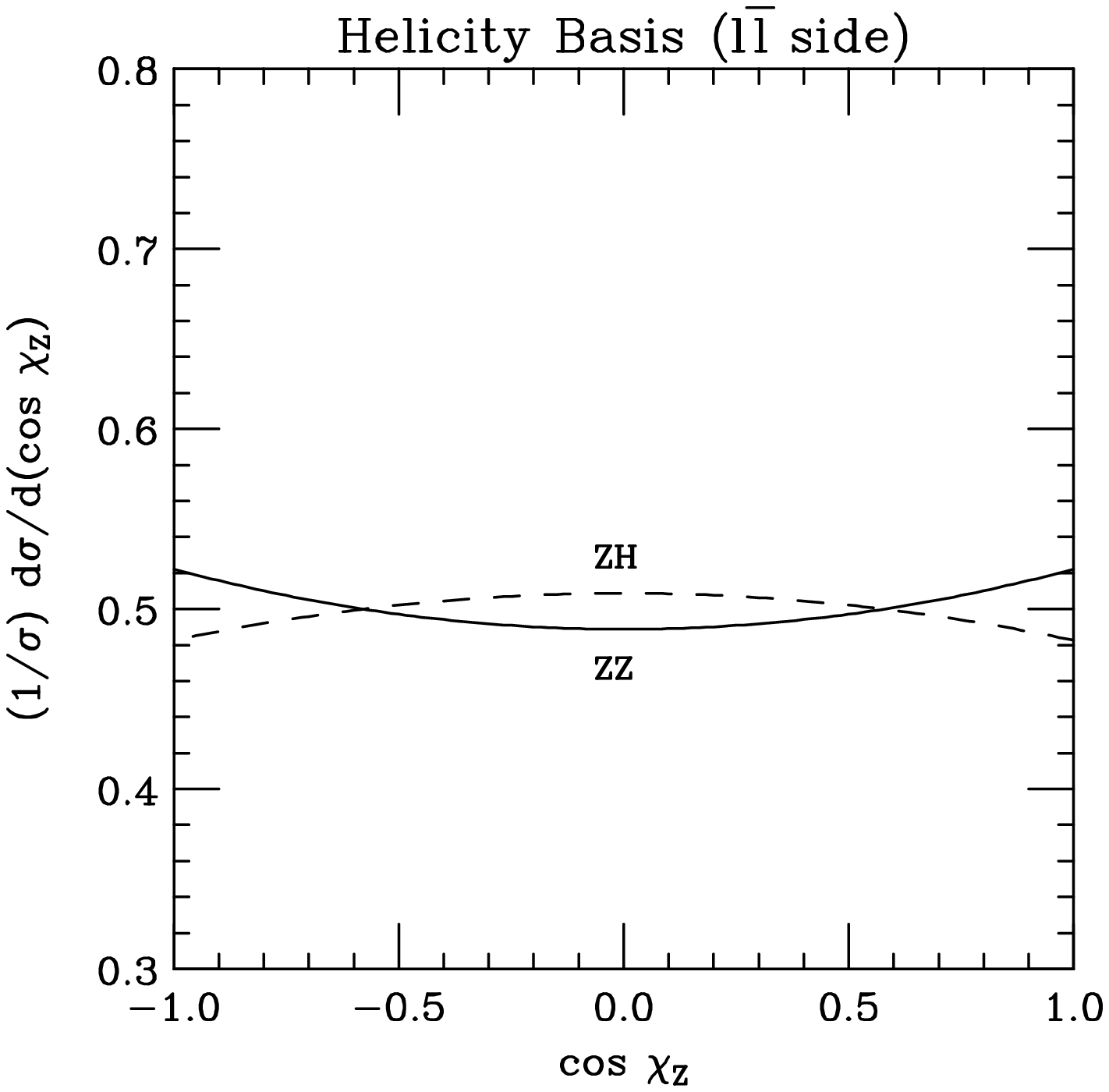}
\includegraphics{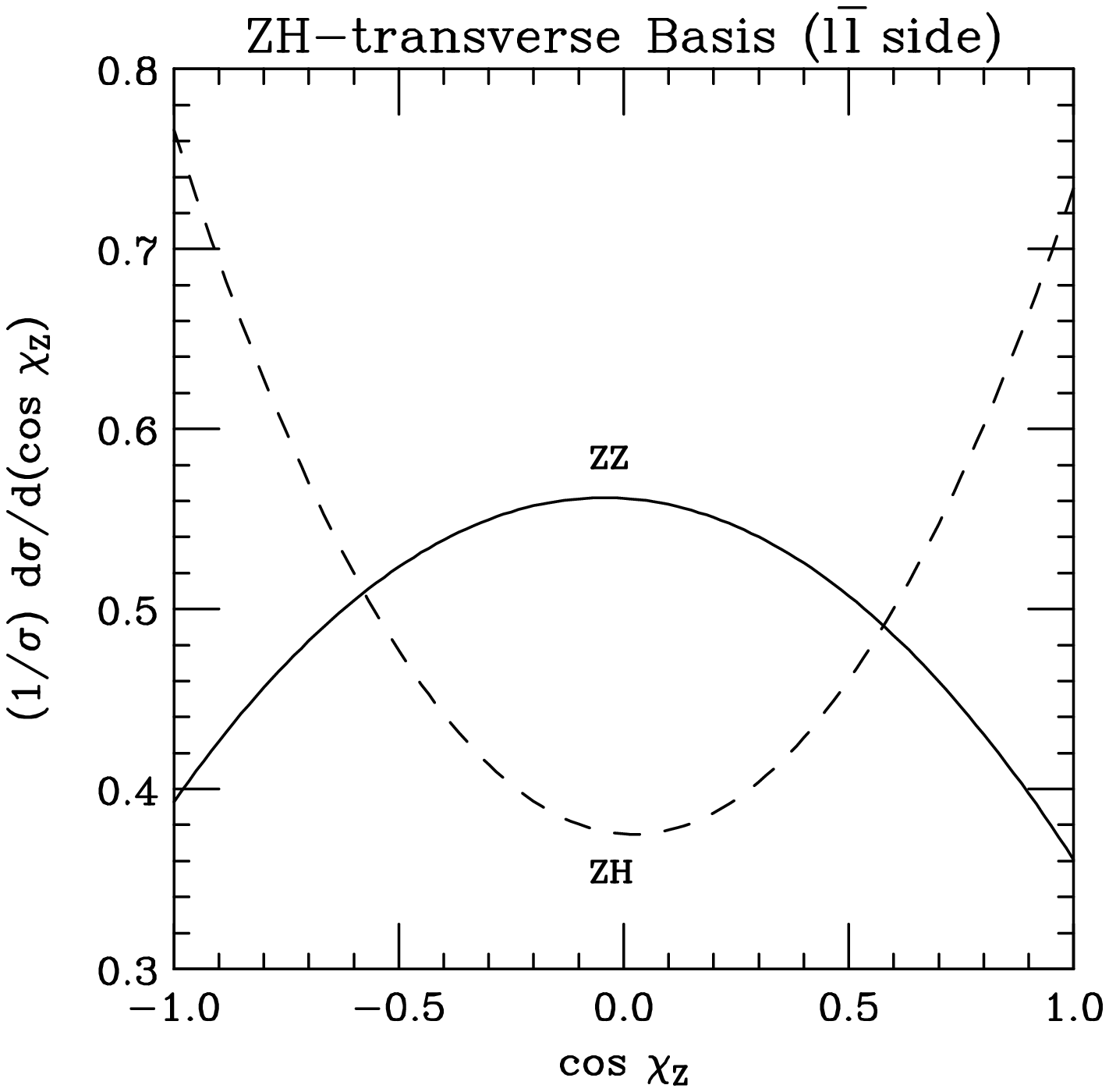}
\includegraphics{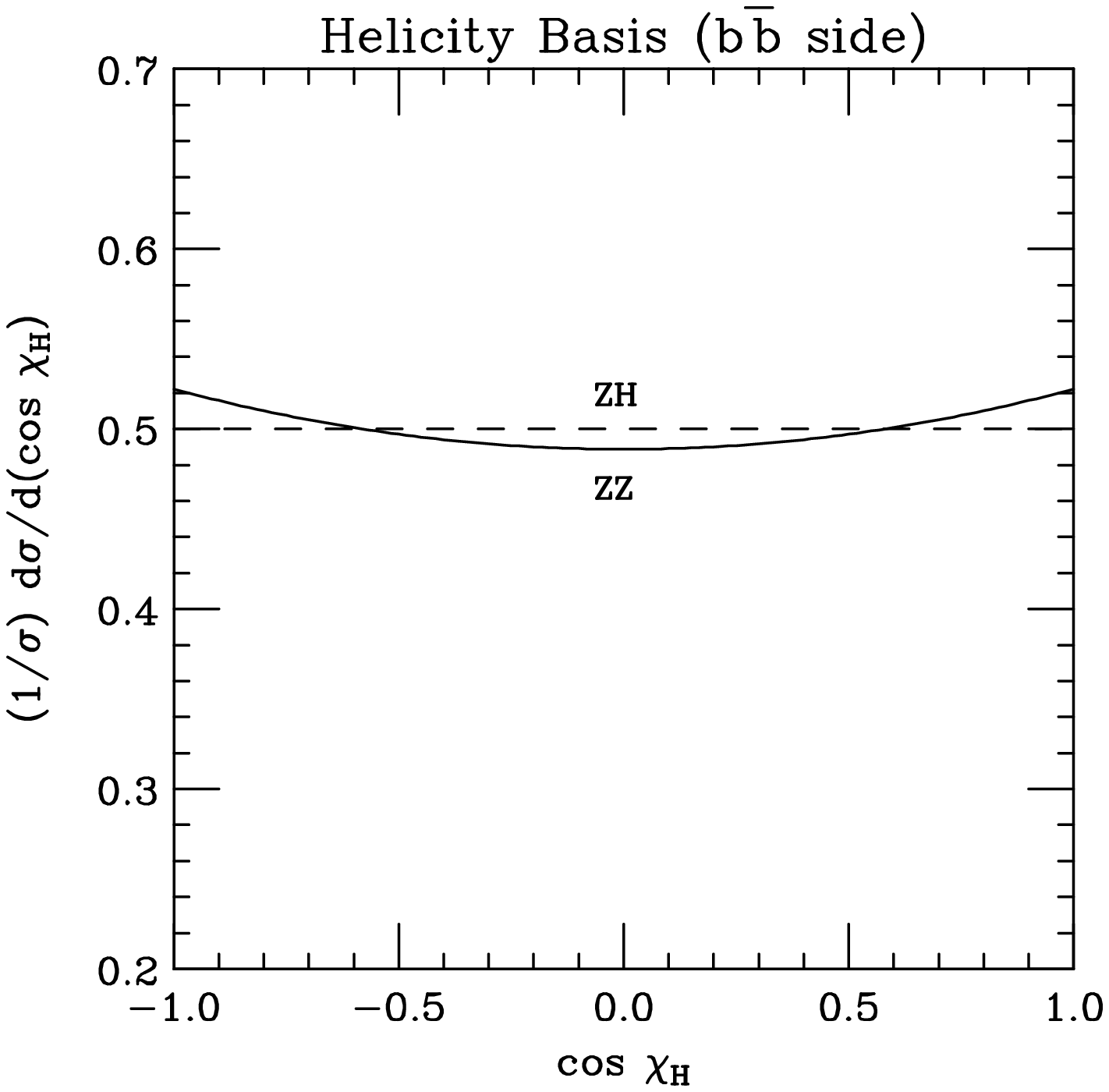}
\includegraphics{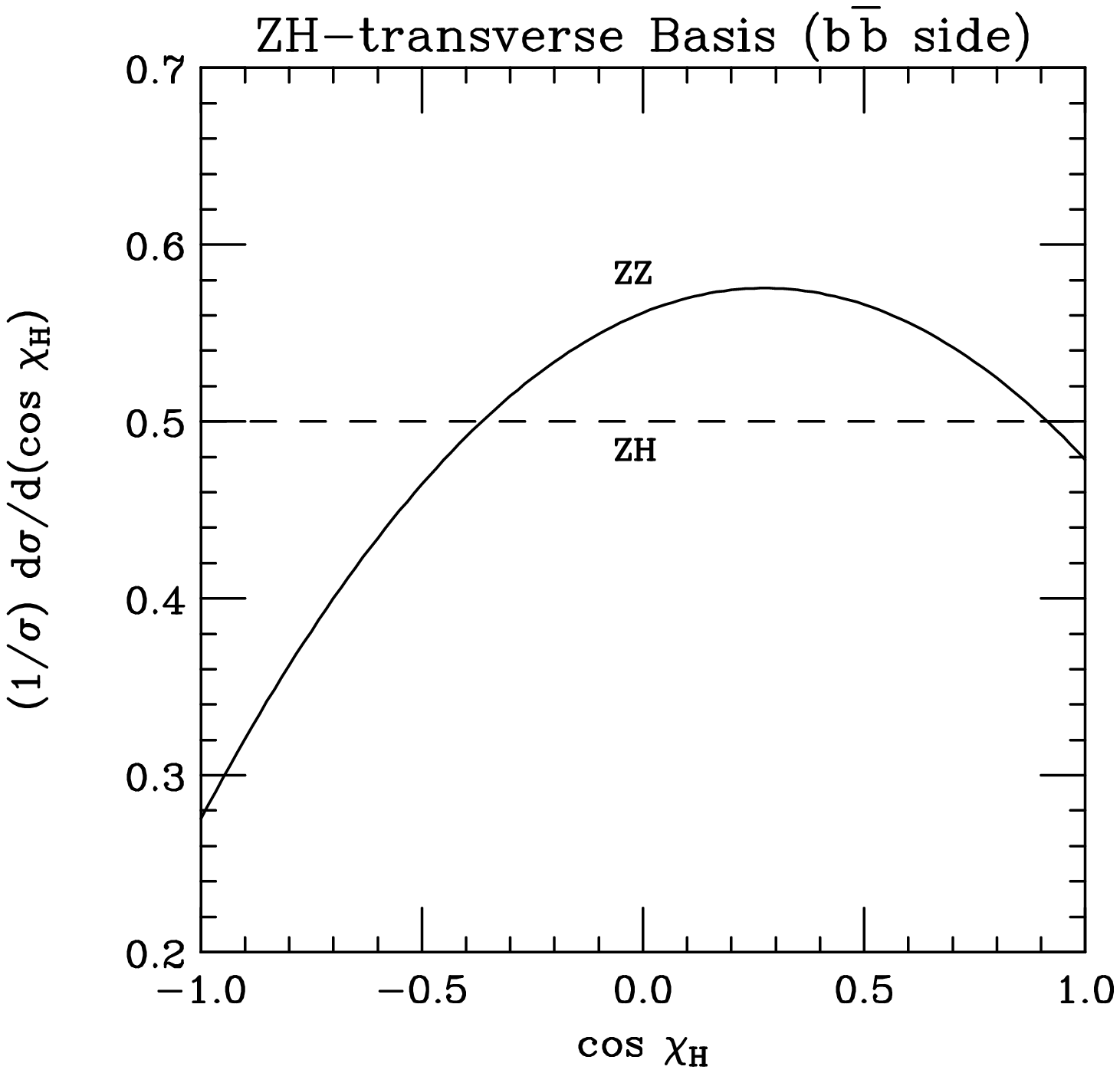}
\vspace{4.0cm}

\caption[]{One dimensional differential decay
distributions for the processes
$e^{+}e^{-} \rightarrow ZH/ZZ \rightarrow \ell\bar\ell b \bar{b}$
in the helicity and \ZHtrans\ bases, using $M_H=M_Z$.  
$\chi_Z$ ($\chi_H$) is the angle between the lepton 
($b$ jet) and the spin axis,
as viewed in the $\ell\bar\ell$ ($b\bar{b}$) rest frame.
In the absence of charge identification for the $b$'s,
these plots should be folded about $\cos\chi_H = 0$.
}
\label{ZZvsZH1D}
\end{figure}


\vspace*{1cm}

\begin{figure}[h]

\vspace*{15cm}
\includegraphics{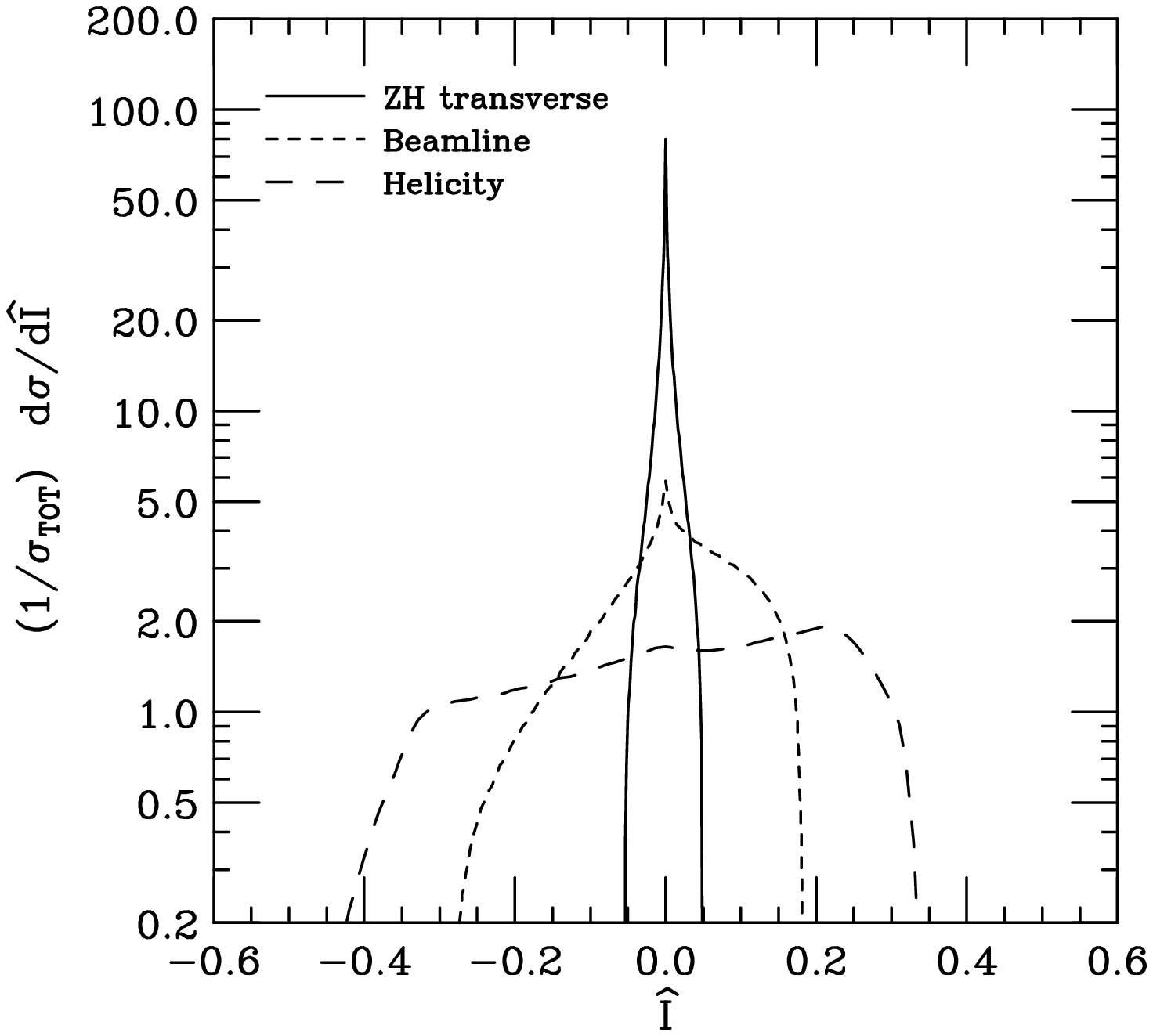}
\vspace{1.0cm}

\caption[]{The relative importance of the interference terms
in the \ZHtrans, beamline and helicity bases in 
$\eebar\rightarrow ZH \rightarrow \mu^{+}\mu^{-}b\bar{b}$
at $\protect\sqrt{s}=192\GeV$.  Plotted is the differential
distribution in $\Ihat$, the value of the interference term
normalized to the square of the total matrix element.
}
\label{ZHinf}
\end{figure}


\vspace*{1cm}

\begin{figure}[h]

\vspace*{15cm}
\includegraphics{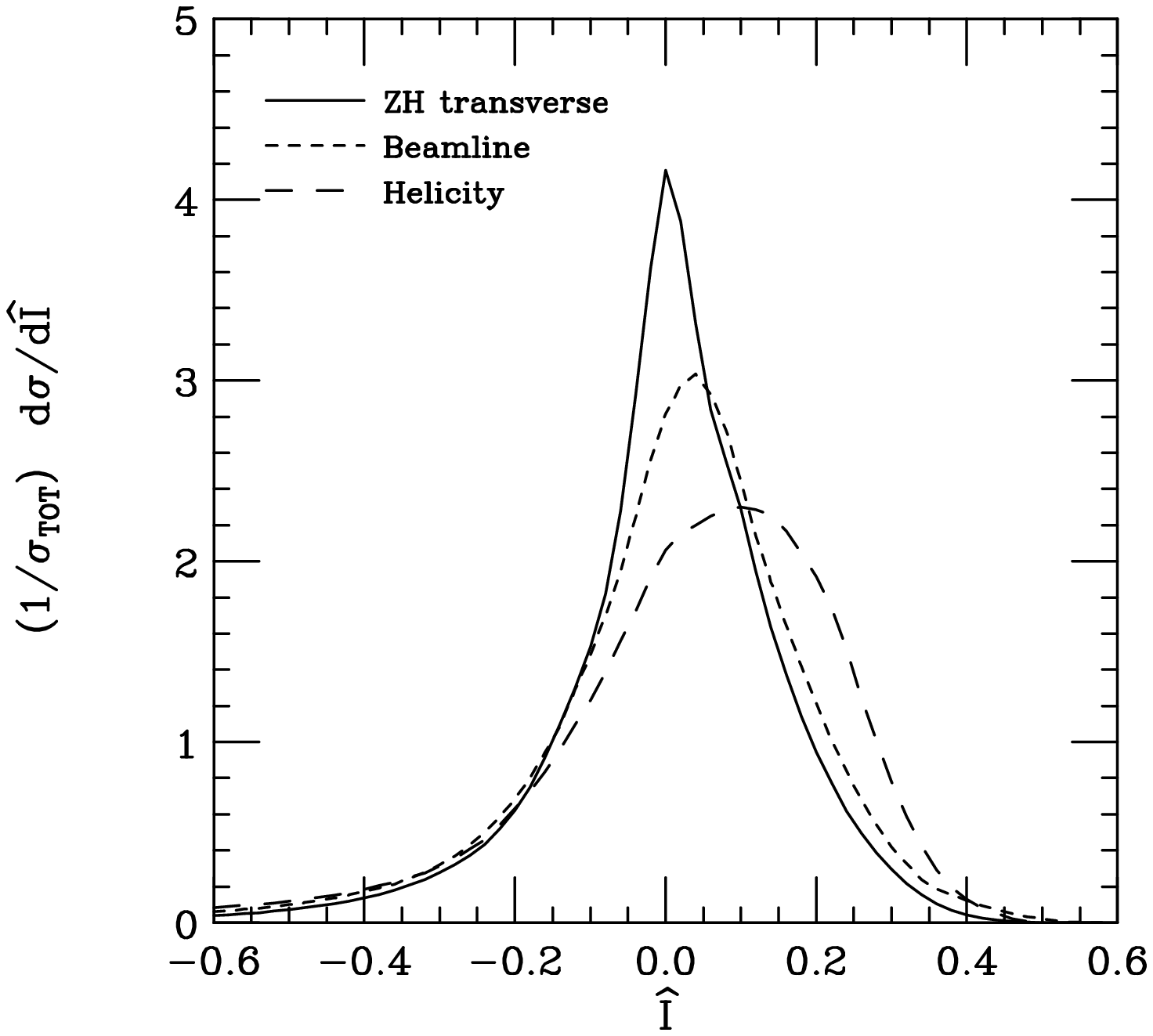}
\vspace{1.0cm}

\caption[]{The relative importance of the interference terms
in the \ZHtrans, beamline and helicity bases in 
$\eebar\rightarrow ZZ \rightarrow \mu^{+}\mu^{-}b\bar{b}$
at $\protect\sqrt{s}=192\GeV$.  Plotted is the differential
distribution in $\Ihat$, the value of the interference term
normalized to the square of the total matrix element.
}
\label{ZZinf}
\end{figure}


\vspace*{1cm}

\begin{figure}[h]

\vspace*{15cm}
\includegraphics{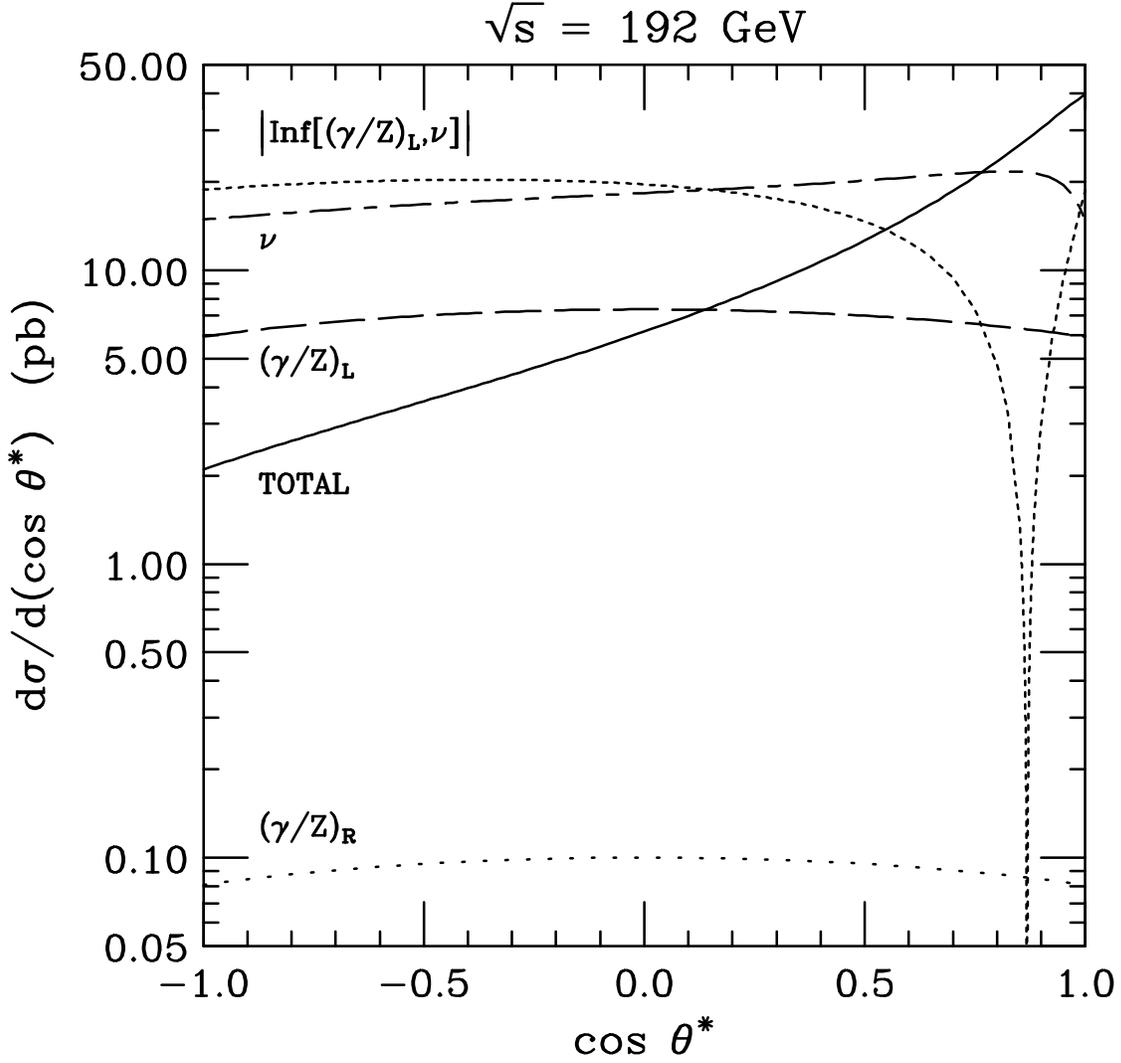}
\vspace{1.0cm}

\caption[]{Distribution in production angle of the 
$\eebar\rightarrow\WWbar$
cross section at $\sqrt{s}=192\GeV$.  Plotted
are the $(\gamma/Z)_{\rm R}$, $(\gamma/Z)_{\rm L}$ and neutrino
contributions, plus the absolute value of the interference term
between $(\gamma/Z)_{\rm L}$ and $\nu$.  To the left (right) of the 
zero at $\cos\thetas \approx 0.87$,
the interference term is negative (positive).  
}
\label{INFZplot}
\end{figure}


\vspace*{1cm}

\begin{figure}[h]

\vspace*{15cm}
\includegraphics{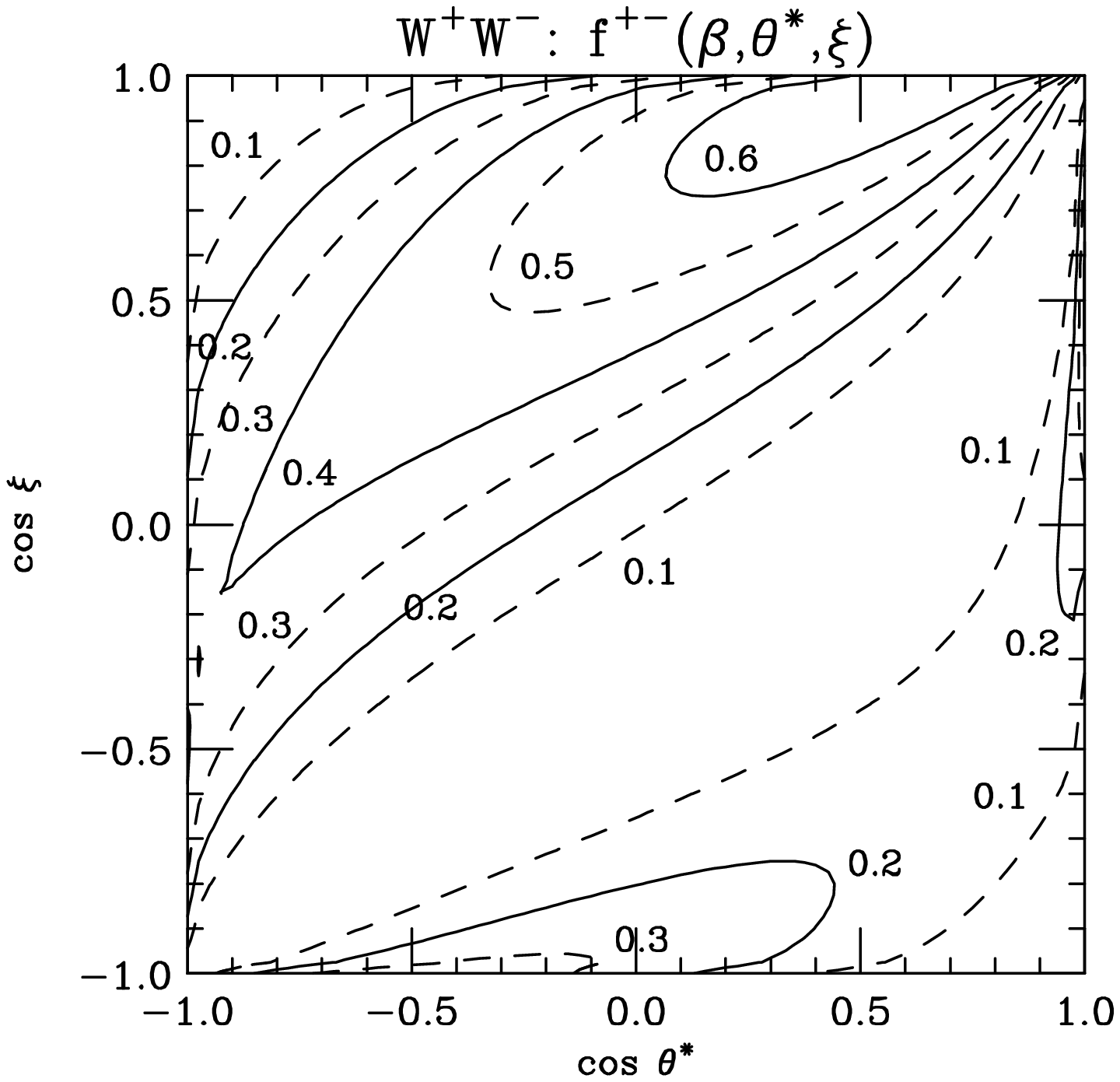}
\includegraphics{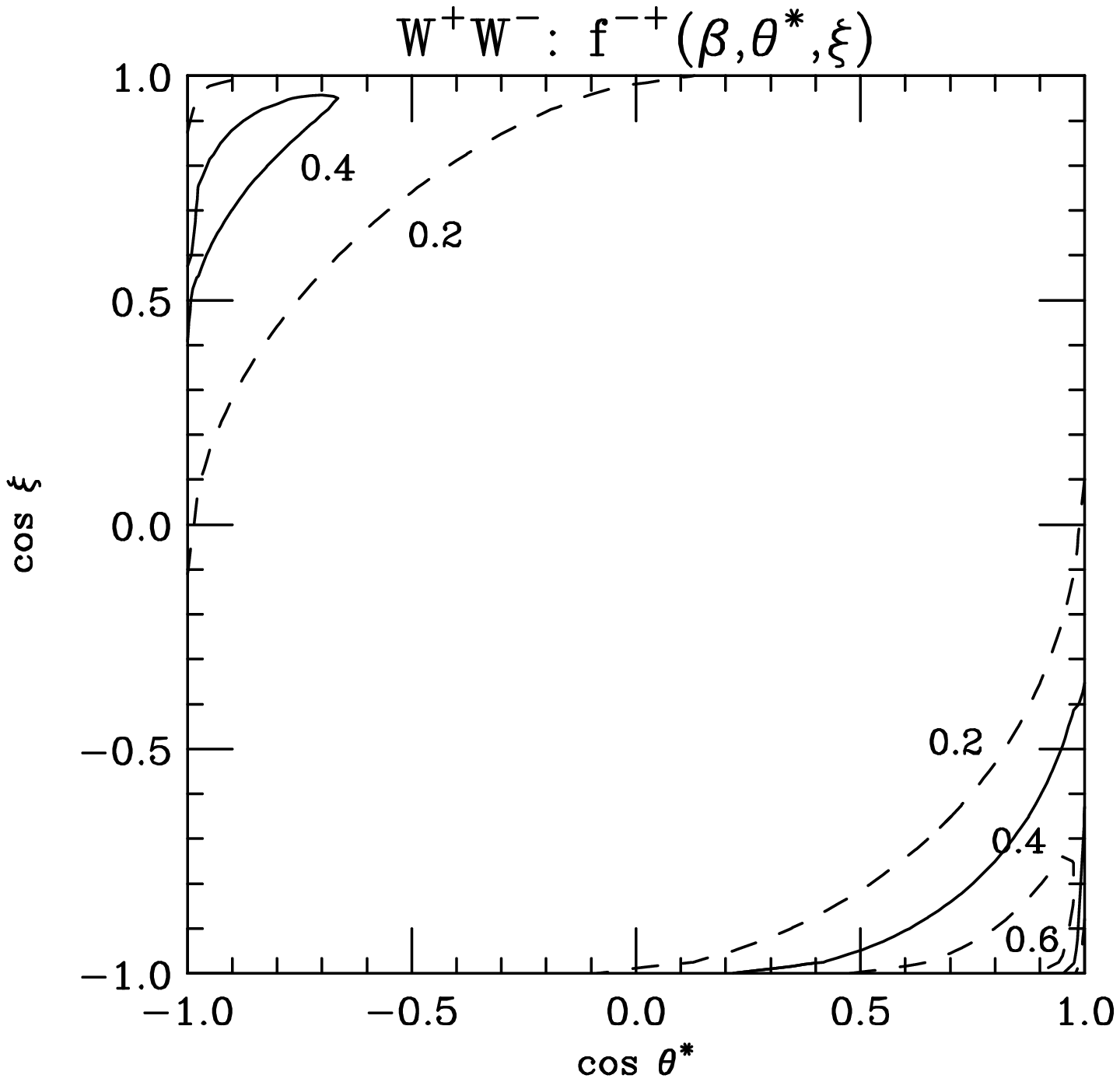}
\vspace{3.0cm}

\caption[]{Structure of the $\eebar\rightarrow \WWbar$
polarized production amplitudes in the $\cos\thetas$-$\cos\xi$
plane for $\sqrt{s}=192\GeV$ ($\beta=0.548$).  Plotted is 
$f^{\lambda\bar\lambda}(\beta,\thetas,\xi)$,
the fraction of the total amplitude squared in each spin 
configuration [see Eq.~(\protect\ref{fraction-def})].
In all of the plots, $\sin\xi \ge 0$.
}
\label{PRODcontours}
\end{figure}
\eject

\vspace*{1cm}

\begin{figure}[h]

\vspace*{15cm}
\includegraphics{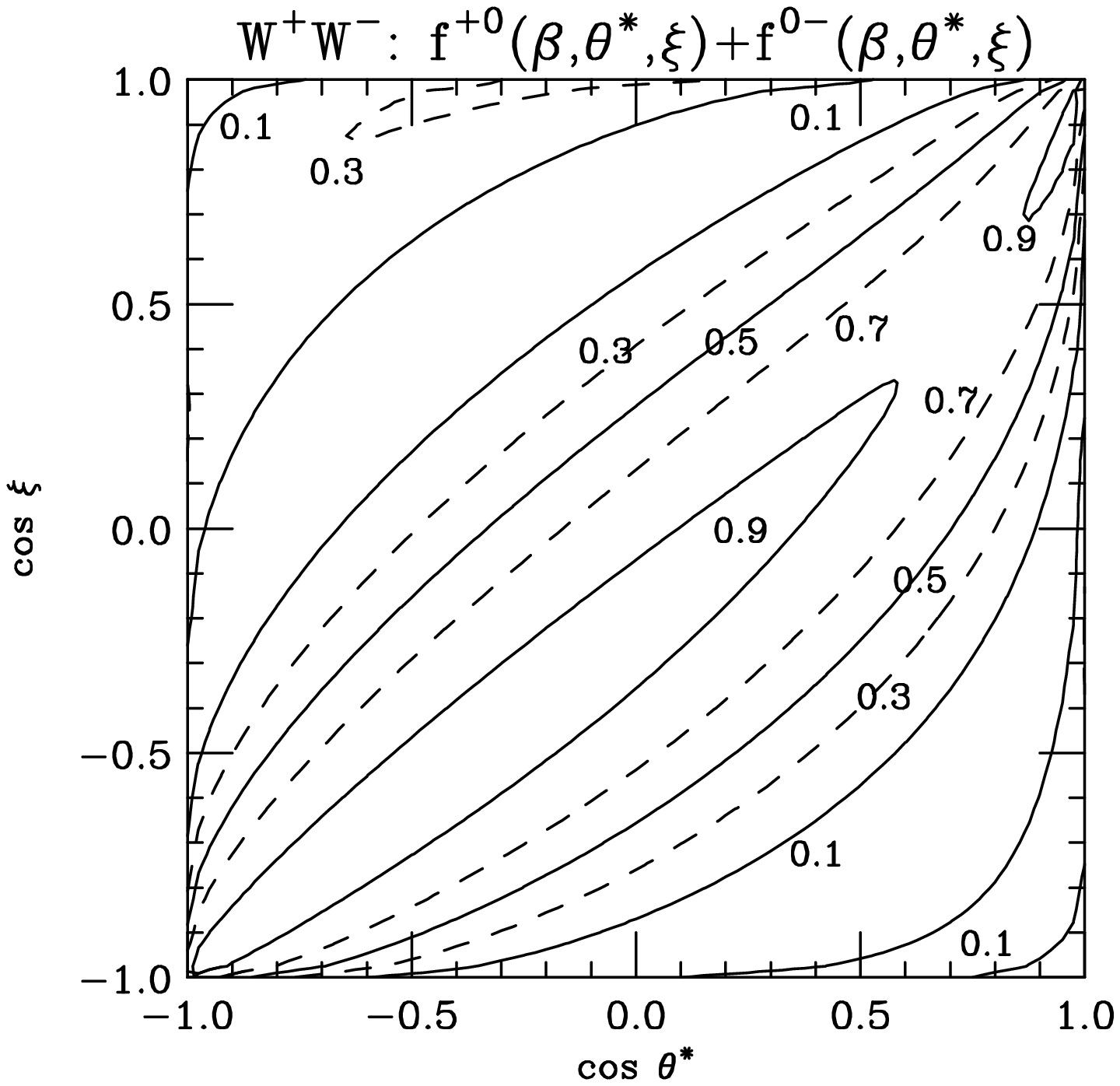}
\includegraphics{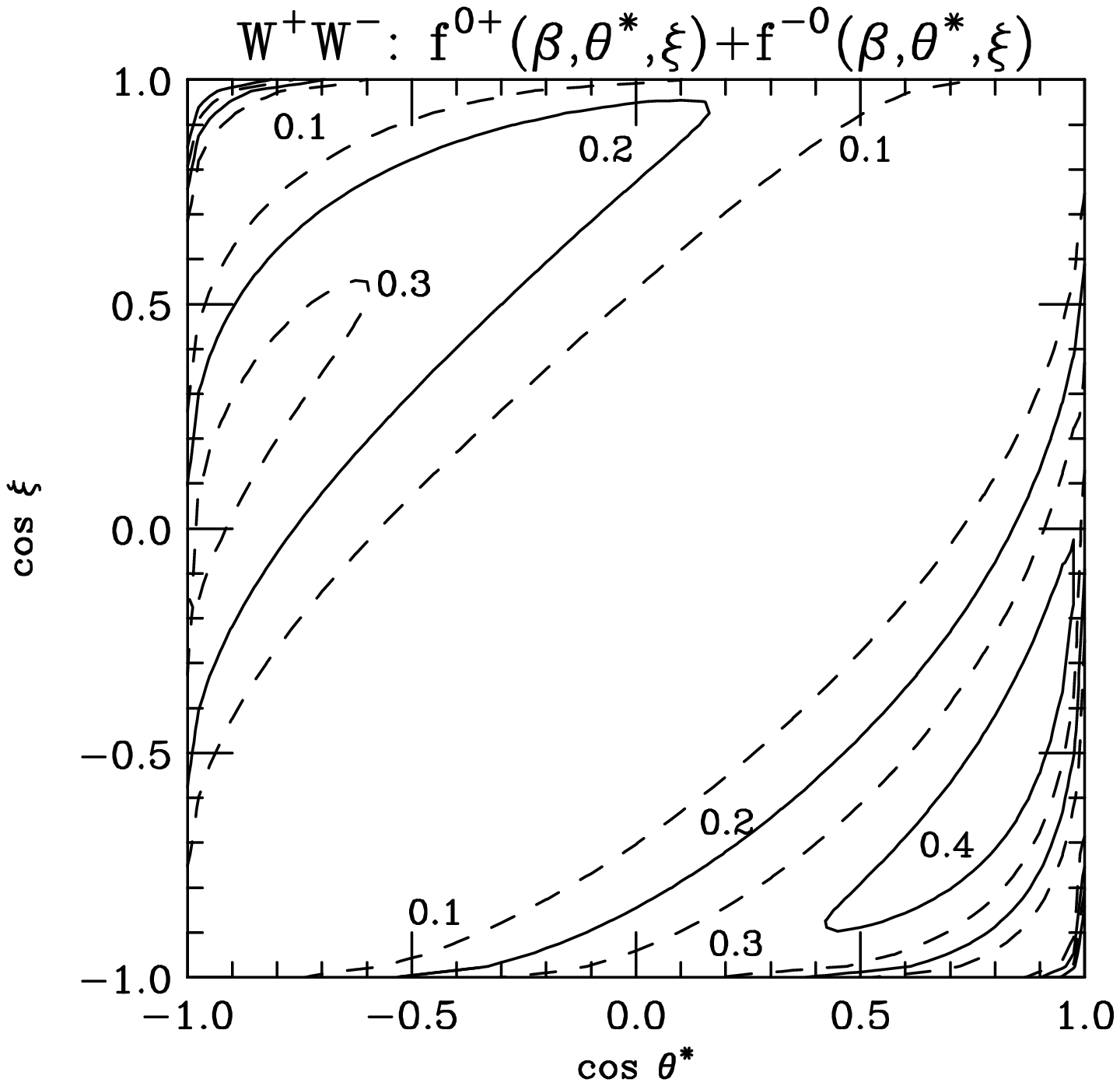}
\vspace{3.0cm}

\end{figure}
\vfill{\hfill FIG. \ref{PRODcontours}. (continued)\hfill}
\eject

\vspace*{1cm}

\begin{figure}[h]

\vspace*{15cm}
\includegraphics{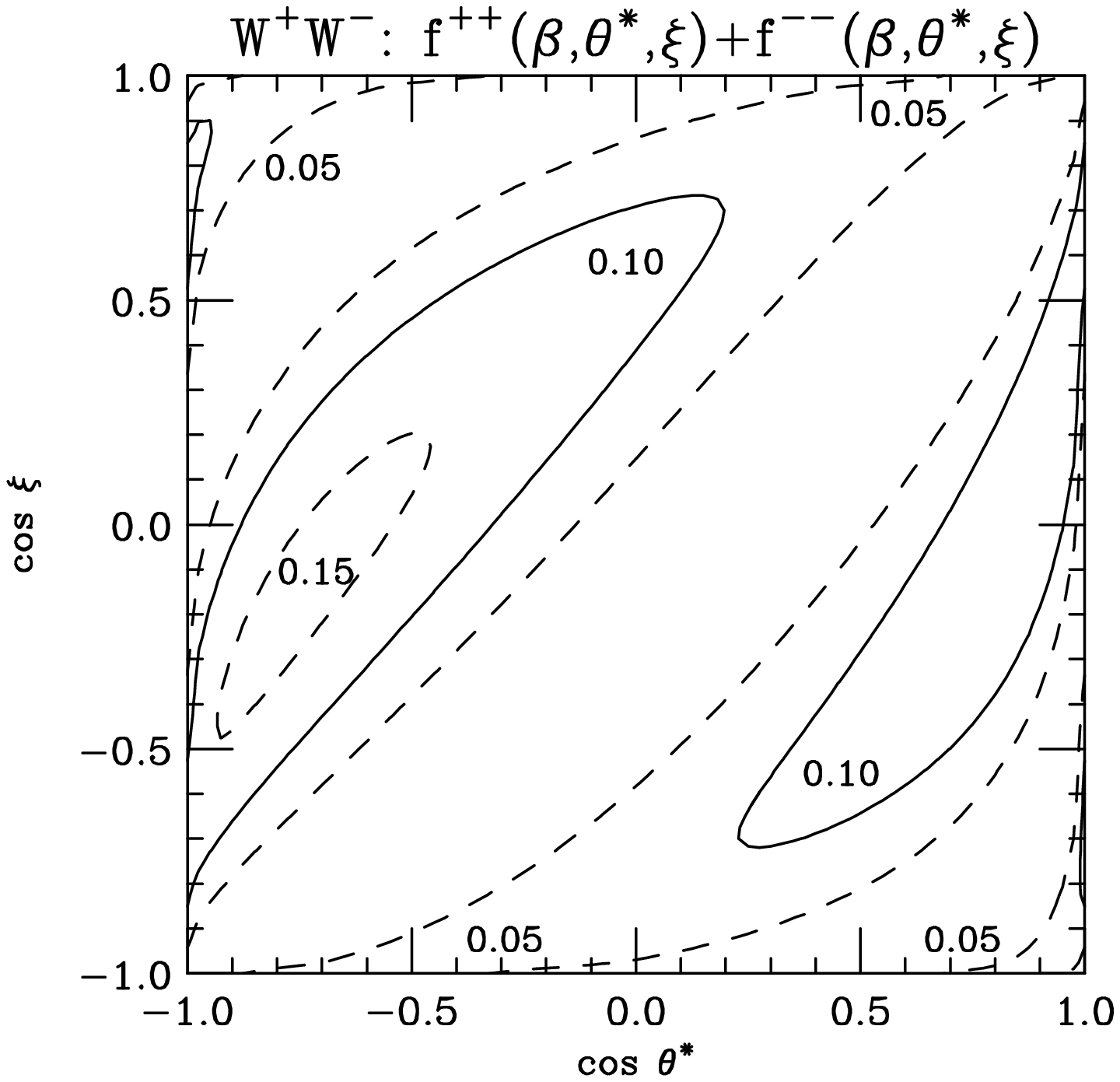}
\includegraphics{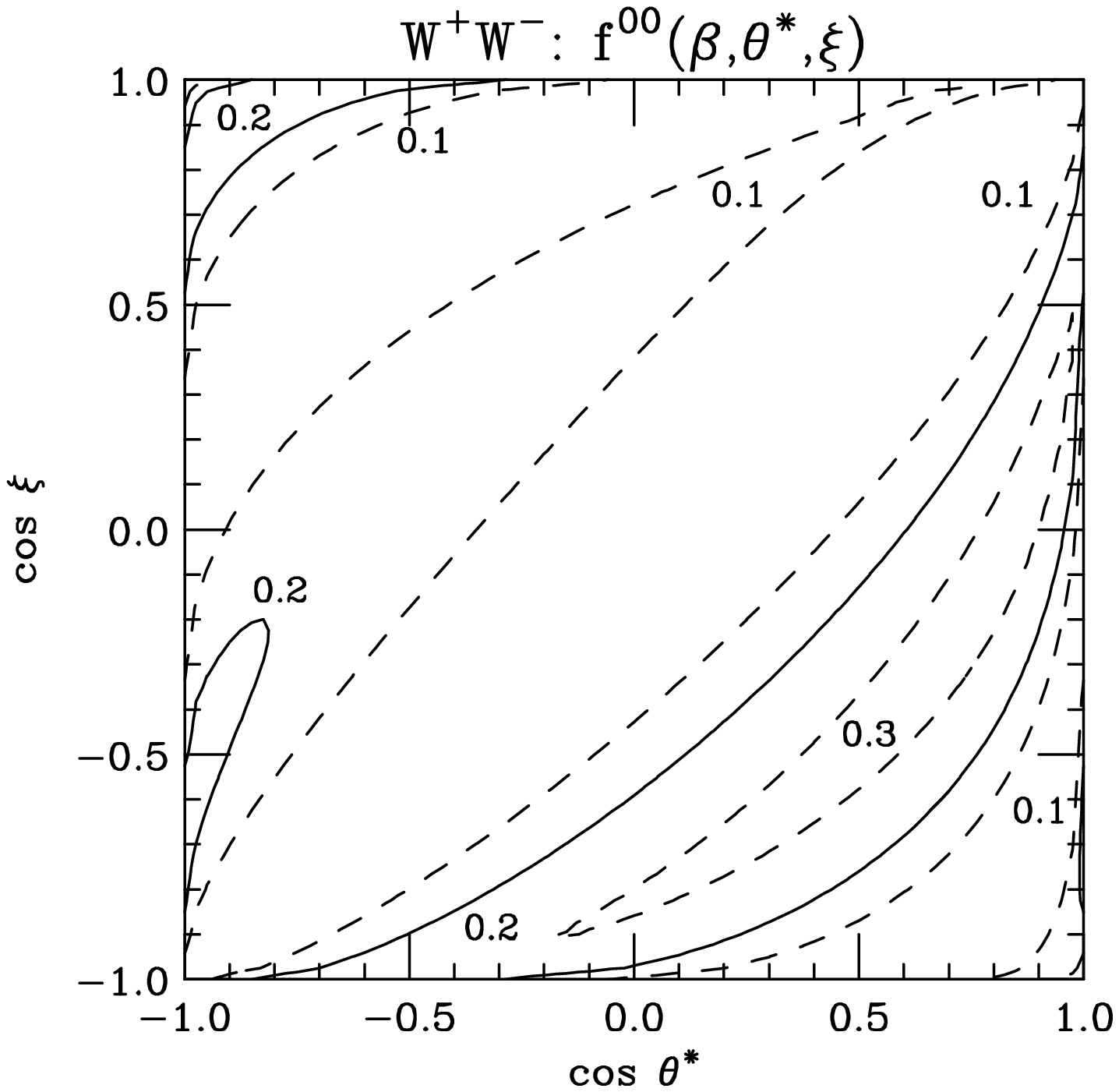}
\vspace{3.0cm}

\end{figure}
\vfill{\hfill FIG. \ref{PRODcontours}. (continued)\hfill}
\eject


\vspace*{1cm}

\begin{figure}[h]

\vspace*{15cm}
\includegraphics{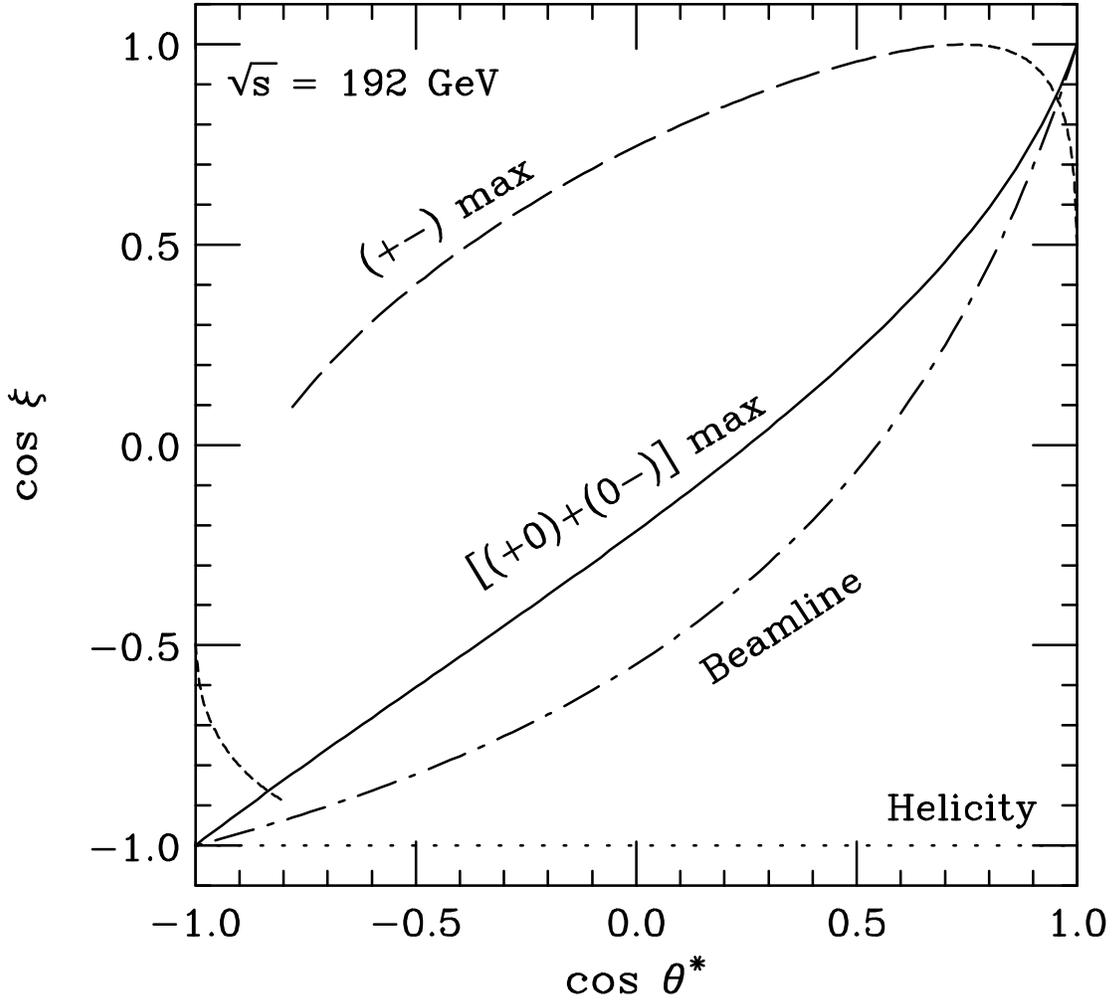}
\vspace{1.0cm}

\caption[]{The dependence of the spin angle $\xi$ on the scattering
angle $\thetas$ for the helicity, \AB,  beamline, and \A00B\ bases
for $W$ pairs produced by a 192 GeV $\eebar$ collider.
$\sin\xi\ge 0$ everywhere except for that portion of the \AB\ curve
drawn with short dashes.}
\label{WWsolns}
\end{figure}


\vspace*{1cm}

\begin{figure}[h]

\vspace*{15cm}
\includegraphics{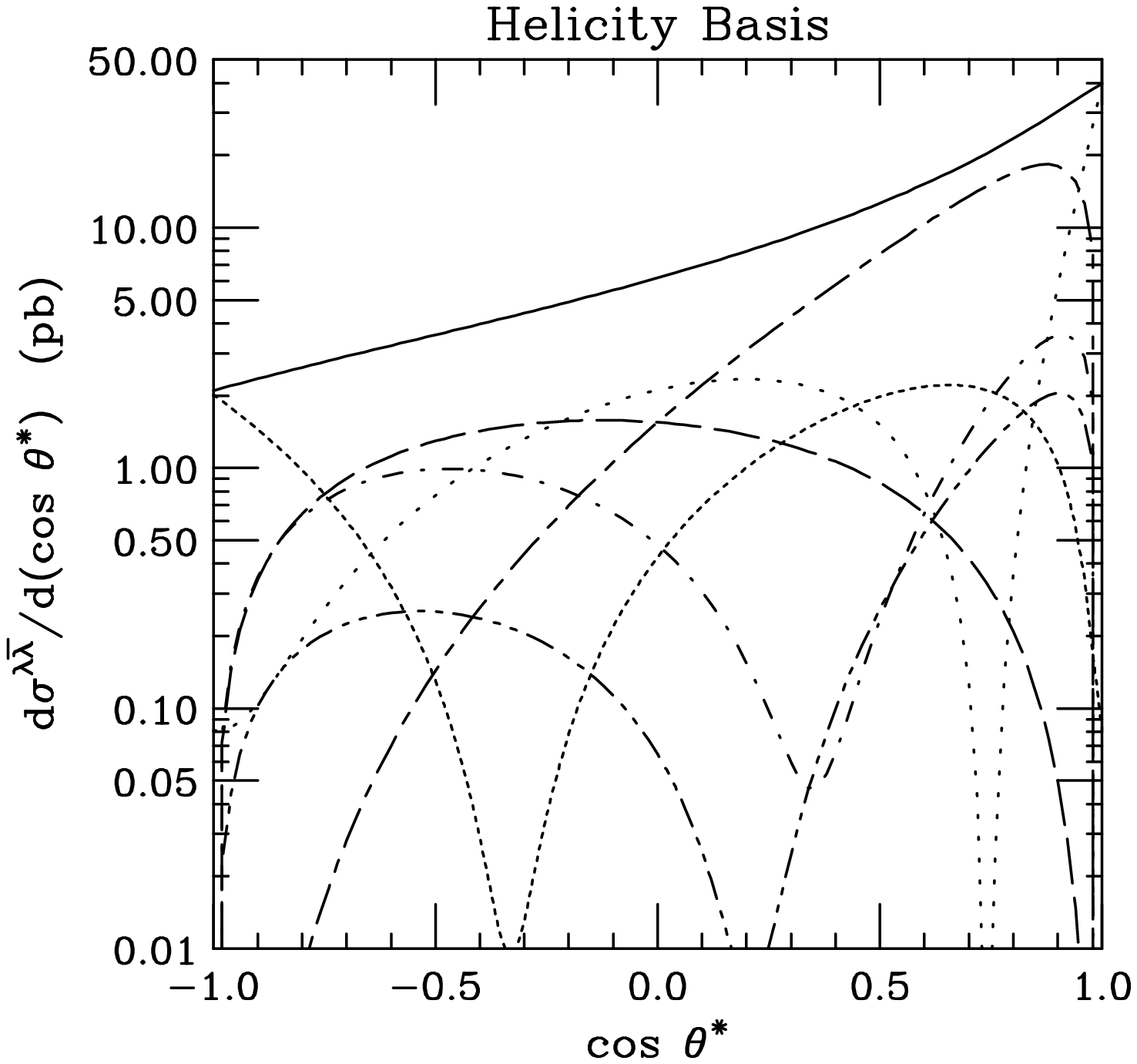}
\includegraphics{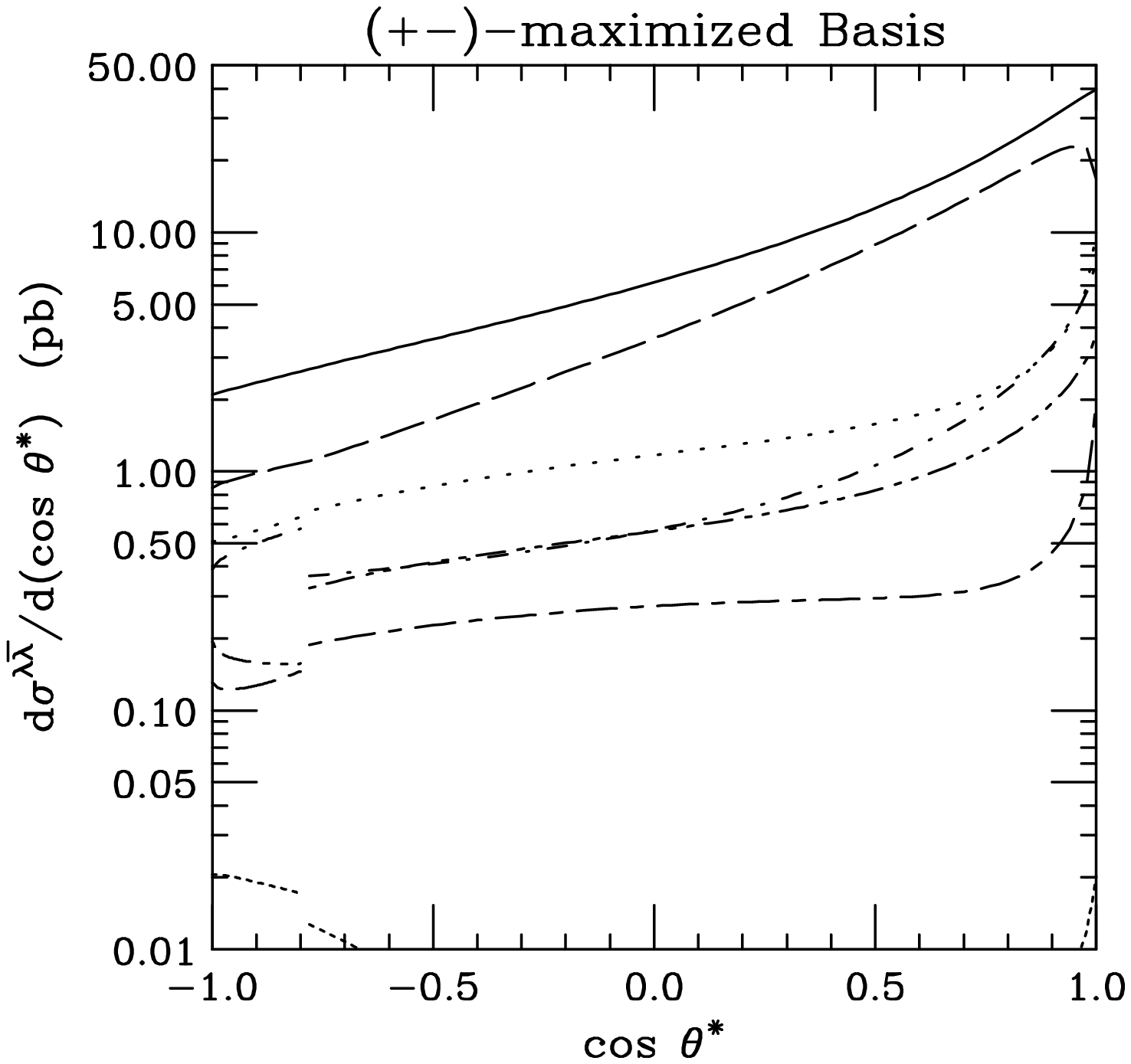}
\includegraphics{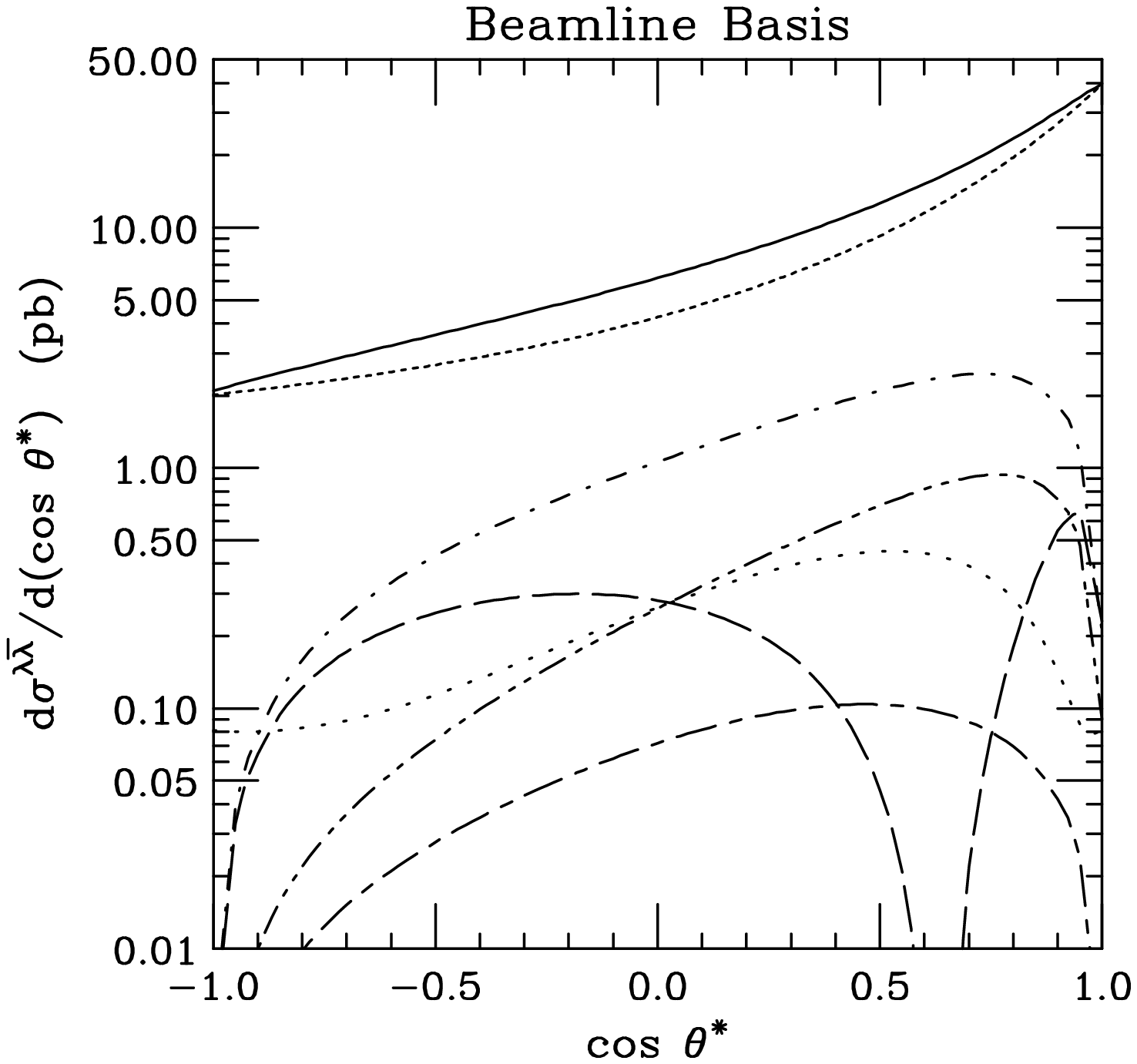}
\includegraphics{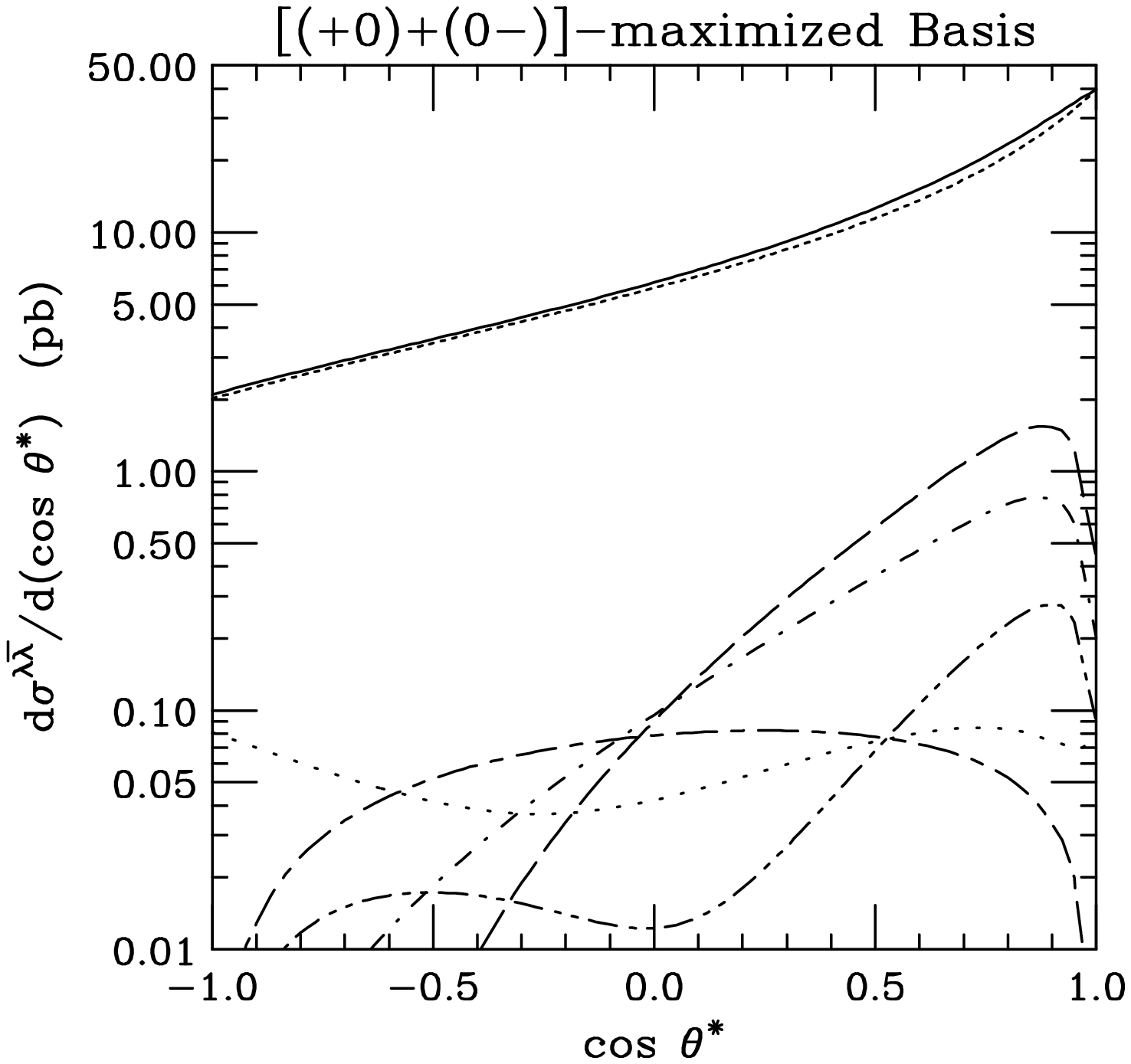}
\includegraphics{KEY7.ps}
\vspace{4.0cm}

\caption[]{Distribution in production angle of the $\eebar\rightarrow
W\overline{W}$ cross section at $\sqrt{s}=192\GeV$, broken down
into the six independent spin combinations for the helicity,
\AB, beamline, and \A00B\ bases.
}
\label{WWcomposCTH}
\end{figure}


\vspace*{1cm}

\begin{figure}[h]

\vspace*{15cm}
\includegraphics{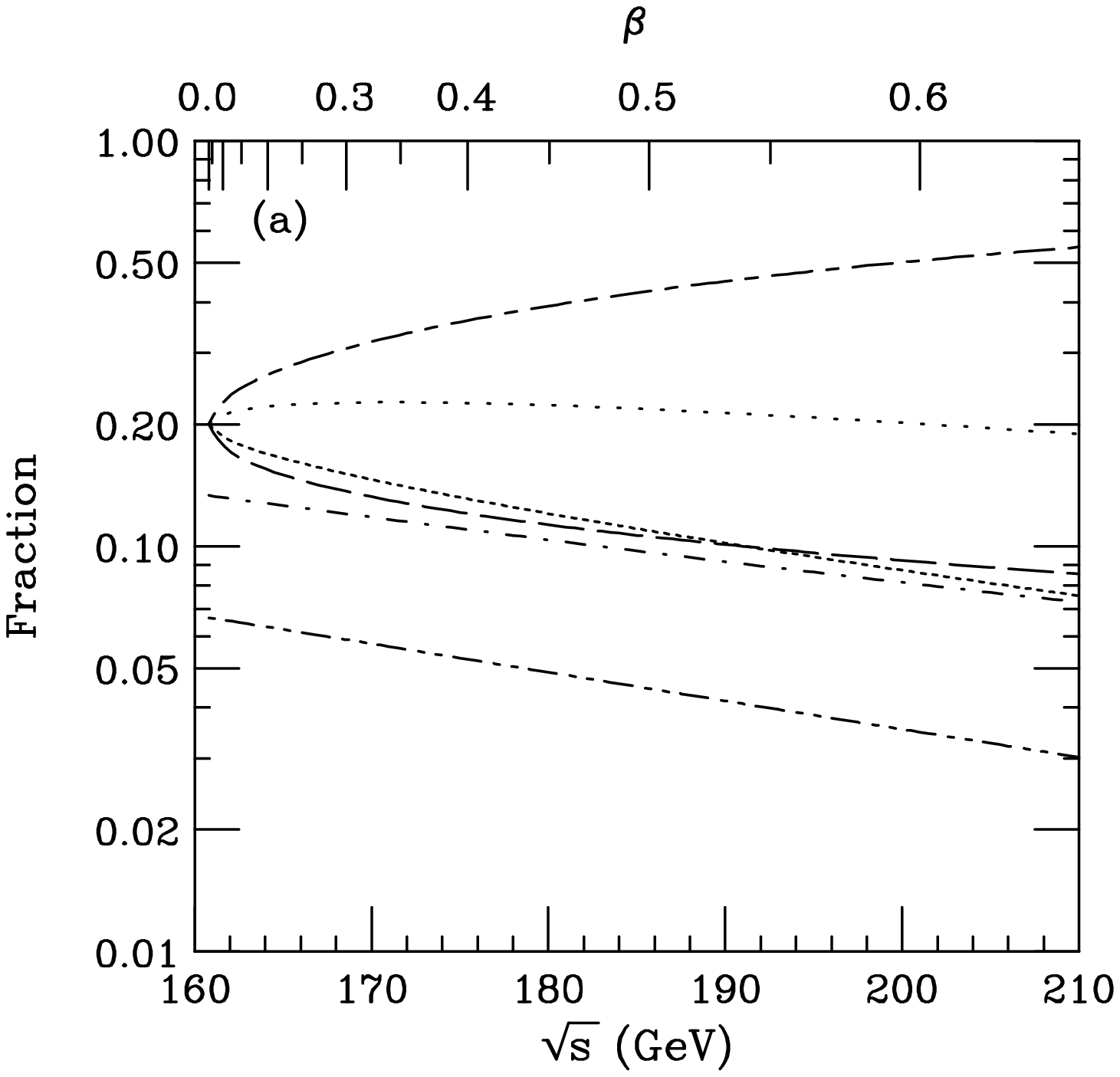}
\includegraphics{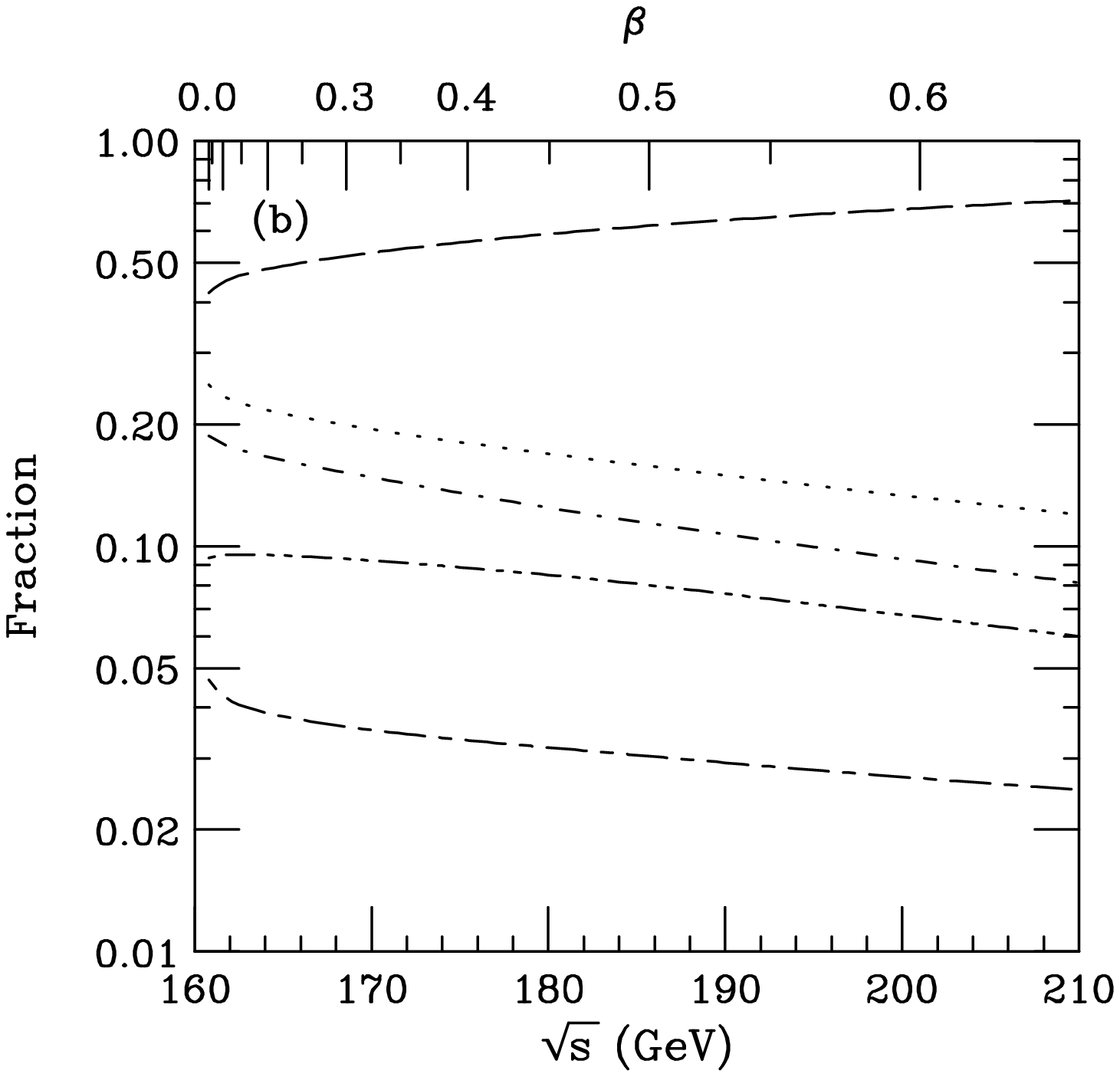}
\includegraphics{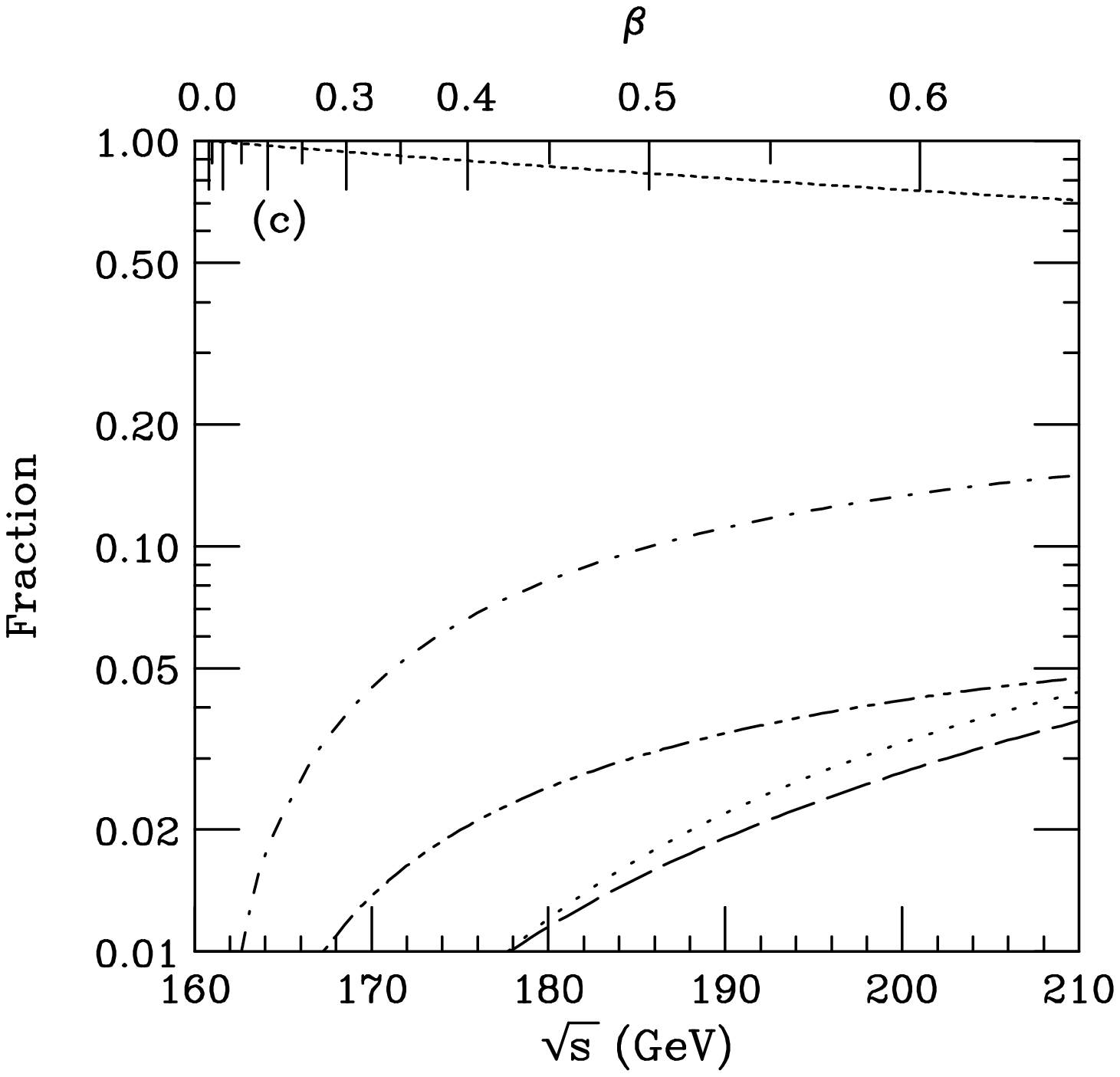}
\includegraphics{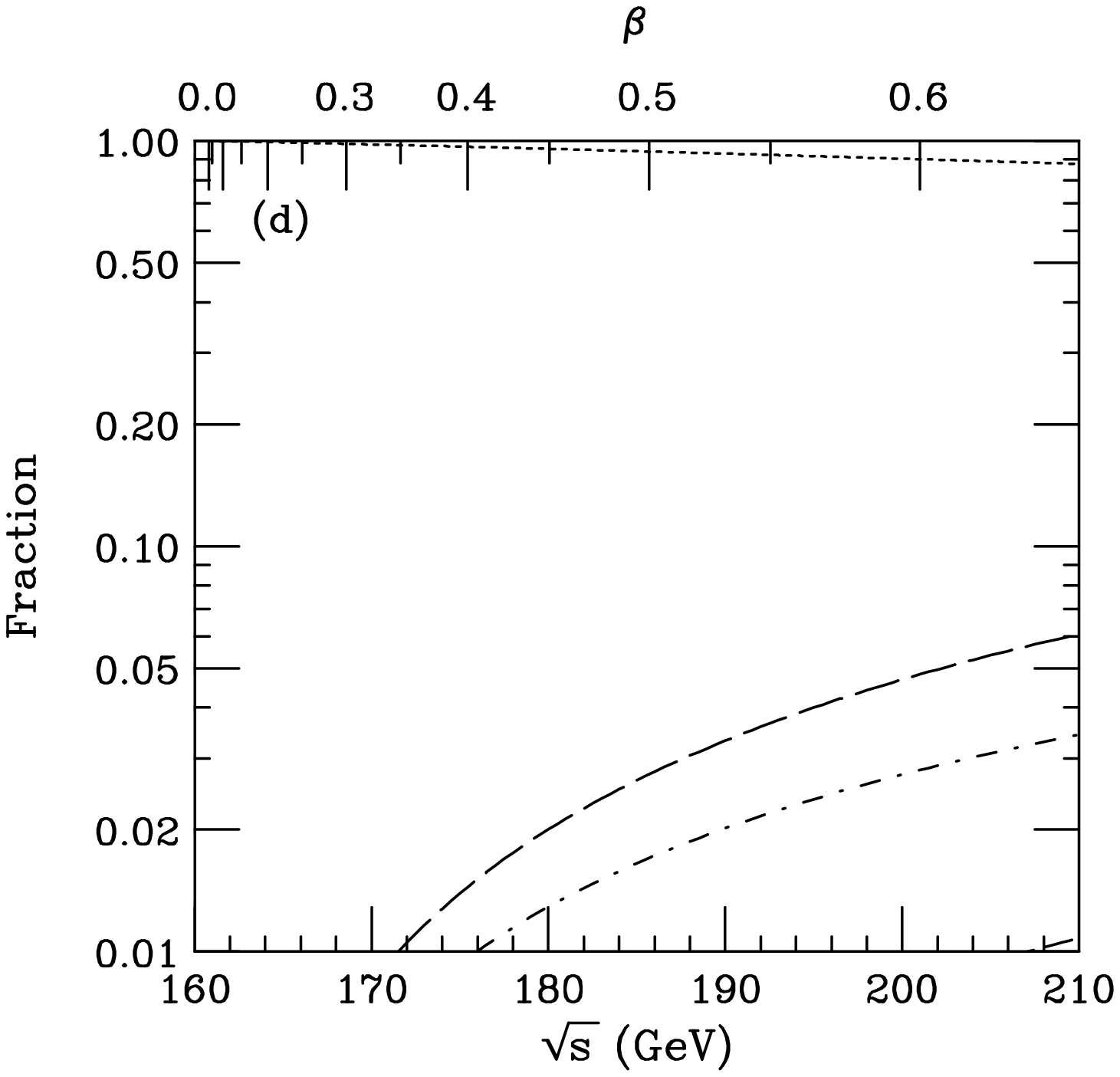}
\includegraphics{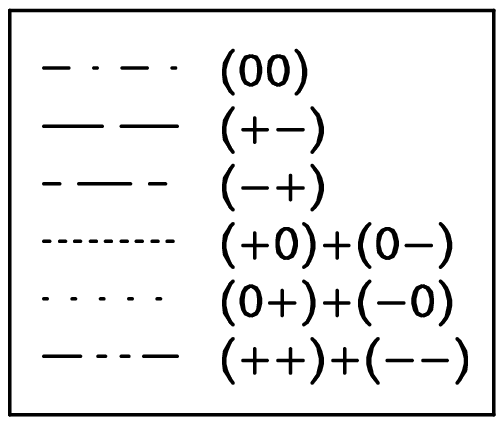}
\vspace{4.0cm}

\caption[]{Spin decomposition of the $e^{+}e^{-}\rightarrow W^{+}W^{-}$
cross section as a function of the machine energy $\sqrt{s}$.
Shown are the fractions of
the total cross section in each of the six independent spin states
for the (a) helicity , (b) \AB, (c) beamline,
and (d) \A00B\ bases.
The scale along the top of these plots measures the ZMF speed
$\beta$ of the $W$ bosons.
}
\label{WWcomposROOTS}
\end{figure}


\vspace*{1cm}

\begin{figure}[h]

\vspace*{15cm}
\includegraphics{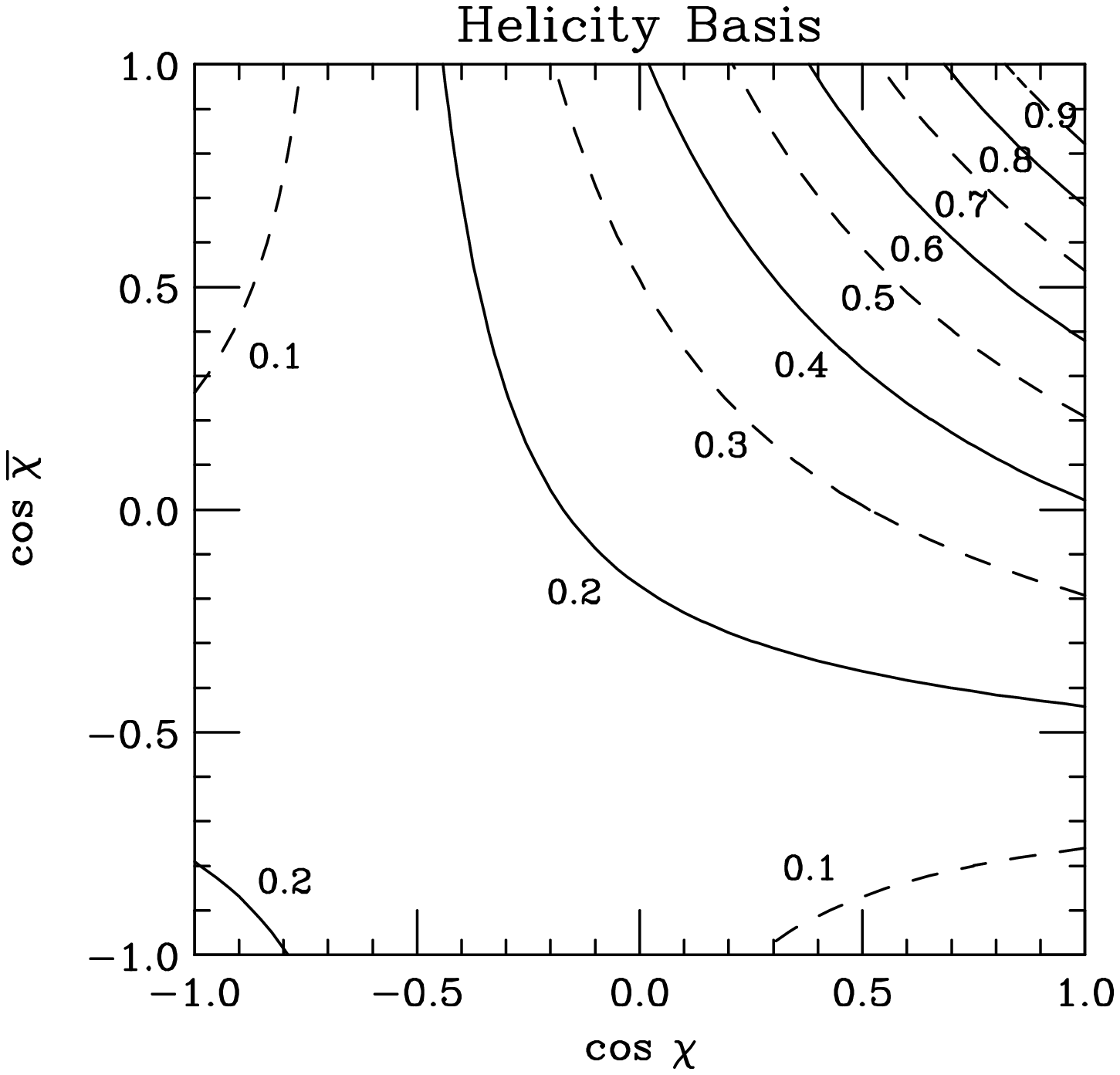}
\includegraphics{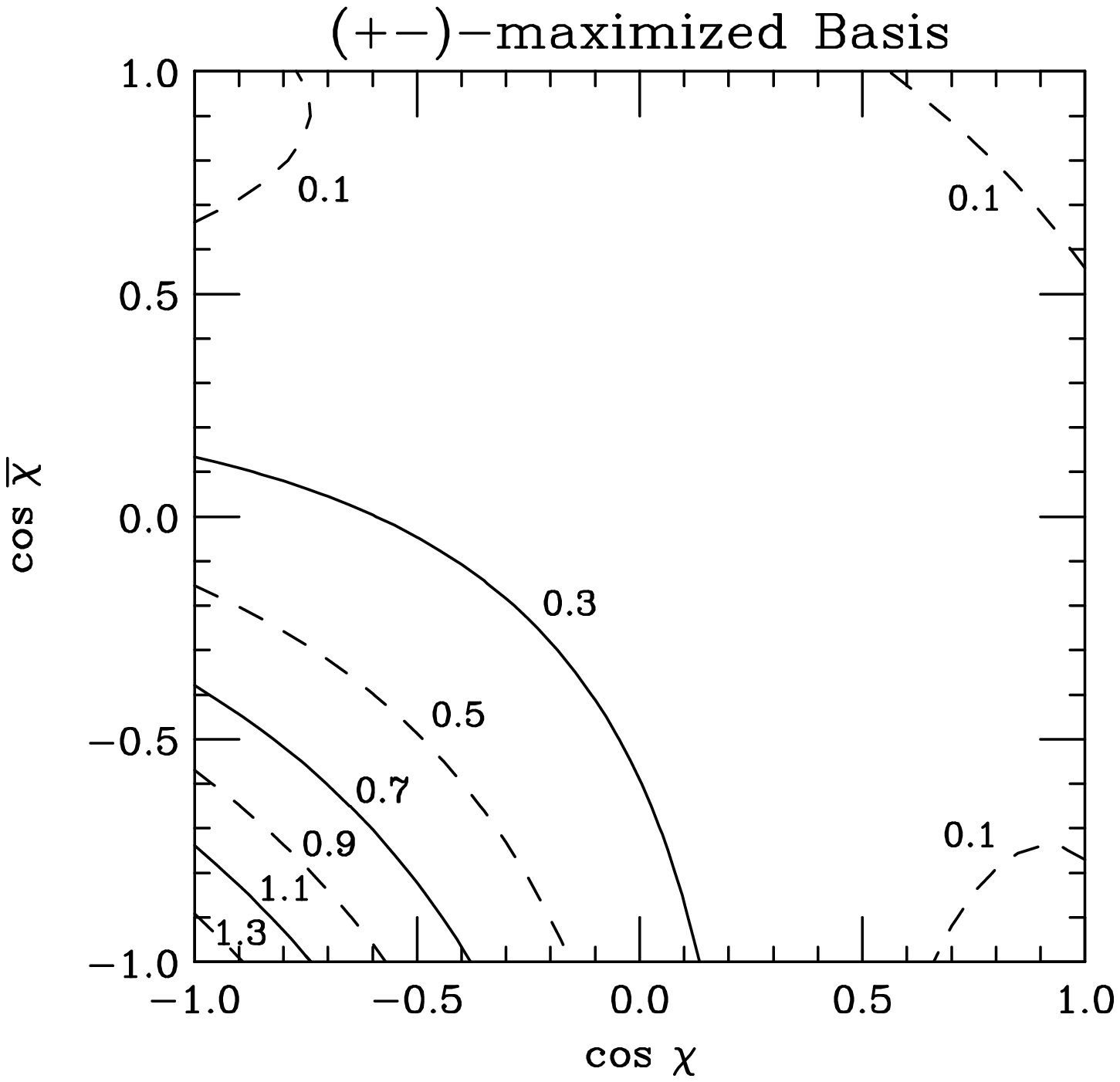}
\includegraphics{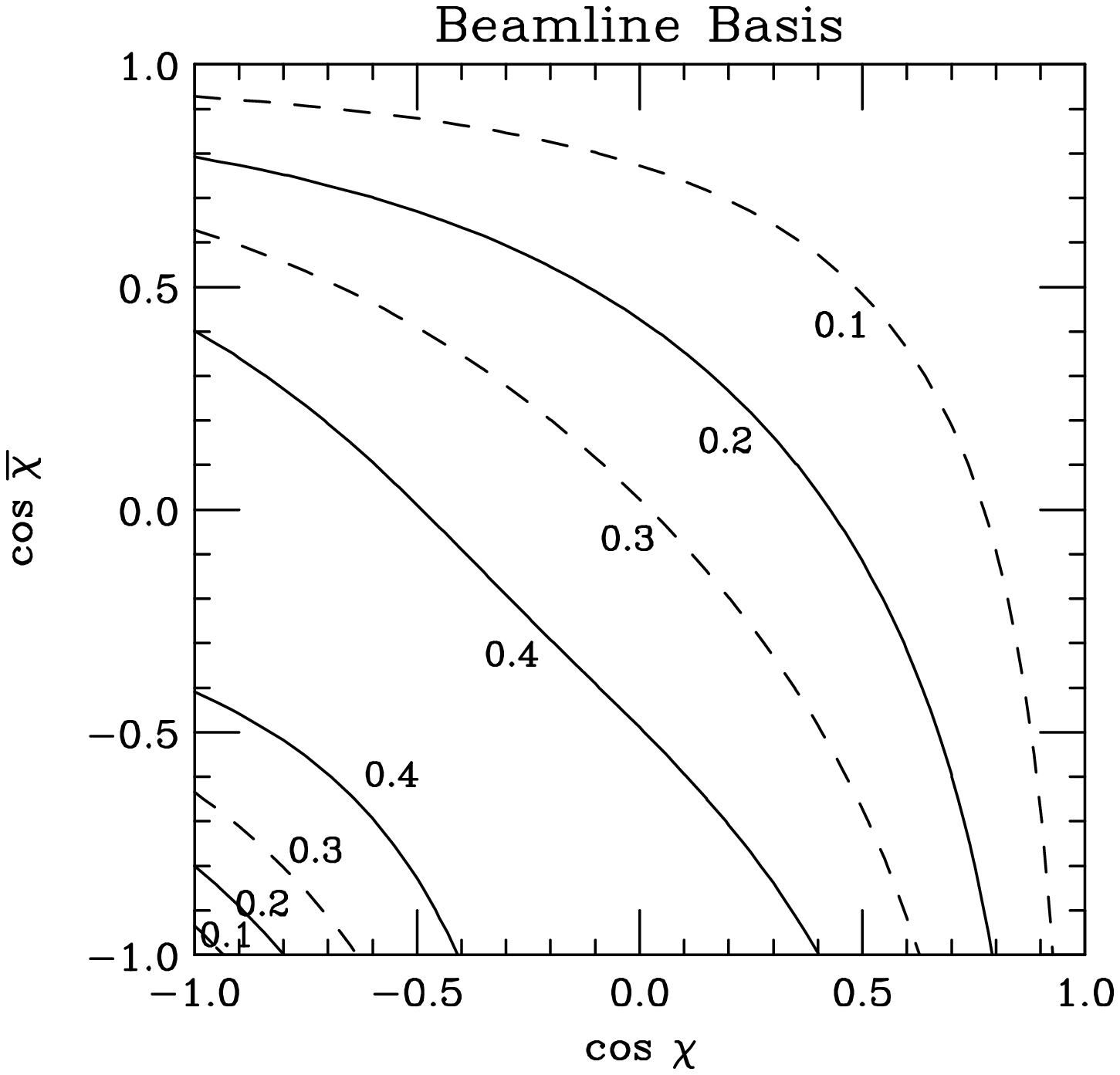}
\includegraphics{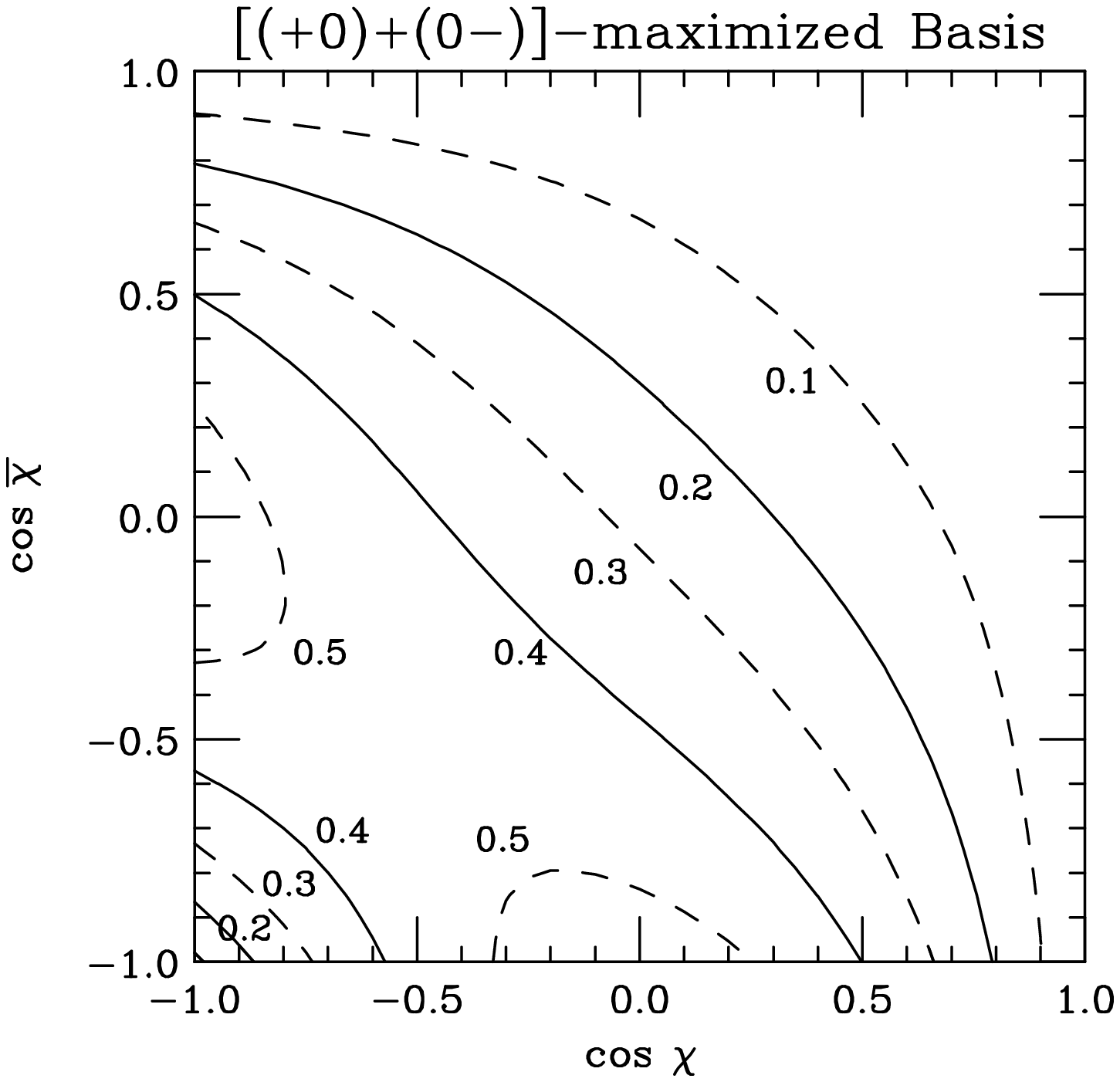}
\vspace{3.0cm}

\caption[]{Double differential decay distributions 
$(1/\sigma)\ts d^2\sigma/[d(\cos\chi)d(\cos\bar\chi)]$
for the processes
$e^{+}e^{-} \rightarrow W^{+}W^{-} 
\rightarrow \mu^{+}\nu_\mu\mu^{-}\bar\nu_\mu$
in the helicity, \AB, beamline, and \A00B\ bases.
$\chi$ ($\bar\chi$) is the angle between the $\mu^{-}$
($\mu^{+}$) and the spin axis,
as viewed in the $W^{-}$ ($W^{+}$) rest frame.
For completely uncorrelated decays,
this distribution would have a uniform value of 1/4.
}
\label{WWdecayCONTOURS}
\end{figure}


\vspace*{1cm}

\begin{figure}[h]

\vspace*{15cm}
\includegraphics{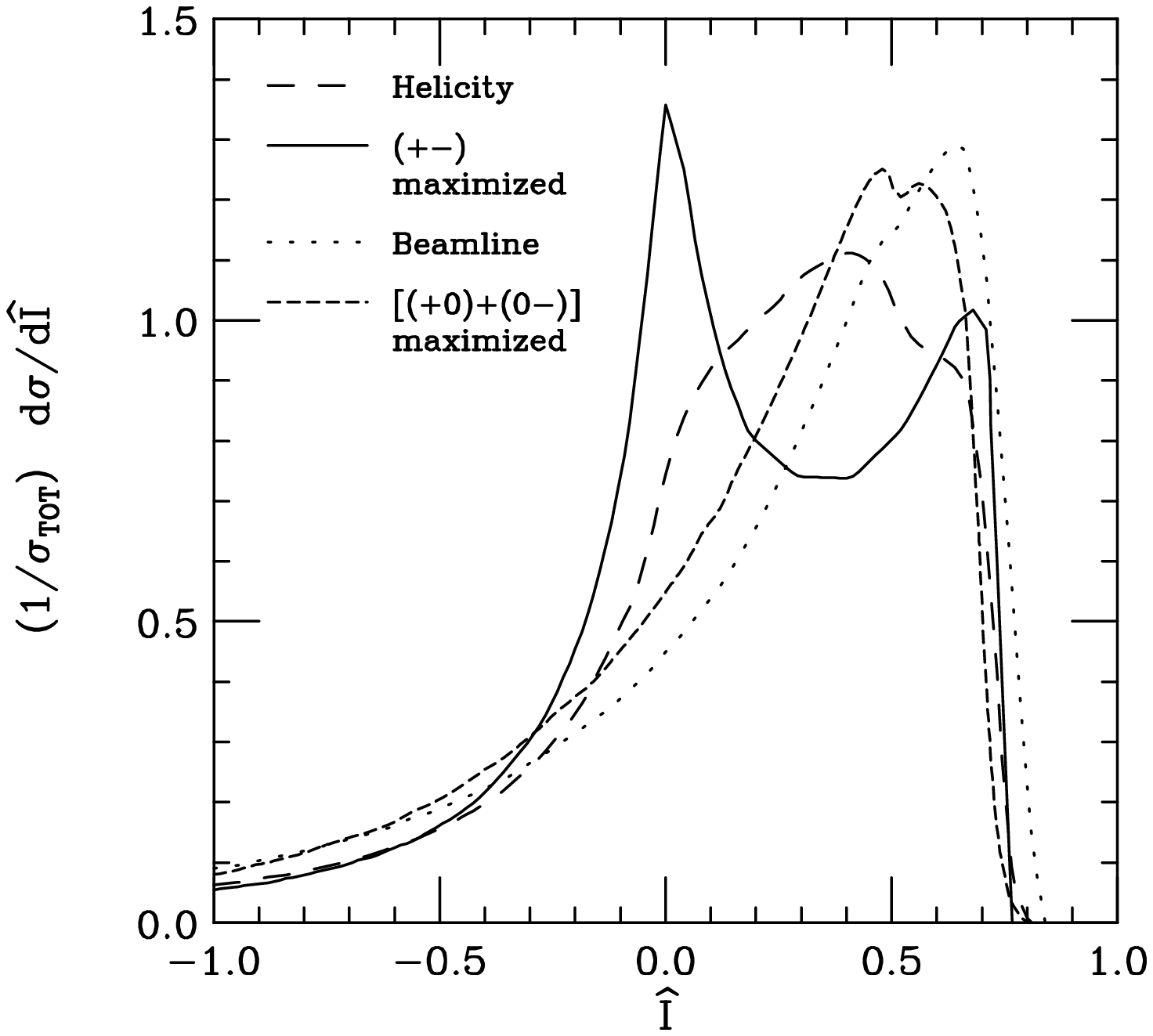}
\vspace{1.0cm}

\caption[]{The relative importance of the interference terms
in the helicity, \AB, beamline, and \A00B\ bases in 
$\eebar\rightarrow \WWbar \rightarrow 
\mu^{+}\nu_\mu \mu^{-}\bar\nu_\mu$
at $\protect\sqrt{s}=192\GeV$.  Plotted is the differential
distribution in $\Ihat$, the value of the interference term
normalized to the square of the total matrix element.
}
\label{WWinf}
\end{figure}


\begin{table}
\caption{Spin decompositions in selected
bases for $\eebar\rightarrow ZH$ at
$\protect\sqrt{s}=192\GeV$ and $M_H = M_Z$.
\label{ZH-Breakdowns}}
\begin{tabular}{ccddddc}
&Spin          & Helicity & Beamline  & \ZHtrans & \ZHlong & \\
&configuration & basis    & basis     & basis    & basis   & \\[0.05in]
\tableline
&$(0)$        &  35.6\%  &   3.5\%   &    0.0\%\tablenotemark[1] 
                                                 & 51.8\% &  \\    
&$(+)$        &  32.2\%  &  55.4\%   &   57.4\%  &  24.1\% & \\    
&$(-)$        &  32.2\%  &  41.2\%   &   42.6\%  &  24.1\% & \\    
\end{tabular}
\tablenotetext[1] {This contribution is exactly zero, by construction.}
\end{table}


\begin{table}
\caption{Spin decompositions in selected
bases for $\eebar\rightarrow ZZ$ at
$\protect\sqrt{s}=192\GeV$  
and $M_H = M_Z$.
The entries in the second part of
the table are the inclusive
fractions obtained when we sum over all possible
spins of the second $Z$.  
\label{ZZ-Breakdowns}}
\begin{tabular}{ccddddc}
&Spin          & Helicity & Beamline  & \ZHtrans & \ZHlong & \\
&configuration & basis    & basis     & basis    & basis   & \\[0.05in]
\tableline
&$(00)$        &  11.7\%  &   5.6\%   &    0.0\%\tablenotemark[1] 
                                                 &   0.1\% & \\    
&$(+-)$        &  22.6\%  &   4.2\%   &    0.3\% &  25.7\% & \\    
&$(-+)$        &  22.6\%  &   2.4\%   &    0.2\% &  25.7\% & \\    
&$(++)+(--)$   &   5.8\%  &   2.8\%   &    0.0\%\tablenotemark[1] 
                                            &   0.0\%\tablenotemark[1] & \\ 
&$(+0)+(0-)$   &  18.7\%  &  54.6\%   &   64.1\% &  24.2\% & \\    
&$(0+)+(-0)$   &  18.7\%  &  30.4\%   &   35.4\% &  24.2\% &\\    
\tableline
&$(0+)+(00)+(0-)$  &  30.4\%  &  48.1\%   &   49.7\% &  24.4\% & \\    
&$(++)+(+0)+(+-)$  &  34.8\%  &  32.9\%   &   32.4\% &  37.8\% & \\    
&$(-+)+(-0)+(--)$  &  34.8\%  &  19.0\%   &   17.9\% &  37.8\% & \\    
\end{tabular}
\tablenotetext[1] {This component is small, but nonzero.}
\end{table}


\begin{table}
\caption{Coefficients for $Z$ decay.
The numerical values are for $\SSQW=0.2315$.
The final entry is for a jet of undetermined charge.
\label{ZDecayTable}}
\begin{tabular}{ccccdcc}
&& fermion &   $\alpha_f$  &   value && \\[0.05in]
\tableline
&& $u,c$ &    
   $ \displaystyle{ {(3-4\SSQW)^2}\over{9-24\SSQW+32\SSQWSQ} } $
   & 0.834 && \\[0.13in]
&& $d,s,b$ &  
   $ \displaystyle{ {(3-2\SSQW)^2}\over{9-12\SSQW+8\SSQWSQ} } $
   & 0.968 && \\[0.13in]
&& $e,\mu,\tau$ &
   $ \displaystyle{ {(1-2\SSQW)^2}\over{1-4\SSQW+8\SSQWSQ} } $
   &  0.574 && \\[0.13in]
&& $\nu$ &       
   $1$    
   & 1.000 && \\[0.13in]
&& $j$ &       
   $1\over2$    
   & 0.500 && \\
\end{tabular}
\end{table}


\begin{table}
\caption{Breakdown of the total cross section for the process
$\eebar\longrightarrow \WWbar$
at tree level with $\protect\sqrt{s}=192\GeV$.
\label{DIAGcontribs}}
\begin{tabular}{cccddcc}
&& diagram(s)           &  contribution  &    fraction &&\\[0.05in]
\tableline
&& $(\gamma/Z)_{\rm R}$ &        0.2 pb  &       1.0\% &&\\
&& $(\gamma/Z)_{\rm L}$ &       13.8 pb  &      70.6\% &&\\
&& $\nu$                &       36.7 pb  &     188.0\% &&\\
&& Inf[$(\gamma/Z)_{\rm L}$,$\nu$]
                        &    $-$31.1 pb  &  $-$159.6\% &&\\[0.05in]
\tableline
&& total                & 19.5 pb  &          &&\\
\end{tabular}
\end{table}


\begin{table}
\caption{Spin decompositions in selected
bases for the total of all of the $\eebar\rightarrow\WWbar$ diagrams
at $\protect\sqrt{s}=192\GeV$.  
\label{WW-Breakdowns}}
\begin{tabular}{ccddddc}
&Spin          & Helicity & $(+-)$    & Beamline  & $(+0)+(0-)$ & \\
&configuration & basis    & maximized & basis     &  maximized  & \\[0.05in]
\tableline
&$(00)$        &   8.9\%  &  10.4\%   &  11.6\%   &    2.2\%    & \\
&$(+-)$        &   9.9\%  &  64.6\%   &   2.1\%   &    3.6\%    & \\
&$(-+)$        &  46.2\%  &   2.9\%   &   0.6\%   &    0.6\%    & \\
&$(++)+(--)$   &   4.0\%  &   7.5\%   &   3.6\%   &    0.6\%    & \\
&$(+0)+(0-)$   &   9.9\%  &   0.1\%   &  79.7\%   &   92.5\%    & \\
&$(0+)+(-0)$   &  21.1\%  &  14.6\%   &   2.4\%   &    0.6\%    & \\
\end{tabular}
\end{table}


\begin{table}
\caption{Spin decompositions in selected
bases for the sum of the squares of the $(\gamma/Z)_{\rm L}$
and $(\gamma/Z)_{\rm R}$ contributions 
to $\eebar\rightarrow\WWbar$ at $\protect\sqrt{s}=192\GeV$. 
As indicated in Table~\protect\ref{DIAGcontribs}, the
sum of the entries in each column is 71.6\%.
\label{gammaZ-Breakdowns}}
\begin{tabular}{ccddddc}
&Spin          & Helicity & $(+-)$    & Beamline  & $(+0)+(0-)$ & \\
&configuration & basis    & maximized & basis     &  maximized  & \\[0.05in]
\tableline
&$(00)$        &  26.8\%  &  11.0\%   &   1.9\%   &    2.1\%    & \\
&$(+-)$        &   0.0\%\tablenotemark[1]
                          &  16.5\%   &  30.6\%   &   42.1\%    & \\
&$(-+)$        &   0.0\%\tablenotemark[1]
                          &   1.2\%   &   0.6\%   &    0.6\%    & \\
&$(++)+(--)$   &   3.6\%  &   1.9\%   &   3.6\%   &    7.3\%    & \\
&$(+0)+(0-)$   &  20.6\%  &  36.6\%   &  32.5\%   &   18.6\%    & \\
&$(0+)+(-0)$   &  20.6\%  &   4.4\%   &   2.4\%   &    0.8\%    & \\
\end{tabular}
\tablenotetext[1] {This component is exactly zero.}
\end{table}


\begin{table}
\caption{Spin decompositions in selected
bases for the square of the neutrino contribution
to $\eebar\rightarrow\WWbar$ at
$\protect\sqrt{s}=192\GeV$.  As indicated in 
Table~\protect\ref{DIAGcontribs}, the sum of the entries
in each column is 188.0\%.
\label{nu-Breakdowns}}
\begin{tabular}{ccddddc}
&Spin          & Helicity & $(+-)$    & Beamline  & $(+0)+(0-)$ & \\
&configuration & basis    & maximized & basis     &  maximized  & \\[0.05in]
\tableline
&$(00)$        &  43.2\%  &   6.3\%   &   10.8\%  &    0.1\%    & \\
&$(+-)$        &   9.9\%  & 110.2\%   &   24.5\%  &   63.9\%    & \\
&$(-+)$        &  46.2\%  &   5.6\%   &   0.0\%\tablenotemark[1]
                                            &    0.0\%\tablenotemark[2]
                                                                & \\
&$(++)+(--)$   &   7.5\%  &   7.6\%   &   0.0\%\tablenotemark[1]
                                                  &    6.0\%    & \\
&$(+0)+(0-)$   &  32.0\%  &  36.5\%   & 152.7\%   &  118.0\%    & \\
&$(0+)+(-0)$   &  49.2\%  &  21.8\%   &   0.0\%\tablenotemark[1]
                                      &    0.8\%    & \\
\end{tabular}
\tablenotetext[1] {This component is exactly zero.}
\tablenotetext[2] {This component is small, but non-zero.}
\end{table}


\begin{table}
\caption{Spin decompositions in selected
bases for the the interference between the $(\gamma/Z)_{\rm L}$
and neutrino contributions to $\eebar\rightarrow\WWbar$ at
$\protect\sqrt{s}=192\GeV$.  As indicated in
Table~\protect\ref{DIAGcontribs}, the sum of the entries in
each column is $-$159.6\%.
\label{INFZ-Breakdowns}}
\begin{tabular}{ccddddc}
&Spin          & Helicity & $(+-)$    & Beamline  & $(+0)+(0-)$ & \\
&configuration & basis    & maximized & basis     &  maximized  & \\[0.05in]
\tableline
&$(00)$        &$-$61.1\% & $-$6.9\%  &  $-$1.1\% &   $-$0.0\%\tablenotemark[2] &  \\
&$(+-)$        &    0.0\%\tablenotemark[1]
                          & $-$62.2\% & $-$53.1\% & $-$102.5\% & \\
&$(-+)$        &   0.0\%\tablenotemark[1]
                          & $-$3.9\%  &     0.0\%\tablenotemark[1]
                                                  &     0.0\%\tablenotemark[2] & \\
&$(++)+(--)$   & $-$7.1\% & $-$2.0\%  &     0.0\%\tablenotemark[1]
                                                  &  $-$12.7\% &  \\
&$(+0)+(0-)$   &$-$42.8\% & $-$73.0\% &$-$105.4\% &  $-$44.1\% & \\
&$(0+)+(-0)$   &$-$48.7\% & $-$11.6\% &     0.0\%\tablenotemark[1]
                                                  &   $-$0.3\% & \\
\end{tabular}
\tablenotetext[1] {This component is exactly zero.}
\tablenotetext[2] {This component is small, but non-zero, and has
the indicated sign.}
\end{table}

\end{document}